\documentstyle[12pt]{article}
\newcommand{\be}{\begin{equation}}
\newcommand{\ee}{\end{equation}}
\newcommand{\beq}{\begin{eqnarray}}
\newcommand{\eeq}{\end{eqnarray}}

\newcommand{\k}{\kappa}

\newcommand{\g}[1]{\gamma_{#1}}
\newcommand{\rot}[1]{{\rm \cal R}({#1})}
\newcommand{\tran}[2]{{\rm \cal T}_{({#1},{#2})}}
\newcommand{\gt}[1]{\tilde{\gamma}_{#1}}

\newcommand{\Ft}{\tilde{F}}
\newcommand{\strobe}{{\rm \cal F}}
\newcommand{\strobec}[1]{{\rm \cal F}_{{#1}}}

\newcommand{\per}{B_P}

\newcommand{\ext}{\tilde{\Omega}}
\newcommand{\rtwo}{{{\rm \bf R}^2}}

\newcommand{\gef}[1]{{\Phi}_{{#1}}}
\newcommand{\ef}[1]{\phi({{#1}})}

\newcommand{\crit}[1]{E^{c}({#1})}
\newcommand{\eig}[2]{E_{{#1}}({#2})}

\newcommand{\irred}[2]{\tilde{V}_{({#1},{#2})}}
\newcommand{\irreds}[1]{\tilde{V}_{#1}}

\newcommand{\sgd}{D_4}

\newcommand{\tor}{T^2}

\newcommand{\euc}{{\rm \cal E}(2)}
\newcommand{\otwo}{{\rm  O}(2)}

\newcommand{\semi}{D_4\dot{+} T^2}

\newcommand{\rperp}{\vec{r}}

\newtheorem{prop}{Proposition}[section]
\newtheorem{theorem}{Theorem}[section]
\newtheorem{lemma}{Lemma}[section]
\newtheorem{corollary}{Corollary}[section]
\begin{document}
\baselineskip=14pt    \begin{titlepage}   \vspace*{0.0cm}

\begin{center}
{\bf $\semi$ Mode Interactions and Hidden Rotational Symmetry}\end{center}
\begin{center}
John David Crawford \\
Department of Physics and Astronomy\\
University of Pittsburgh\\
Pittsburgh, PA 15260 \\
\end{center}
\vspace{0.5cm}

\centerline{\bf ABSTRACT}
Bifurcation problems in which periodic boundary conditions or Neumann
boundary conditions are imposed often involve partial differential equations
that have Euclidean symmetry. As a result the normal form equations for the
bifurcation may be constrained by the ``hidden'' Euclidean symmetry of the
equations, even though this symmetry is broken by the boundary conditions. The
effects of such hidden rotation symmetry on $\semi$ mode interactions are
studied by analyzing when a $\semi$ symmetric normal form $\Ft$ can be extended
to a vector field $\strobe$ with Euclidean symmetry. The fundamental case of
binary mode interactions between two irreducible representations of $\semi$ is
treated in detail. Necessary and sufficient conditions are given that permit
$\Ft$ to be extended when the Euclidean group $\euc$ acts irreducibly. When the
Euclidean action is reducible, the rotations do not impose any constraints on
the normal form of the binary mode interaction.  In applications, this
dependence on the representation of $\euc$ implies that the effects of hidden
rotations are not present if the critical eigenvalues are imaginary.
Generalization of these results to more complicated mode interactions is
discussed.

\begin{center}
keywords: bifurcation, symmetry, boundary
conditions, hidden symmetry
\end{center}
\vfill
\begin{center}
November 12, 1993
\end{center}
\end{titlepage}
\tableofcontents
\newpage
\baselineskip=24pt

\section{Introduction}
When the governing equations for a system have Euclidean symmetry then
bifurcations can reflect this symmetry even when the boundary conditions for
the system are defined for a domain that is invariant only under a subgroup
$\Gamma\subset{\rm \cal E}(n)$ of the Euclidean
group. In particular, the normal form describing bifurcation from a
$\Gamma$-invariant equilibrium may have extra symmetry or special
structure not typically associated with a $\Gamma$-symmetric bifurcation. These
``non-generic'' features of the normal form can be explained by appropriately
incorporating the ``hidden'' Euclidean symmetry into the normal form
construction. Bifurcation with hidden symmetry in this sense has been
investigated in reaction-diffusion systems,\cite{fuj}-\cite{gomes} the
Kuramoto-Shivashinsky equation,\cite{ash}
and fluid dynamical models of convection and surface
waves.\cite{irw}-\cite{cgl} In addition, Field {\em et al} discuss the
occurrence of hidden symmetries in more general settings.\cite{fgs}

Recent research has concentrated on the effects of hidden translation symmetry
for systems with Neumann or Dirichlet boundary conditions at rigid sidewalls.
The effects of hidden rotation symmetry are less well understood and have not
been systematically studied either theoretically or experimentally. This paper
analyzes such hidden rotation symmetry in a specific context that occurs often
in applications: bifurcation
from a $\semi$-symmetric equilibrium described by a $\semi$-symmetric normal
form. For systems described by Euclidean symmetric equations on $\rtwo$, this
type of bifurcation can arise in at least two ways:
\begin{itemize}
\item a symmetry breaking bifurcation from an $\euc$-invariant equilibrium
often leads to spatially periodic states and one can analyze such patterns and
their
dynamics by posing the problem on a periodic lattice in the plane.\cite{dg} For
a square lattice, the symmetry of the bifurcation problem is $\semi$ and one
would like to understand the effect of hidden rotation symmetry on the $\semi$
normal form;
\item the physical system (e.g. a fluid) may occupy a volume of square cross
section with boundary conditions at the sidewalls that allow mathematical
extension of solutions to a larger square domain such that the extended
solutions are spatially periodic.\cite{craw1} The bifurcation problem for the
extended solutions then has $\semi$ symmetry. In typical examples, it is often
the presence of Neumann or Dirichlet boundary conditions in the original
problem that allows the extension to a bifurcation problem with periodic
boundary conditions and $\semi$ symmetry.  If the normal form for the extended
problem is constrained by hidden rotation symmetry this can in turn modify the
normal form for the original problem.
\end{itemize}

In such examples, one studies bifurcations described by an $\euc$-equivariant
evolution equation
\be
\frac{\partial\psi}{\partial t}=G(\psi)\label{eq:eveqn}
\ee
subject to periodic boundary conditions (PBC):
\beq
\psi(-\pi,y,t)&=&\psi(\pi,y,t)\nonumber\\
\psi(x,-\pi,t)&=&\psi(x,\pi,t)\label{eq:pbc}
\eeq
on a square domain $\ext=[-\pi,\pi]\times[-\pi,\pi]$ in $\rtwo$.
Here $\psi(\vec{r},t)$ represents the field or multiplet of fields that
describes the physical state. Any dependence of $\psi$ on additional spatial
coordinates is suppressed. In some cases, the dynamics (\ref{eq:eveqn}) may be
given by a map
$\psi_{j+1}=G(\psi_j)$
describing the state of the system at discrete times; our discussion of hidden
symmetry applies to such systems as well.

The Euclidean group $\euc\,:\,\rtwo\rightarrow\rtwo$ is generated by rotations
$\rot\theta$, reflection $\g 2$, and translations $\tran ab$ which we denote by
\be
\rot\theta\cdot(x,y)\rightarrow(x',y')\label{eq:rotdef}
\ee
where
\be
\left(\begin{array}{c}x'\\ y'\end{array}\right) =
\left(\begin{array}{cc}\cos\theta &-\sin\theta\\ \sin\theta & \cos\theta
\end{array}\right)\left(\begin{array}{c}x\\ y\end{array}\right),\nonumber
\ee
and
\beq
\g 2\cdot(x,y)&\rightarrow&(y,x),\label{eq:g2def}\\
\tran ab\cdot(x,y)&\rightarrow&(x+a,y+b).\label{eq:transdef}
\eeq
A transformation $\gamma\in\euc$ acts on the state $\psi$ in the usual way:
\be
(\gamma\cdot\psi)(\vec{r})= \psi(\gamma^{-1}\cdot\vec{r});\label{eq:eucact}
\ee
in systems described by a multiplet of fields this action may involve a
prefactor on the right hand side that rearranges the members of the multiplet.
For simplicity of notation we focus on the case without this prefactor; if it
is present the transformation in (\ref{eq:eucact}) is more complicated but the
analysis of normal forms and hidden symmetry is not affected.

The governing equation (\ref{eq:eveqn}) commutes with this $\euc$ action:
$\gamma\cdot G(\psi)=G(\gamma\cdot\psi).$
In addition, we assume that solutions satisfying PBC (\ref{eq:pbc}) may be
extended to solutions of (\ref{eq:eveqn}) on all of $\rtwo$ by periodic
replication. In this case, the PBC solutions form a subset of the solutions on
$\rtwo$ and can be precisely identified by their invariance under the
subgroup\footnote{The subgroup is specified by listing its generators.}
$\per = \{\tran {\pm 2\pi}0, \tran 0{\pm 2\pi}\}.$
It follows easily from the Euclidean symmetry of $G(\psi)$ that this subset
of PBC solutions is dynamically invariant.

The subgroup of $\euc$ that maps the set of $\per$-invariant solutions into
itself is the semi-direct product $\semi$  where $\tor$ denotes the
translations $\tran ab$ with $(a,b)$ taken to be $2\pi$-periodic variables and
$\sgd=\{\g{1},\g{2}\}$
is generated by reflection in $x$:
\be
\g{1}:(x,y)\rightarrow(-x,y),\label{eq:g1def}
\ee
and diagonal reflection $\g{2}$ (\ref{eq:g2def}).

We assume an $\euc$-invariant equilibrium $G(\psi_0)=0$ which provides a
solution for the PBC problem on $\ext$ with full $\semi$ symmetry. The linear
stability of $\psi_0$ is determined by the spectrum of $DG(\psi_0)$:
\be
DG(\psi_0)\gef{\vec{k}}=\lambda(|\vec{k}|)\gef{\vec{k}}.\label{eq:spec}
\ee
The eigenvector
\be
\gef{\vec{k}}(\vec{r})=\ef{|\vec{k}|}\,e^{i\vec{k}\cdot\vec{r}}\label{eq:gefdef}
\ee
satisfies PBC on $\ext$ if $\vec{k}$ has integer components:
$(k_x,k_y)\in{\cal Z}^2$. The Euclidean invariance of the equilibrium $\psi_0$
implies that $DG(\psi_0)$ commutes with the action of $\euc$,
$\gamma\cdot DG(\psi_0)=DG(\gamma\cdot\psi_0);$
hence the eigenspaces for PBC define representations of $\semi$.

At a bifurcation from $\psi_0$, the normal form,
$\Ft: \crit\ext\rightarrow\crit\ext,$
commutes with the $\semi$ representation carried by $\eig c\ext$, the
center subspace defined by the critical modes. The $\semi$ equivariance
of $\Ft$ is the most
readily noted consequence of the $\euc$ symmetry of $G(\psi)$ and $\psi_0$. In
addition, $\Ft$ may be further constrained by symmetries in $\euc-\semi$ that
do {\em not} leave $\crit\ext$ invariant. Our problem is to characterize these
these constraints.

The nature of these more subtle constraints appears when we consider the form
of $\euc$-equivariant vector fields,
$\strobe:\crit\rtwo\rightarrow\crit\rtwo,$
on the infinite-dimensional space $\crit\rtwo$ defined by the critical modes
$\gef{\vec{k}}(\vec{r})$ without regard to PBC. This space carries a
representation of $\euc$ induced from the action (\ref{eq:eucact}) on
functions. The space $\crit{\ext}$ of the critical modes subject to PBC is a
subspace of $\crit{\rtwo}$. The normal form $\Ft$ for the PBC bifurcation
problem is defined on $\crit{\ext}$; the consistency requirement that $\Ft$
reflect all constraints on the PBC problem due to the $\euc$ symmetry of
$G(\psi)$ is formalized by requiring that $\Ft$ arise as the restriction of a
Euclidean symmetric vector field. More precisely, there should exist an
$\euc$-symmetric $\strobe:\crit{\rtwo}\rightarrow \crit{\rtwo}$
such that
\be
\Ft=\strobe|_{\crit{\ext}}.\label{eq:pbcres}
\ee
This requirement ensures that $\Ft$ is $\semi$-equivariant, but it can impose
further constraints on $\Ft$. These additional constraints, if present, are
said to reflect hidden rotation symmetry in $\Ft$ arising from the manifest
rotation symmetry of $\strobe$. Because the effect of the $\semi$ symmetry
depends on the representation carried by $\crit{\ext}$, the role of hidden
rotation symmetry  depends on the specific representations of $\euc$ and
$\semi$ under consideration.

For $\semi$, the irreducible representations $\irred{l}{n}$ may be classified
by two mode numbers $l\geq n\geq0$ physically representing the wave vector
components of $\vec{k}=(l,n)$ for an eigenvector. Thus each representation is
characterized in part by $\k=|\vec{k}|$, the length of the associated wave
vectors,
\be
\k^2=l^2+n^2.\label{eq:kapdef}
\ee
The center subspace for PBC decomposes into a finite direct sum of these
irreducible representations,
\be
\crit{\ext}=\irred{l_1}{n_1}\oplus\irred{l_2}{n_2}\cdots
\oplus\irred{l_j}{n_j};\label{eq:irrsum}
\ee
in the simplest bifurcations $\crit{\ext}$ is $\semi$-irreducible and there is
only one term in this sum. In these cases, it is known that the rotations do
not constrain the form of $\Ft$.\cite{craw2}
\begin{theorem} Assume $\crit{\ext}$ carries an irreducible representation of
$\semi$, and
$\Ft:{\crit{\ext}}\rightarrow{\crit{\ext}}$
is a $\semi$-equivariant vector field, then there is an $\euc$-equivariant
vector field
$\strobe:{\crit{\rtwo}}\rightarrow{\crit{\rtwo}}$
such that
$\Ft = \strobe |_{{\crit{\ext}}}.$
\end{theorem}

The goal of this paper is to generalize this theorem to bifurcations in which
the $\semi$ representation carried by ${\crit{\ext}}$ is reducible, i.e. when
there is a mode interaction.\cite{craw} There are many possible mode
interactions involving different
representations (\ref{eq:irrsum}), however to characterize the influence of the
hidden rotation symmetry we need only focus on  the set of values
$\{\k_1,\k_2,\ldots,\k_j\}$ defined by (\ref{eq:kapdef}) - (\ref{eq:irrsum})
and the structure of the $\euc$ representation on $\crit{\rtwo}$. For a mode
interaction (\ref{eq:irrsum}) defined by a specific reducible representation of
$\semi$, let
\be
\crit{\rtwo}=\eig{\k_1}{\rtwo}\oplus\eig{\k_2}{\rtwo}\cdots
\oplus\eig{\k_\nu}{\rtwo}\label{eq:eucirrsum}
\ee
give the corresponding decomposition of $\crit{\rtwo}$ into $\euc$-irreducible
subspaces, each of which is labeled by a value of $\k$ appearing the
decomposition of $\crit{\ext}$. In general, this decomposition of
$\crit{\rtwo}$ cannot contain more subspaces than occurred in the original mode
interaction (\ref{eq:irrsum}); i.e.
\be
\nu\leq j.
\ee

There are several possibilities:
\begin{enumerate}
\item $\crit{\rtwo}$ is irreducible so that $\nu=1$. This can only happen if
$\k_1=\k_2=\cdots=\k_j$ in (\ref{eq:irrsum}), and the critical modes must
correspond to a real eigenvalue.
\item $\crit{\rtwo}$ is reducible so that $\nu>1$; then there are two
sub-cases:
\begin{enumerate}
\item The case $\nu=j$ so that the decomposition of $\crit{\rtwo}$ parallels
the decomposition of $\crit{\ext}$; in this circumstance the eigenvalues of the
critical modes may be real or complex. When $\nu=j$ the decomposition of
$\crit{\ext}$ is ``forced'' by the decomposition of $\crit{\rtwo}$. A
particular example occurs when all of the $\k_i$ in (\ref{eq:irrsum}) are
distinct:
$\k_i=\k_m\hspace{0.5in} \mbox{\rm if and only if} \hspace{0.5in} i=m.$
\item The case $\nu<j$ which can be analyzed as a hybrid of cases (1) and (2a).
\end{enumerate}
\end{enumerate}
The normal form $\Ft$ is modified by the hidden rotation symmetry in cases (1)
and (2b); in case (2a) there is no effect.  This situation may be summarized
by saying that the hidden rotations are important whenever the decomposition of
$\crit{\ext}$ in (\ref{eq:irrsum}) is not completely forced by the
decomposition of $\crit{\rtwo}$ in (\ref{eq:eucirrsum}).

When the hidden symmetries constrain $\Ft$ the mechanism is always the same:
the occurrence of hidden rotational symmetries on subspaces of ${\crit{\ext}}$.
Since ${\crit{\ext}}$ is a subspace of ${\crit{\rtwo}}$, for a given rotation
$\rot\theta\in\euc$ we may consider the rotated subspace
$[\rot\theta{\crit{\ext}}]$; if $\rot\theta\not\in\semi$ and
$[\rot\theta{\crit{\ext}}]$ intersects ${\crit{\ext}}$ in a nontrivial
subspace and  then $\rot\theta$ is a hidden symmetry on a subspace of
${\crit{\ext}}$. The $\euc$ symmetry of $\strobe$ implies that $\Ft$ in
(\ref{eq:pbcres}) must commute with $\rot\theta$ on the subspace
$E^c(\ext)\,\cap\,[\rot{-\theta}\,E^c(\ext)]$. Our main results show that all
the constraints imposed on $\Ft$ due to hidden rotation symmetry are of this
kind.

In normal form theory $\Ft$ is a formal power series, and the terms at any
finite order define a homogeneous vector field. Thus without loss of generality
we can assume $\Ft$ is a smooth homogeneous vector field on $E^c(\ext)$.
Section II summarizes our notation for the irreducible representations of
$\euc$ and $\semi$ and characterizes $\euc$-equivariant vector fields on
${\crit{\ext}}$. Section III forms the main part of the paper; the case of a
binary $\semi$ mode interaction with $\euc$ acting irreducibly on
${\crit{\ext}}$ is treated in detail. Necessary and sufficient conditions for
$\Ft$ to extend to $\strobe$ are obtained. Section IV analyzes binary mode
interactions with a reducible $\euc$ representation.

\section{Representations of $\euc$ and $\semi$}
\subsection{$\euc$ on $\crit{\rtwo}$}
We first define the irreducible representations $\eig\k\rtwo$ of $\euc$
associated with a given irreducible representation of $\semi$. For $\irred l n$
we have $\vec{k}=(l,n)$ and $\k^2=l^2+n^2$. Let $\gef{\vec{k}}(\vec{r})$ denote
a PBC eigenvector in $\irred l n$ with wave vector $\vec{k}$; the Euclidean
action
(\ref{eq:eucact}) acts on such modes (\ref{eq:gefdef}) by
\beq
(\tran ab\cdot \gef{\vec{k}})(\vec{r})&=& e^{-i(ak_x+bk_y)}
\gef{\vec{k}}(\vec{r}),\label{eq:pwtrans}\\
(\rot\theta\cdot\gef{\vec{k}})(\vec{r})&=&\gef{\vec{k}'}(\vec{r})
\label{eq:pwrot}\\
&\mbox{\rm where}&\; \vec{k}'=\rot\theta\cdot\vec{k}\;\;\; \mbox{\rm from}\;
(\ref{eq:rotdef}), \nonumber\\
(\gamma_2\cdot\gef{\vec{k}})(\vec{r})&=&\gef{\vec{k}'}(\vec{r})\;\;\;\mbox{\rm
where} \; \vec{k}'=(k_y,k_x).\label{eq:pwref}
\eeq
The Euclidean symmetry of the problem implies that all of the modes
$\gef{\vec{k}'}(\vec{r})$ obtained from $\gef{\vec{k}}(\vec{r})$ under $\euc$
are also eigenvectors for the same eigenvalue, although they may not satisfy
PBC. The wave vectors $\vec{k}'$ obtained in this way cover $A(\k)$, the circle
in $\vec{k}$-space of radius $\k$,
\be
A(\k)=\{\vec{k}\in\rtwo\;|\;\k=|\vec{k}|\}.\label{eq:ak}
\ee

Sums of these $\euc$-related modes
\be
\Phi(\vec{r})=\sum_{\vec{k}\in A(\k)}
a(\vec{k})\,\gef{\vec{k}}(\vec{r})\label{eq:genef}
\ee
define $\eig\k\rtwo$; more precisely we introduce a norm
\be
\parallel \Phi(\vec{r})\parallel\equiv \sum_{\vec{k}\in A(\k)} |a(\vec{k})|,
\label{eq:norm}
\ee
and assume in (\ref{eq:genef}) that $a(\vec{k})=0$ for all but a finite set of
points in $A(\k)$ to assure convergence. The finite sums in (\ref{eq:genef})
form a linear vector space whose closure with respect to (\ref{eq:norm}) gives
us an infinite-dimensional space $\eig{\k}{\rtwo}$ such that
$\irred l n\subset\eig{\k}{\rtwo}.$

The action of $\euc$ on the basic modes (\ref{eq:pwtrans}) - (\ref{eq:pwref})
defines a representation of $\euc$ on $\eig{\k}{\rtwo}$.
Melbourne has shown that these representations are absolutely irreducible.
\begin{theorem} {\rm (Melbourne)} Let
$L:\eig{\k}{\rtwo} \rightarrow\eig{\k}{\rtwo}$
be a linear map which commutes with the representation of $\euc$ in {\rm
(\ref{eq:pwtrans}) - (\ref{eq:pwref})}, then $L$ must be a scalar multiple of
the identity
$L=\sigma(\k)\, I.$
\end{theorem}
\noindent {\em {\bf Proof}.} \begin{quote}See the Appendix.
{\bf $\Box$}\end{quote}

This construction extends easily to a mode interaction (\ref{eq:irrsum})
involving multiple irreducible representations. We replace (\ref{eq:genef}) by
\be
\Phi(\vec{r})=\sum_{i=1}^j\sum_{\vec{k}\in A(\k_i)}
a_i(\vec{k})\,\gef{\vec{k}}(\vec{r})\label{eq:rgenef}
\ee
where each $\gef{\vec{k}}(\vec{r})$ is a critical mode. The collection of all
such finite sums forms a linear vector space whose closure with respect to
\be
\parallel \Phi(\vec{r})\parallel\equiv \sum_{i=1}^j \sum_{\vec{k_i}\in A(\k)}
|a_i(\vec{k})|
\label{eq:norm2}
\ee
defines the space $\crit{\rtwo}$.  In general the representation of $\euc$ on
$\crit{\rtwo}$ will not be irreducible and $\crit{\rtwo}$
is a direct sum of the irreducible representations $\eig\k\rtwo$:
\be
\crit{\rtwo}=\eig{\k_1}{\rtwo}\oplus\eig{\k_2}{\rtwo}\cdots
\oplus\eig{\k_\nu}{\rtwo}.\label{eq:eucirrsum2}
\ee
Here each irreducible representation is labeled by a value of $\k$ appearing in
the initial decomposition of $\crit{\ext}$ into $\semi$-irreducibles. If
$l^2+n^2=l'^2+n'^2=\k^2$, then
different representations of $\semi$ in (\ref{eq:irrsum}), $\irred ln$ and
$\irred{l'}{n'}$, corresponding to a single eigenvalue may both be subspaces of
$\eig{\k}{\rtwo}$. Hence the number of $\euc$ irreducibles in $\crit{\rtwo}$
can be smaller than the number of $\semi$-irreducibles in $\crit{\ext}$:
$\nu\leq j.$

\subsection{$\semi$ on $\crit{\ext}$}

The center subspace $\crit{\ext}$ for PBC is a subspace of $\crit{\rtwo}$ and
corresponds to the fixed point subspace for $\per$.
Since the intersection of the circle $A(\k)$ with the integer lattice ${\cal
Z}^2$,
\be
\tilde{A}(\k)=A(\k)\;\cap\;{\cal Z}^2,\label{eq:akint}
\ee
contains a finite number of points, $\crit{\ext}$ will always be
finite-dimensional.

The (non-trivial) irreducible representations $\irred{l}{n}$ of $\semi$ have
dimension four or eight and may be classified by two mode numbers
$\vec{k}=(l,n)$ with $l\geq n\geq 0$ giving the components of a specific wave
vector in $\tilde{A}(\k)$.
\begin{enumerate}
\item There are two sets of distinct four-dimensional irreducible
representations:
\begin{enumerate}
\item For $l=n>0$, $\tilde{A}(\k)=\{\pm\vec{k_1},\pm\vec{k_2}\}$ where
$\vec{k_1}=(l,l)$ and $\vec{k_2}=(l,-l)$.  Then $\k^2=2l^2$ and
\be
\irred ll= \{z_1\gef{\vec{k_1}}(\vec{r}) + z_2\gef{\vec{k_2}}(\vec{r}) +
cc)|\;(z_1,z_2)\in {\bf C}^2\},\label{eq:asubtll}
\ee
carries the representation of $\semi$ generated by
\beq
\g{1}\cdot(z_1,z_2)&=&(\overline{z}_2,\overline{z}_1)\label{eq:4drepa}\\
\g{2}\cdot(z_1,z_2)&=&(z_1,\overline{z}_2)\label{eq:4drepb}\\
\tran{a}{b}\cdot(z_1,z_2)&=&
(e^{-il(a+b)}z_1,e^{-il(a-b)}z_2). \label{eq:4drepc}
\eeq

\item For $l>n=0$, $\tilde{A}(\k)=\{\pm\vec{k_1},\pm\vec{k_2}\}$ where
$\vec{k_1}=(l,0)$ and $\vec{k_2}=(0,l)$. Then $\k^2=l^2$ and
\be
\irred l0= \{z_1\gef{\vec{k_1}}(\vec{r}) + z_2\gef{\vec{k_2}}(\vec{r}) +
cc)|\;(z_1,z_2)\in {\bf C}^2\},\label{eq:asubtl0}
\ee
carries the representation of $\semi$ generated by
\beq
\g{1}\cdot(z_1,z_2)&=&(\overline{z}_1,{z}_2)\label{eq:l0repa}\\
\g{2}\cdot(z_1,z_2)&=&(z_2,{z}_1)\label{eq:l0repb}\\
\tran{a}{b}\cdot(z_1,z_2)&=&
(e^{-ila}z_1,e^{-ilb}z_2). \label{eq:l0repc}
\eeq
\end{enumerate}
Note that $\tran{\pi/l}{\pi/l}$ is in the kernel of type 1 representations but
not type 2 representations, hence they must be inequivalent.

\item The eight-dimensional representations correspond to $l>n>0$ with
$\tilde{A}(\k)=\{\pm\vec{k_1},\pm\vec{k_2},\pm\vec{k_3},\pm\vec{k_4}\}$ where
$\vec{k_1}=(l,n)$, $\vec{k_2}=(l,-n)$, $\vec{k_3}=(n,l)$, and
$\vec{k_4}=(n,-l)$. Now $\k^2=l^2+n^2$ and
\be
\irred ln = \{(z_1\gef{\vec{k_1}}(\vec{r}) + z_2\gef{\vec{k_2}}(\vec{r}) +
z_3\gef{\vec{k_3}}(\vec{r}) + z_4\gef{\vec{k_4}}(\vec{r}) +
cc)|\;(z_1,z_2,z_3,z_4)\in {\bf C}^4\}\label{eq:subt}
\ee
transforms under the following representation of $\semi$:
\beq
\g{1}\cdot z&=&
(\overline{z}_2,\overline{z}_1,\overline{z}_4,\overline{z}_3)\label{eq:8drepa}\\
\g{2}\cdot z&=&
(z_3,\overline{z}_4,z_1,\overline{z}_2)\label{eq:8drepb}\\
\tran{a}{b}\cdot z&=&
(e^{-i(la+nb)}z_1,e^{-i(la-nb)}z_2,
e^{-i(na+lb)}z_3,e^{-i(na-lb)}z_4).\label{eq:8drepc}
\eeq
where $z=(z_1,z_2,z_3,z_4)$.
\end{enumerate}

\subsection{$\euc$-equivariant vector fields on $\crit{\rtwo}$}

Next we describe a result that characterizes the nonlinear $\euc$-equivariant
vector fields on $\crit{\rtwo}=\eig{\k_1}{\rtwo}\oplus\eig{\k_2}{\rtwo}\cdots
\oplus\eig{\k_\nu}{\rtwo}$. For a given representation of $\euc$ we define
\be
A_c=A(\k_1)\cup A(\k_2)\cup\ldots\cup A(\k_\nu),
\ee
and let
\be
\strobe:\crit{\rtwo}\rightarrow\crit{\rtwo}\label{eq:strobe}
\ee
be a Euclidean-symmetric polynomial vector field of degree p. That is we assume
\be
\strobe (\alpha\Phi)=\alpha^p\strobe (\Phi)\,\;\;\;{\rm for}\;{\rm
any}\;\alpha\in {\rm R}\label{eq:homo}
\ee
and
\be
\gamma\cdot \strobe (\Phi)=\strobe (\gamma\cdot \Phi) \,\;\;\;{\rm for}\;{\rm
any}\;\gamma\in\euc.\label{eq:eucN}
\ee
When $\crit{\rtwo}$ is reducible ($\nu>1$), then $\strobe$ has multiple
components
\be
\strobe=\strobec 1+\strobec 2\ldots+\strobec \nu\label{eq:strobec}
\ee
where $\strobec i:\crit{\rtwo}\rightarrow\eig{\k_i}{\rtwo}$
for $i=1,\ldots,\nu$.
If we represent $\Phi$ as in (\ref{eq:rgenef}), then the homogeneity of
$\strobe$ (\ref{eq:homo}) implies $\strobe_i$ must have the form
\be
\strobe_i (\Phi)=\sum_{\vec{k}\in
A(\k_i)}\,a_i'(\vec{k})\,\gef{\vec{k}}(\vec{r})\label{eq:pnonlin}
\ee
where
\be
a_i'(\vec{k})=\sum_{l_1\leq\ldots\leq l_p}
\left[\sum_{\vec{k_1}\in A(\k_{l_1})}\sum_{\vec{k_2}\in A(\k_{l_2})}
\ldots\sum_{\vec{k_p}\in A(\k_{l_p})}
a_{l_1}(\vec{k_1})\,a_{l_2}(\vec{k_2})\,\ldots
a_{l_p}(\vec{k_p})\,P_i(\vec{k},\vec{k_1},\ldots, \vec{k_p})\right].
\label{eq:akprime}
\ee

\begin{theorem} $\strobe$ will have Euclidean symmetry {\rm (\ref{eq:eucN})}
if each $P_i(\vec{k},\vec{k_1},\vec{k_2},\ldots,\vec{k_p})$ satisfies two
conditions. Given $p+1$ vectors,
$\{\vec{k},\vec{k_1},\vec{k_2},\ldots,\vec{k_p}\}$, such that $\vec{k}\in
A(\k_i)$ and $\vec{k_j}\in A_c$ for $j=1,\ldots,p$ then
\be
P_i(\gamma\cdot\vec{k},\gamma\cdot\vec{k_1},\gamma\cdot\vec{k_2},
\ldots,\gamma\cdot\vec{k_p})= P_i(\vec{k},\vec{k_1},\vec{k_2},\ldots,\vec{k_p})
\;\;\mbox{\rm for all}\;\; \gamma\in \otwo;\label{eq:cond1}
\ee
and
\be
P_i(\vec{k},\vec{k_1},\vec{k_2},\ldots,\vec{k_p})=0\;\;\mbox{\rm if}\;\;
\vec{k}\neq\vec{k_1}+\vec{k_2}+\ldots+\vec{k_p}.\label{eq:cond2}
\ee
In addition, if $\strobe_i $ is real-valued then $P_i$ must be real-valued.
\end{theorem}
\noindent {\em {\bf Proof}.}\begin{quote}  The proof is an easy generalization
of Melbourne's result for the case $\nu=1$; see the Appendix.{\bf $\Box$}
\end{quote}

\section{\hspace{0.125in}Binary Mode interactions with $\crit{\rtwo}$
irreducible}
Irreducibility of $\crit{\rtwo}$ means that $E^c(\rtwo)=\eig\k\rtwo,$
so the critical modes correspond to a single eigenvalue $\sigma(\k)$ in Theorem
II.1. If the original problem is described by real fields, so that complex
eigenvalues must come in conjugate pairs, then $\sigma(\k)$ must be real.  In
addition, $E^c(\rtwo)=\eig\k\rtwo$ implies that in the decomposition
\be
\crit{\ext}=\irred{l_1}{n_1}\oplus\irred{l_2}{n_2}\cdots
\oplus\irred{l_j}{n_j}\label{eq:modeint}
\ee
we will find
$\k^2=l_1^2+n_1^2=l_2^2+n_2^2=\cdots=l_j^2+n_j^2.$
In (\ref{eq:modeint})  a given representation $\irred{l}{n}$ appears at most
once;  a so-called ``accidental'' degeneracy cannot occur. This feature allows
us to order the mode indices $l_1>l_2>\cdots>l_j$ and $n_1<n_2<\cdots<n_j$
in (\ref{eq:modeint}) without loss of generality.

The form of an $\euc$-symmetric vector field $\strobe$ of degree p is given by
(\ref{eq:strobe})-(\ref{eq:akprime}); for a vector
\be
\Phi(\vec{r})=\sum_{\vec{k}\in A(\k)}
a(\vec{k})\,\gef{\vec{k}}(\vec{r})\label{eq:genef2}
\ee
in $E^c(\rtwo)$, $\strobe(\Phi)$ is given by
\be
\strobe(\Phi)=\sum_{\vec{k}\in
A(\k)}\,a'(\vec{k})\,\gef{\vec{k}}(\vec{r})\label{eq:irrnonlin}
\ee
with
\be
a'(\vec{k})=\sum_{\vec{k_1}\in A(\k)}\sum_{\vec{k_2}\in A(\k)}
\ldots\sum_{\vec{k_p}\in A(\k)} a(\vec{k_1})\,a(\vec{k_2})\,\ldots
a(\vec{k_p})\,P(\vec{k},\vec{k_1},\ldots, \vec{k_p}).\label{eq:irrakprime}
\ee
The function $P$ satisfies
\beq
P(\vec{k},\vec{k_1},\vec{k_2},\ldots,\vec{k_p})&=&0\;\;\mbox{\rm if}\;\;
\vec{k}\neq\vec{k_1}+\vec{k_2}+\ldots+\vec{k_p};\label{eq:irrcond1}\\
P(\gamma\cdot\vec{k},\gamma\cdot\vec{k_1},\gamma\cdot\vec{k_2},
\ldots,\gamma\cdot\vec{k_p})&=&P(\vec{k},\vec{k_1},\vec{k_2},\ldots,\vec{k_p})
\;\;\mbox{\rm for all}\;\; \gamma\in \otwo\label{eq:irrcond2}
\eeq
where $\vec{k}\in A(\k)$ and $\vec{k_j}\in A(\k)$ for $j=1,\ldots,p$.
The restriction of $\strobe$ to $E^c(\ext)$
\be
\Ft=\strobe|_{E^c(\ext)}\label{eq:Ftrestrict}
\ee
is accomplished by choosing $a(\vec{k})$ in (\ref{eq:genef2}) so that $\Phi\in
E^c(\ext)$; the appropriate form of $a(\vec{k})$ depends on the specific mode
interaction.

We consider binary mode interactions comprised of two irreducible
representations of $\semi$, $\irred {l_1}{n_1}$ and $\irred {l_2}{n_2}$, such
that
\be
\k^2=l_1^2+n_1^2= l_2^2+n_2^2\label{eq:rotdeg}
\ee
and $l_1>l_2$. For simplicity write $\irreds {i}{ }$ for $\irred {l_i}{n_i}$ so
that
\be
{E^c(\ext)}=\irreds {1}{}\oplus\irreds {2}{}.\label{eq:binary}
\ee
There are three possibilities in (\ref{eq:binary}) which we refer to as the
$[4,8]$ mode interaction, the $[8,8]$ mode interaction, and the $[8,4]$ mode
interaction.  The first integer gives the dimension of $\irreds {1}$ and the
second integer gives the dimension of $\irreds {2}$. Table 1 gives an example
of each type. Note that the condition (\ref{eq:rotdeg}) does not allow a
$[4,4]$ mode interaction between two four-dimensional representations.
\begin{table}
\begin{center}
Table 1.  Examples of Binary Mode Interactions
\end{center}
\vspace{7mm}
\begin{center}
\begin{tabular}{lccc}
\underline{$\k^2$}
&\underline{$({l_1},{n_1})$}
&\underline{$({l_2},{n_2})$}
&\underline{Type}\\
\\
$25$ & (5,0) & $(4,3)$ & [4,8]\\
\\
$50$ &(7,1) & (5,5) & [8,4]\\
\\
$65$ & (8,1)&(7,4) & [8,8]\\
\end{tabular}
\par
\vspace{7 mm}
\protect \vspace*{\fill}
\end{center}
\end{table}

The relevant hidden rotations connect the representations $\irreds {1}{ }$ and
$\irreds {2}{ }$. For example, in terms of the angles
$\tan\theta_{1}={n_1}/{l_1}$ and $\tan\theta_{2}={n_2}/{l_2}$
characterizing the two representations, we define
$\phi\equiv\theta_2-\theta_1.$
The rotation $\rot{\phi}$ is not an element of $\semi$ and it matches our
earlier description of a hidden symmetry on a subspace:
\be
E^c(\ext)\,\cap\,[\rot{\phi}\,E^c(\ext)]\neq\{0\}.\label{eq:intersect}
\ee
Thus there are non-zero vectors $\Phi\in E^c(\ext)$ such that $\rot{\phi}\cdot
\Phi\in E^c(\ext)$. It turns out that for the binary mode interactions
considered here, the subspace in (\ref{eq:intersect}) is the fixed point
subspace for an isotropy subgroup of $\semi$ and therefore any
$\semi$-equivariant vector field $\Ft$ leaves
$E^c(\ext)\,\cap\,[\rot{\phi}\,E^c(\ext)]$ invariant. Our main result is that
the necessary and sufficient condition for $\Ft$ to extend to a Euclidean
symmetric vector field is simply that $\Ft$ commutes with $\rot{\theta}$ on the
subspace $E^c(\ext)\,\cap\,[\rot{-\theta}\,E^c(\ext)]$ for all
$\rot{\theta}\in\euc$. Obviously this requirement is nontrivial only for those
rotations not in $\semi$ for which
$E^c(\ext)\,\cap\,[\rot{-\theta}\,E^c(\ext)]\neq\{0\}$.

\begin{theorem} Assume ${E^c(\rtwo)}$ is $\euc$-irreducible and
${E^c(\ext)}=\irreds {1}\oplus\irreds {2}$ carries an reducible representation
of $\semi$
with $l_1>l_2$ and $\k^2=l_1^2+n_1^2= l_2^2+n_2^2$. For a homogeneous
$\semi$-equivariant vector field
$\Ft:{E^{c}(\ext)}\rightarrow{E^{c}(\ext)},$
there exists an $\euc$-equivariant vector field,
$\strobe:{E^{c}(\rtwo)}\rightarrow{E^{c}(\rtwo)}$,
such that
\be
\Ft = \strobe |_{{E^{c}(\ext)}}\label{eq:irrmi}
\ee
if and only if, for all $\rot{\theta}\in\euc$, $\Ft$ satisfies
\be
\rot {\theta}\cdot\Ft(\Phi)=\Ft(\rot {\theta}\cdot\Phi)\label{eq:ns}
\ee
for $\Phi\in E^c(\ext)\,\cap\,[\rot{-\theta}\,E^c(\ext)]$.
\end{theorem}
\noindent {\em {\bf Proof}.}
\begin{quote}  The necessity of (\ref{eq:ns}) will be proved here. Sufficiency
follows by analyzing the three types of mode interaction separately. This is
done in the following subsections; see Lemmas III.3, III.6, and III.9. Since
$\gamma\cdot\strobe(\Phi)=\strobe(\gamma\cdot\Phi)$ for all $\gamma\in\euc$ and
$\Phi\in{E^c(\rtwo)}$, then
\be
\rot {\theta}\cdot\strobe(\Phi)= \strobe(\rot {\theta}\cdot\Phi).
\label{eq:rotsym}
\ee
For $\Phi\in E^c(\ext)\,\cap\,[\rot{-\theta}\,E^c(\ext)]$ then $\rot
{\theta}\cdot\Phi\in E^c(\ext)$ so if $\Ft$ denotes the restriction of
$\strobe$ to $E^c(\ext)$ we can write:
\beq
\strobe(\Phi)&=&\Ft(\Phi)\\
\strobe(\rot {\theta}\cdot\Phi)&=&\Ft(\rot {\theta}\cdot\Phi),
\eeq
and substitute in (\ref{eq:rotsym}).
Hence (\ref{eq:rotsym}) becomes (\ref{eq:ns}).
{\bf $\Box$}\end{quote}

\subsection{\hspace{0.125in}The $[4,8]$-mode interaction}

In the notation of section II.B, the four-dimensional representation
$(l_1,n_1)=(l_1,0)$
\be
\irreds {1}=\{z_1\gef{\vec{q_1}}(\vec{r}) + z_2\gef{\vec{q_2}}(\vec{r}) +
cc)|\;(z_1,z_2)\in {\bf C}^2\},\label{eq:irredl0}
\ee
has wave vectors $\vec{q_1}=(l_1,0)$ and $\vec{q_2}=(0,l_1)$. The
eight-dimensional representation $(l_2,n_2)$
\be
\irreds {2}=\{(w_1\gef{\vec{p_1}}(\vec{r}) + w_2\gef{\vec{p_2}}(\vec{r}) +
w_3\gef{\vec{p_3}}(\vec{r}) + w_4\gef{\vec{p_4}}(\vec{r}) +
cc)|\;(w_1,w_2,w_3,w_4)\in {\bf C}^4\}\label{eq:irredln}
\ee
has wave vectors $\vec{p_1}=(l_2,n_2)$, $\vec{p_2}=(l_2,-n_2)$,
$\vec{p_3}=(n_2,l_2)$ and $\vec{p_4}=(n_2,-l_2)$. The reducible representation
of $\semi$ on the twelve-dimensional space $\irreds {1}\oplus\irreds {2}$ is
generated by
\beq
\g{1}\cdot(z,w)&=&(\overline{z}_1,{z}_2,
\overline{w}_2,\overline{w}_1,\overline{w}_4,\overline{w}_3)\label{eq:48repa}\\
\g{2}\cdot(z,w)&=&(z_2,{z}_1,w_3,\overline{w}_4,w_1,\overline{w}_2)\label{eq:48repb}\\
\tran{a}{b}\cdot(z,w)&=&
(e^{-il_1a}z_1,e^{-il_1b}z_2,\nonumber\\
&&\hspace{0.5in}e^{-i(l_2a+n_2b)}w_1,e^{-i(l_2a-n_2b)}w_2,
e^{-i(n_2a+l_2b)}w_3,e^{-i(n_2a-l_2b)}w_4) \label{eq:48repc}
\eeq
where $z\equiv(z_1,z_2)$ and $w\equiv(w_1,w_2,w_3,w_4)$. We will also refer to
the reflection $\g 3\equiv\g 2\g 1\g 2$ which acts by
\be
\g 3\cdot(z_1,z_2,w_1,w_2,w_3,w_4)=(z_1,\overline{z}_2,w_2,w_1,w_4,w_3).
\label{eq:48g3}
\ee

Our discussion is based on a convenient representation for vector fields that
commute with this representation.
\begin{prop}
$\Ft$ is $\semi$-equivariant if and only if it has the form
\be
\Ft(z,w)=\left(
\begin{array}{c}
f(z,w)\\
\\
f(\g 2\cdot(z,w))\\
\\
h(z,w)\\
\\
h(\g 3\cdot(z,w))\\
\\
h(\g 2\cdot(z,w))\\
\\
h(\g 2\g 1\cdot(z,w))
\end{array}
\right)=\left(
\begin{array}{c}
f(z_1,z_2,w_1,w_2,w_3,w_4)\\
\\
f(z_2,z_1,w_3,\overline{w}_4,w_1,\overline{w}_2)\\
\\
h(z_1,z_2,w_1,w_2,w_3,w_4)\\
\\
h(z_1,\overline{z}_2,w_2,w_1,w_4,w_3)\\
\\
h(z_2,z_1,w_3,\overline{w}_4,w_1,\overline{w}_2)\\
\\
h(\overline{z}_2,z_1,w_4,\overline{w}_3,w_2,\overline{w}_1)
\end{array}
\right)\label{eq:48map}
\ee
where $f(z,w)$ and $h(z,w)$ are complex-valued functions satisfying following
conditions:
\begin{enumerate}
\item $f(\overline{z},\overline{w})=\overline{f(z,w)}$ and
$h(\overline{z},\overline{w})=\overline{h(z,w)}$
\item $f(z,w)$ is $\g 3$-invariant:
\be
f(z_1,z_2,w_1,w_2,w_3,w_4)=f(z_1,\overline{z}_2,w_2,w_1,w_4,w_3)\label{eq:fg3}
\ee
\item $\overline{z}_1\,f(z,w)$ and $\overline{w}_1\,h(z,w)$ are invariant under
the translations $\tor$.
\end{enumerate}
\end{prop}
\noindent {\em {\bf Proof}.}\begin{quote} See the appendix. {\bf
$\Box$}\end{quote}
If $f(z,w)$ and $h(z,w)$ are functions satisfying the conditions of this
theorem, we write
\be
\Ft=[f,h]\label{eq:maprep}
\ee
to indicate the corresponding $\semi$ symmetric vector field (\ref{eq:48map}).

For $\semi$-symmetric vector fields $\Ft=[f,h]$, we can assume $f$ and $h$ are
homogeneous functions with no essential loss of
generality. A $\{\tor,\g 3\}$-invariant function such as $\overline{z}_1 f$ can
always be written in the form $M(z,w)+M(\g 3\cdot(z,w))$ where $M(z,w)$ is
$\tor$-invariant. For homogeneous $f$ we can reduce to the case where
$M(z,w)=\overline{z}_1 m(z,w)$ is a monomial and
\be
f(z,w)=m(z,w)+m(\g 3\cdot(z,w)).\label{eq:fmon}
\ee
Similarly when $h$ is homogeneous, then we can reduce to the case of single
monomial
\be
h(z,w)=m'(z,w)\label{eq:hmon}
\ee
where $M(z,w)=\overline{w}_1 m'(z,w)$ is a $\tor$-invariant monomial.
Henceforth we shall assume $f$ and $h$ have the simple forms given in
(\ref{eq:fmon}) and (\ref{eq:hmon}) respectively.

The $\tor$-invariant monomials are important in characterizing the vector field
$\Ft$.
For the representation (\ref{eq:48repc}) there are the elementary quadratic
invariants given by
\beq
\sigma_i&=&|z_i|^2\hspace{0.5in}\label{eq:zelinv}\\
\rho_i&=&|w_i|^{2}.\hspace{0.5in}\label{eq:welinv}
\eeq
For the general case it is convenient to introduce the notation
\beq
\Omega_i^{\mu_i}&\equiv&\left\{
\begin{array}{c}
z_i^{\mu_i}\;\;\mbox{\rm if $\mu_i\geq0$}\\
{\overline{z}_i}^{|\mu_i|}\;\;\mbox{\rm if $\mu_i<0$}
\end{array}\right.\\
&&\nonumber\\
\omega_i^{\nu_i}&\equiv&\left\{\begin{array}{c}w_i^{\nu_i}\;\;\mbox{\rm if
$\nu_i\geq0$}\\ {\overline{w}_i}^{|\nu_i|}\;\;\mbox{\rm if $\nu_i<0$}.
\end{array}\right.
\eeq
\begin{prop}
A $\tor$-invariant monomial $M(z,w)$ can always be written in the form
\be
M(z,w)=(\sigma_1^{\mu_1'}\sigma_2^{\mu_2'}
\rho_1^{\nu_1'}\rho_2^{\nu_2'}\rho_3^{\nu_3'}\rho_4^{\nu_4'})\,
\Omega_1^{\mu_1}\Omega_2^{\mu_2}
\omega_1^{\nu_1}\omega_2^{\nu_2}\omega_3^{\nu_3}\omega_4^{\nu_4}\label{eq:t2mon}
\ee
where $(\mu_1,\mu_2,\nu_1,\nu_2,\nu_3,\nu_4)$ are integers satisfying
\beq
l_1\mu_1+l_2(\nu_1+\nu_2)+n_2(\nu_3+\nu_4)&=&0\label{eq:th1}\\
l_1\mu_2+n_2(\nu_1-\nu_2)+l_2(\nu_3-\nu_4)&=&0\label{eq:th2}
\eeq
and the integers $(\mu_1',\mu_2',\nu_1',\nu_2',\nu_3',\nu_4')$ are
non-negative.
\end{prop}
\noindent {\em {\bf Proof}.}\begin{quote}See the Appendix.
{\bf $\Box$}\end{quote}

The geometry of the wave vector set (\ref{eq:akint})
\be
\tilde{A}(\k)=\{\pm\vec{q_1},\pm\vec{q_2},\pm\vec{p_1},\pm\vec{p_2},\pm\vec{p_3},\pm\vec{p_4}\}\label{eq:wavev}
\ee
is important for our analysis, see Fig. 1. Since $\theta_1=0$ in this case, we
have
$\phi=\theta_2=\arctan({n_2}/{l_2});$
for this mode interaction it is sufficient to restrict attention to $\phi$.
Let $\alpha$ denote the angle between $\vec{p_3}$ and $\vec{p_1}$, a notable
feature of Fig. 1 is that $\alpha$ and $\phi$ are unequal. Since
$\cos\phi=l_2/\k$ and $\cos\alpha=2l_2n_2/\k^2$, these angles are only equal
when $2n_2=\k$ which would imply $3n_2^2=l_2^2$. Since $l_2$ and $n_2$ are
integers, this is impossible. Because $\phi$ and $\alpha$ are unequal, the set
$\tilde{A}(\k)$ satisfies
\beq
\tilde{A}(\k)\cap[\rot{\phi}\tilde{A}(\k)]&=&\{\pm\vec{q_1},\pm\vec{q_2},\pm\vec{p_1},\pm\vec{p_4}\}\label{eq:wavevint1}\\
\tilde{A}(\k)\cap[\rot{-\phi}\tilde{A}(\k)]&=&\{\pm\vec{q_1},\pm\vec{q_2},\pm\vec{p_2},\pm\vec{p_3}\}.\label{eq:wavevint2}
\eeq

The non-zero intersections (\ref{eq:wavevint1}) and (\ref{eq:wavevint2}) imply
that $\rot{\pm\phi}E^c$ intersects $E^c$ along eight-dimensional
subspaces\footnote{The abbreviated notation $(z_1,z_2,w_1,0,0,w_4)$ denotes the
subspace $\{(z_1\gef{\vec{q_1}}(\vec{r})+z_2\gef{\vec{q_2}}(\vec{r})+
w_1\gef{\vec{p_1}}(\vec{r})+w_4\gef{\vec{p_4}}(\vec{r})
+cc)\,|\,(z_1,z_2,w_1,w_4)\in{\bf C}^4\}.$}:
\beq
E^c(\ext)\,\cap\,[\rot{\phi}\,E^c(\ext)]&=&(z_1,z_2,w_1,0,0,w_4)
\label{eq:sub1}\\
E^c(\ext)\,\cap\,[\rot{-\phi}\,E^c(\ext)]&=&(z_1,z_2,0,w_2,w_3,0).
\label{eq:sub2}
\eeq
These subspaces are related by reflection symmetry; for example since $\g
3\cdot(z_1,z_2,w_1,0,0,w_4)=(z_1,\overline{z}_2,0,w_1,w_4,0)$ we have
\be
\g 3:E^c(\ext)\,\cap\,[\rot{\phi}\,E^c(\ext)]\rightarrow
E^c(\ext)\,\cap\,[\rot{-\phi}\,E^c(\ext)].
\ee
\begin{prop}
For the representation \mbox{\rm (\ref{eq:48repa})-(\ref{eq:48repc})}, the
subspaces $E^c(\ext)\,\cap\,[\rot{\phi}\,E^c(\ext)]$ and
$E^c(\ext)\,\cap\,[\rot{-\phi}\,E^c(\ext)]$ are fixed point subspaces for
isotropy subgroups of $\semi$; hence they are invariant under any $\semi$
symmetric vector field $\Ft=[f,h]$. This invariance implies
\be
h(z_1,z_2,0,w_2,w_3,0)=0.\label{eq:fpspace}
\ee
\end{prop}
\noindent {\em {\bf Proof}.}\begin{quote} Since
$E^c(\ext)\,\cap\,[\rot{\phi}\,E^c(\ext)]$ and
$E^c(\ext)\,\cap\,[\rot{-\phi}\,E^c(\ext)]$ are related by reflection, if
$E^c(\ext)\,\cap\,[\rot{-\phi}\,E^c(\ext)]$ is the fixed point subspace
$\mbox{\rm Fix}(\Sigma)$ for an isotropy subgroup $\Sigma$ then
$E^c(\ext)\,\cap\,[\rot{\phi}\,E^c(\ext)]$ is the fixed point subspace for $\g
3\Sigma\g 3$. Thus it is sufficient to consider (\ref{eq:sub2}) with $\Sigma$
the isotropy subgroup for $(z_1,z_2,0,w_2,w_3,0)$.
Let $\tran ab\in\Sigma$, then $\tran
ab\cdot(z_1,z_2,0,w_2,w_3,0)=(z_1,z_2,0,w_2,w_3,0)$, or
\beq
e^{-il_1a}&=&1\;\;\;\;\;e^{-il_1b}=1\\
e^{-i(l_2a-n_2b)}&=&1\;\;\;\;\;e^{-i(n_2a+l_2b)}=1.
\eeq
The $(z_1,z_2)$ equations require
\be
(a,b)=(\frac{2\pi a'}{l_1},\frac{2\pi b'}{l_1})\label{eq:ab}
\ee
where
$(a',b')$ are arbitrary integers, and the $(w_2,w_3)$ equations require
\beq
\frac{(l_2 a'-n_2 b')}{l_1}&=&s_1\label{eq:xcomp}\\
&&\nonumber\\
\frac{n_2 a'+l_2 b'}{l_1}&=&s_2\label{eq:ycomp}
\eeq
where $(s_1,s_2)$ are integers. Since ${l_1}^2=\k^2$ and
$(\cos\phi,\sin\phi)=(l_2/\k,n_2/\k)$, if we regard $(a',b')$ and $(s_1,s_2)$
as
vectors then (\ref{eq:xcomp})-(\ref{eq:ycomp}) are equivalent to
$\rot\phi\cdot(a',b')=(s_1,s_2).$
If we seek solutions in $\tilde{A}(\k)$ then $(a',b')\in
\tilde{A}(\k)\cap[\rot{-\phi}\tilde{A}(\k)]$ or
\be
(a',b')\in\{\pm\vec{q_1},\pm\vec{q_2},\pm\vec{p_2},\pm\vec{p_3}\}
\ee
from (\ref{eq:wavevint2}). The translation corresponding to $(a',b')=\vec{p_2}$
provides a necessary condition for $(z,w)\in \mbox{\rm Fix}(\Sigma)$: $\tran
{2\pi l_2/l_1}{-2\pi n_2/l_1}\cdot(z,w)=(z,w)$ or
\be
e^{-i2\pi ({l_2}^2-{n_2}^2)/l_1}w_1=w_1\;\;\;\;\;\;\; e^{-i2\pi
(2{l_2}{n_2})/l_1}w_4=w_4.\label{eq:nec}
\ee
Since neither $({l_2}^2-{n_2}^2)/l_1$ nor $(2{l_2}{n_2})/l_1$ can be an
integer, the necessary condition (\ref{eq:nec}) implies $w_1=0$ and $w_4=0$. By
definition this is also a sufficient condition for $(z,w)\in \mbox{\rm
Fix}(\Sigma)$ so $\mbox{\rm Fix}(\Sigma)=(z_1,z_2,0,w_2,w_3,0)$ or equivalently
\be
\mbox{\rm Fix}(\Sigma)=E^c(\ext)\,\cap\,[\rot{-\phi}\,E^c(\ext)].
\ee
The invariance of fixed point subspaces is a well known property and for
$E^c(\ext)\,\cap\,[\rot{-\phi}\,E^c(\ext)]=(z_1,z_2,0,w_2,w_3,0)$ this
invariance implies (\ref{eq:fpspace}) from (\ref{eq:sub2}) and
(\ref{eq:48map}).{\bf $\Box$} \end{quote}

When $\Ft$ is obtained by restriction from $\strobe$ there are additional
conditions on $f$ and $h$. For example we know from (\ref{eq:ns}) that $\Ft$
commutes with $\rot{\phi}$ on $E^c(\ext)\,\cap\,[\rot{-\phi}\,E^c(\ext)]$. This
implies a crucial relationship between $f$ and $h$.
\begin{prop}
For a $\semi$-equivariant vector field $\Ft=[f,h]$, the condition
\be
\rot {\phi}\cdot\Ft(\Phi)=\Ft(\rot {\phi}\cdot\Phi)\label{eq:rotcond0}
\ee
for all $\Phi\in E^c(\ext)\,\cap\,[\rot{-\phi}\,E^c(\ext)]$ is satisfied if and
only if
\be
f(z_1,z_2,0,w_2,w_3,0)=h(w_2,w_3,z_1,0,0,\overline{z}_2),\label{eq:rotcond}
\ee
or equivalently
\be
h(z_1,z_2,w_1,0,0,w_4)=f(w_1,\overline{w}_4,0,z_1,z_2,0).\label{eq:rotcondeqv}
\ee
\end{prop}
\noindent {\em {\bf Proof}.}\begin{quote} We will verify that
(\ref{eq:rotcond}) is necessary and sufficient; condition (\ref{eq:rotcondeqv})
follows from (\ref{eq:rotcond}) by relabeling arguments. In coordinates,
$\Phi\in E^c(\ext)\,\cap\,[\rot{-\phi}\,E^c(\ext)]$ is represented as
$\Phi=(z_1,z_2,0,w_2,w_3,0)$
and $\rot {\phi}\cdot\Phi=(w_2,w_3,z_1,0,0,\overline{z}_2).$
Since $h(z_1,z_2,0,w_2,w_3,0)=0$ from (\ref{eq:fpspace}) we have
\be
\rot {\phi}\cdot\Ft(\Phi)=\left(
\begin{array}{c}
h(z_1,\overline{z}_2,w_2,0,0,w_3)\\
\\
h(z_2,z_1,w_3,0,0,\overline{w}_2)\\
\\
f(z_1,z_2,0,w_2,w_3,0)\\
\\
0\\
\\
0\\
\\
f(\overline{z}_2,\overline{z}_1,\overline{w}_3,0,0,{w}_2)
\end{array}
\right)\label{eq:lhs}
\ee
and
\be
\Ft(\rot {\phi}\cdot\Phi)=\left(
\begin{array}{c}
f(w_2,w_3,z_1,0,0,\overline{z}_2)\\
\\
f(w_3,w_2,0,z_2,z_1,0)\\
\\
h(w_2,w_3,z_1,0,0,\overline{z}_2)\\
\\
0\\
\\
0\\
\\
h(\overline{w}_3,w_2,\overline{z}_2,0,0,\overline{z}_1)
\end{array}
\right)\label{eq:rhs}
\ee
from Propositions {\rm III.1} and {\rm III.3}. Comparing (\ref{eq:lhs}) and
(\ref{eq:rhs}) we find equality if and only if
\beq
h(z_1,\overline{z}_2,w_2,0,0,w_3)&=&f(w_2,w_3,z_1,0,0,\overline{z}_2)
\label{eq:c1}\\
h(z_2,z_1,w_3,0,0,\overline{w}_2)&=&f(w_3,w_2,0,z_2,z_1,0)\label{eq:c2}\\
f(z_1,z_2,0,w_2,w_3,0)&=&h(w_2,w_3,z_1,0,0,\overline{z}_2)\label{eq:c3}\\
f(\overline{z}_2,\overline{z}_1,\overline{w}_3,0,0,{w}_2)&=&
h(\overline{w}_3,w_2,\overline{z}_2,0,0,\overline{z}_1).\label{eq:c4}
\eeq
These four conditions are not independent: (\ref{eq:c4}) is obtained from
(\ref{eq:c1}) by simply relabeling arguments, similarly (\ref{eq:c2}) is
obtained from (\ref{eq:c3}). Furthermore (\ref{eq:c1}) is equivalent to
(\ref{eq:c3}); this follows from the $\g 3$-invariance of $f$ (\ref{eq:fg3})
and a relabeling of arguments. Thus (\ref{eq:c3}) is a necessary and sufficient
condition for (\ref{eq:rotcond0}).
{\bf $\Box$}\end{quote}

{}From (\ref{eq:genef2})-(\ref{eq:irrakprime}) we can represent those
homogeneous $\semi$-symmetric vector fields that arise by restricting
$\euc$-symmetric vector fields $\strobe$. The restriction in
(\ref{eq:Ftrestrict}) is accomplished by choosing $a(\vec{k})$ in
(\ref{eq:genef2}) so that $\Phi\in E^c(\ext)$:
\beq
a(\vec{k})&=&z_1\;\delta_{\vec{k},\vec{q_1}} +
\overline{z}_1\;\delta_{\vec{k},\vec{-q_1}} + z_2\;\delta_{\vec{k},\vec{q_2}} +
\overline{z}_2\;\delta_{\vec{k},\vec{-q_2}}+ w_1\;\delta_{\vec{k},\vec{p_1}} +
\overline{w}_1\;\delta_{\vec{k},\vec{-p_1}}  \nonumber\\
&&\hspace{0.0in} + w_2\;\delta_{\vec{k},\vec{p_2}} +
\overline{w}_2\;\delta_{\vec{k},\vec{-p_2}} + w_3\;\delta_{\vec{k},\vec{p_3}} +
\overline{w}_3\;\delta_{\vec{k},\vec{-p_3}} + w_4\;\delta_{\vec{k},\vec{p_4}} +
\overline{w}_4\;\delta_{\vec{k},\vec{-p_4}}\label{eq:restrict}
\eeq
where $\delta_{\vec{k},\vec{q}}$ equals one if $\vec{k}=\vec{q}$ and zero
otherwise (the two-dimensional Kronecker delta). This determines a
$\semi$-equivariant vector field $\Ft=[f,h]$ on $E^c(\ext)$ such
that $f$ and $h$ are homogeneous functions of degree p:
\beq
f(z,w)&=&\sum_{\vec{k_1}\in \tilde{A}(\k)} \sum_{\vec{k_2}\in \tilde{A}(\k)}
\ldots\sum_{\vec{k_p}\in\tilde{A}(\k)} a(\vec{k_1})\,a(\vec{k_2})\,\ldots
a(\vec{k_p})\,P(\vec{q_1},\vec{k_1},\ldots, \vec{k_p})\label{eq:48f}\\
&&\nonumber\\
h(z,w)&=&\sum_{\vec{k_1}\in\tilde{A}(\k)}\sum_{\vec{k_2}\in \tilde{A}(\k)}
\ldots\sum_{\vec{k_p}\in \tilde{A}(\k)} a(\vec{k_1})\,a(\vec{k_2})\,\ldots
a(\vec{k_p})\,P(\vec{p_1},\vec{k_1},\ldots, \vec{k_p})\label{eq:48h}
\eeq
where $\tilde{A}(\k)$ is given in (\ref{eq:wavev}).
When $f$ and $h$ are determined by monomials as in (\ref{eq:fmon}) and
(\ref{eq:hmon}) then these expressions become
\beq
m(z,w)+m(\g 3\cdot(z,w))&=&\sum_{\vec{k_1}\in \tilde{A}(\k)}
\ldots\sum_{\vec{k_p}\in\tilde{A}(\k)} a(\vec{k_1})\,\ldots
a(\vec{k_p})\,P(\vec{q_1},\vec{k_1},\ldots, \vec{k_p})\label{eq:48fmon}\\
&&\nonumber\\
m'(z,w)&=&\sum_{\vec{k_1}\in\tilde{A}(\k)} \ldots\sum_{\vec{k_p}\in
\tilde{A}(\k)} a(\vec{k_1})\,\ldots a(\vec{k_p})\,P(\vec{p_1},\vec{k_1},\ldots,
\vec{k_p}).\label{eq:48hmon}
\eeq
The essential problem of extension from $\Ft$ to $\strobe$ is formulated in
terms of these equations: if we choose monomials $m$ and $m'$ in
(\ref{eq:fmon}) and (\ref{eq:hmon}), does there exist a function $P$,
satisfying conditions (\ref{eq:irrcond1})-(\ref{eq:irrcond2}), such that
(\ref{eq:48fmon}) and (\ref{eq:48hmon}) will hold? We answer this by analyzing
a specific procedure for constructing $P$.

As a motivating example, consider the case of extending a linear vector field
$\Ft$. For
arbitrary real constants $\lambda$ and $\lambda'$ let
$m(z,w)=\lambda z_1/2$ and $m'(z,w)=\lambda' w_1$
so that $\overline{z}_1 m(z,w)$ and $\overline{w}_1 m'(z,w)$ are
$\tor$-invariant, and from (\ref{eq:fmon}) and (\ref{eq:hmon}), we have
$f(z,w)=\lambda z_1$ and $h(z,w)=\lambda' w_1.$
This determines $\Ft=[f,h]$ the most general {\em linear} $\semi$-equivariant
vector field:
\be
\Ft(z,w)=\left(
\begin{array}{c}
\lambda z_1\\
\lambda z_2\\
\lambda' w_1\\
\lambda' w_2\\
\lambda' w_3\\
\lambda' w_4
\end{array}
\right).\label{eq:48linearmap}
\ee
Note that the condition (\ref{eq:rotcond}) is not satisfied unless we impose
$\lambda=\lambda';$
in which case it is possible to extend $\Ft$ to $\strobe$ by taking
\be
P(\vec{k},\vec{k_1})=\lambda\,\delta_{\vec{k}\cdot\vec{k_1},\k^2}.
\label{eq:plin}
\ee
This choice clearly satisfies the conditions
(\ref{eq:irrcond1})-(\ref{eq:irrcond2}) since $\vec{k}\cdot\vec{k_1}$ is
$\otwo$-invariant and $P(\vec{k},\vec{k_1})=0$ unless $\vec{k}=\vec{k_1}$. It
is also straightforward to verify that we recover $f$ and $h$:
\beq
\sum_{\vec{k_1}\in \tilde{A}(\k)} a(\vec{k_1})\,P(\vec{q_1},\vec{k_1})&=&
\lambda \sum_{\vec{k_1}\in \tilde{A}(\k)} a(\vec{k_1})
\delta_{\vec{q_1}\cdot\vec{k_1},\k^2}\nonumber\\
&=&\lambda a(\vec{q_1})=\lambda z_1=f(z,w)
\eeq
and
\beq
\sum_{\vec{k_1}\in \tilde{A}(\k)} a(\vec{k_1})\,P(\vec{p_1},\vec{k_1})&=&
\lambda \sum_{\vec{k_1}\in \tilde{A}(\k)} a(\vec{k_1})
\delta_{\vec{p_1}\cdot\vec{k_1},\k^2}\nonumber\\
&=&\lambda a(\vec{p_1})=\lambda w_1=h(z,w).
\eeq
In this example the necessary condition (\ref{eq:rotcond}) is also sufficient;
this turns out to be true for the nonlinear case as well.

The procedure of constructing $P$ as in (\ref{eq:plin}) can be generalized and
formalized. The technique is to identify the set of wave vectors characteristic
of a monomial $m(z,w)$ and then use inner products between the wave vectors to
obtain an $\otwo$-invariant description of the monomial. This $\otwo$-invariant
description is formalized in the function $P$. In (\ref{eq:restrict}) each mode
amplitude $(z_1,z_2,w_1,w_2,w_3,w_4)$ and its complex conjugate is associated
with a distinct wave vector in $\tilde{A}(\k)$, so the amplitudes in a monomial
$m(z,w)$ determine an associated wave vector set for that monomial. For example
$m(z,w)=z_1z_2\overline{w}_3$ is associated with
$\{\vec{q_1},\vec{q_2},-\vec{p_3}\}$ and $m(z,w)=z_1|z_1|^2\,z_2\overline{w}_3$
is associated with $\{\pm\vec{q_1},\vec{q_2},-\vec{p_3}\}$. Note that we list
each vector only once. Thus for an arbitrary monomial $m(z,w)$ of degree $p$
the associated wave vector set $\{\vec{c_1},\vec{c_2},\ldots,\vec{c_n}\}$ is a
subset of $\tilde{A}(\k)$ with $n\leq p$.

We shall establish that (\ref{eq:48fmon}) and (\ref{eq:48hmon}) can be
satisfied separately using different functions $P_f$ and $P_h$. Then we show
that it is possible to satisfy (\ref{eq:48fmon}) and (\ref{eq:48hmon})
simultaneously with a single function $P$ only when $f$ and $h$ satisfy the
condition (\ref{eq:rotcond}).

The inner products amongst the vectors in $\tilde{A}(\k)$ are used below to
specify subsets of $\tilde{A}(\k)$. Given a vector $\vec{c_i}\in\tilde{A}(\k)$
denote the reflection in $\euc$ that fixes $\vec{c_i}$ by $\gamma_{\vec{c_i}}$,
i.e.
\be
\gamma_{\vec{c_i}}\cdot\vec{c_i}=\vec{c_i}.\label{eq:ciref}
\ee
For example, $\gamma_{\vec{q_1}}=\g 3$, $\gamma_{\vec{q_2}}=\gt 1$, and
$\gamma_{\vec{p_1}}=\rot{\phi}\g 3\rot{-\phi}.$ Thus $\gamma_{\vec{q_1}}$ and
$\gamma_{\vec{q_2}}$ leave $\tilde{A}(\k)$ invariant, but for ${\vec{p_i}},
\;i=1,2,3,4$ we have:
\beq
\tilde{A}(\k)\cap[\gamma_{\vec{p_1}}\tilde{A}(\k)]&=&
\tilde{A}(\k)\cap[\gamma_{\vec{p_4}}\tilde{A}(\k)]=
\{\pm\vec{p_1},\pm\vec{p_4}\}\label{eq:insect1}\\
\tilde{A}(\k)\cap[\gamma_{\vec{p_2}}\tilde{A}(\k)]&=&
\tilde{A}(\k)\cap[\gamma_{\vec{p_3}}\tilde{A}(\k)]=
\{\pm\vec{p_2},\pm\vec{p_3}\}.\label{eq:insect2}
\eeq
\begin{prop} Let $\{\vec{c_1},\vec{c_2},\ldots,\vec{c_n}\}$ denote a fixed
subset of $\tilde{A}(\k)$. Given $(\vec{k},\vec{k'})$, let
$\{\vec{k_1},\vec{k_2},\ldots,\vec{k_n}\}$ denote a second set of $n$ vectors
in $\tilde{A}(\k)$ required to satisfy the conditions:
\beq
\vec{k}\cdot\vec{k_i}&=&\vec{k'}\cdot\vec{c_i}\;\;\hspace{0.25in} \mbox{\rm for
}i=1,\ldots n\label{eq:cinner1}\\
\vec{k_i}\cdot\vec{k_j}&=&\vec{c_i}\cdot\vec{c_j}\;\;\hspace{0.25in} \mbox{\rm
for }i,j=1,\ldots n.\label{eq:cinner2}
\eeq
\begin{enumerate}
\item If $(\vec{k},\vec{k'})=(\vec{p_1},\vec{p_1})$, then the conditions
\mbox{\rm (\ref{eq:cinner1})-(\ref{eq:cinner2})}
can be satisfied in only one way:
\be
\{\vec{k_1},\vec{k_2},\ldots,\vec{k_n}\}=
\{\vec{c_1},\vec{c_2},\ldots,\vec{c_n}\}\label{eq:pset1}
\ee
unless
\be
\{\vec{c_1},\vec{c_2},\ldots,\vec{c_n}\}\subseteq
\tilde{A}(\k)\cap[\gamma_{\vec{p_1}}\tilde{A}(\k)] \label{eq:insect3}
\ee
in which case there is a second (not necessarily distinct) solution:
\be
\{\vec{k_1},\vec{k_2},\ldots,\vec{k_n}\}= \gamma_{\vec{p_1}}
\cdot\{\vec{c_1},\vec{c_2},\ldots,\vec{c_n}\}.\label{eq:pset2}
\ee
\item If $(\vec{k},\vec{k'})=(\vec{q_1},\vec{q_1})$, then the conditions
\mbox{\rm (\ref{eq:cinner1})-(\ref{eq:cinner2})} have only two (not necessarily
distinct) solutions:
\be
\{\vec{k_1},\vec{k_2},\ldots,\vec{k_n}\}=
\{\vec{c_1},\vec{c_2},\ldots,\vec{c_n}\}\label{eq:set1}
\ee
and
\be
\{\vec{k_1},\vec{k_2},\ldots,\vec{k_n}\}= \gamma_{\vec{q_1}}
\cdot\{\vec{c_1},\vec{c_2},\ldots,\vec{c_n}\}.\label{eq:set2}
\ee
\item If $(\vec{k},\vec{k'})=(\vec{q_1},\vec{p_1})$, then the conditions
\mbox{\rm (\ref{eq:cinner1})-(\ref{eq:cinner2})} have no solutions unless
\be
\{\vec{c_1},\vec{c_2},\ldots,\vec{c_n}\}\subseteq
\tilde{A}(\k)\cap[\rot{\phi}\tilde{A}(\k)].\label{eq:qpcond}
\ee
When \mbox{\rm (\ref{eq:qpcond})} holds then there are two (not necessarily
distinct) solutions:
\be
\{\vec{k_1},\vec{k_2},\ldots,\vec{k_n}\}=
\rot{-\phi}\{\vec{c_1},\vec{c_2},\ldots,\vec{c_n}\}\label{eq:qpset1}
\ee
and
\be
\{\vec{k_1},\vec{k_2},\ldots,\vec{k_n}\}= \g 3 \rot{-\phi} \cdot
\{\vec{c_1},\vec{c_2},\ldots,\vec{c_n}\}.\label{eq:qpset2}
\ee
\item If $(\vec{k},\vec{k'})=(\vec{p_1},\vec{q_1})$, then the conditions
\mbox{\rm (\ref{eq:cinner1})-(\ref{eq:cinner2})} have only two (not necessarily
distinct) solutions:
\be
\{\vec{k_1},\vec{k_2},\ldots,\vec{k_n}\}= \left\{\begin{array}{cc}
\rot{\phi}\cdot\{\vec{c_1},\vec{c_2},\ldots,\vec{c_n}\}&
\mbox{\rm if}\;\;\{\vec{c_1},\ldots,\vec{c_n}\}\subseteq
\tilde{A}(\k)\cap[\rot{-\phi}\tilde{A}(\k)]\\
&\\
\g 3\rot{-\phi}\cdot\{\vec{c_1},\vec{c_2},\ldots,\vec{c_n}\}&
\mbox{\rm if}\;\;\{\vec{c_1},\ldots,\vec{c_n}\}\subseteq
\tilde{A}(\k)\cap[\rot{\phi}\tilde{A}(\k)].
\end{array}\right.
\ee
Otherwise there are no solutions.
\end{enumerate}
\end{prop}
\noindent {\em {\bf Proof}.}\begin{quote} See the appendix. {\bf $\Box$}
\end{quote}

\begin{corollary}
If the monomial $\overline{w}_1 m'(z,w)$ is $\tor$-invariant and the wave
vector set $\{\vec{c_1},\vec{c_2},\ldots,\vec{c_n}\}$ for $m'(z,w)$ satisfies
\be
\{\vec{c_1},\vec{c_2},\ldots,\vec{c_n}\}\subseteq
\tilde{A}(\k)\cap[\gamma_{\vec{p_1}}\tilde{A}(\k)],\label{eq:h1}
\ee
then $m'(z,w)$ must have the form
\be
m'(z,w)=w_1 |w_1|^{2\nu_1'}|w_4|^{2\nu_4'}.\label{eq:cor1}
\ee
This implies
\beq
m'({z}_1,z_2,w_1,w_2,w_3,w_4)&=&m'({z}_1,z_2,w_1,0,0,w_4)\nonumber\\
&=&m'(0,0,w_1,0,0,w_4)\\
&=&m'(0,0,w_1,0,0,\overline{w}_4).\nonumber
\eeq
\end{corollary}
\noindent {\em {\bf Proof}.}
\begin{quote}
Since $\tilde{A}(\k)\cap[\gamma_{\vec{p_1}}\tilde{A}(\k)]=
\{\pm\vec{p_1},\pm\vec{p_4}\}$, the assumption in (\ref{eq:h1}) implies that
$\overline{w}_1 m'(z,w)$ depends only on the amplitudes
$(w_1,\overline{w}_1,w_4,\overline{w}_4)$. Then Proposition \mbox{\rm III.2}
asserts that $(\nu'_{1},\nu'_{4})$ are the only possible non-zero exponents so
that
\be
\overline{w}_1 m'(z,w)=|w_1|^{2\nu'_{1}}|w_4|^{2\nu'_{4}}
\ee
with $\nu'_{1}\geq 1$ and $\nu'_{4}\geq 0$. Relabeling
$\nu'_{1}\rightarrow\nu'_{1}+1$ gives (\ref{eq:cor1}). {\bf $\Box$}
\end{quote}

\begin{corollary}
If the monomial $\overline{z}_1 m(z,w)$ is $\tor$-invariant. Then
$f(z,w)=m(z,w)+m(\g 3\cdot(z,w))$ satisfies
\be
f(z_1,z_2,0,w_2,w_3,0)=\left\{\begin{array}{cc}
2\,m(z,0)&\mbox{\rm if}\;\;m(z,0)\neq0\\
&\\
m(z_1,z_2,0,w_2,w_3,0)&\mbox{\rm if}\;\;m(z,0)=0\;\;\mbox{\rm and}\\
& \{\vec{c_1},\ldots,\vec{c_n}\}\subseteq
\tilde{A}(\k)\cap[\rot{-\phi}\tilde{A}(\k)]\\
&\\
m(\g 3\cdot(z_1,z_2,0,w_2,w_3,0))&\mbox{\rm if}\;\;m(z,0)=0\;\;\mbox{\rm and}\\
&\{\vec{c_1},\ldots,\vec{c_n}\}\subseteq
\tilde{A}(\k)\cap[\rot{\phi}\tilde{A}(\k)]\\
&\\
0&\mbox{\rm otherwise}
\end{array}\right.
\ee
where $\{\vec{c_1},\ldots,\vec{c_n}\}$ is the wave vector set for $m(z,w)$.
\end{corollary}
\noindent {\em {\bf Proof}.} \begin{quote} By definition
\be
f(z_1,z_2,0,w_2,w_3,0)=m(z_1,z_2,0,w_2,w_3,0)+m(\g
3\cdot(z_1,z_2,0,w_2,w_3,0)).
\label{eq:fonh}
\ee
The first term vanishes unless $\{\vec{c_1},\ldots,\vec{c_n}\}\subseteq
\tilde{A}(\k)\cap[\rot{-\phi}\tilde{A}(\k)]$ and the second term vanishes
unless $\{\vec{c_1},\ldots,\vec{c_n}\}\subseteq
\tilde{A}(\k)\cap[\rot{\phi}\tilde{A}(\k)]$. Both terms can be non-zero only if
$m(z,w)$ is independent of $w$ or equivalently $m(z,0)\neq0$; in this case
$m(z,0)=m(\g 3\cdot(z,0))$ so (\ref{eq:fonh}) yields $2m(z,0)$.
{\bf $\Box$}\end{quote}

\begin{lemma}  Let $\overline{z}_1 m(z,w)$ and $\overline{w}_1 m'(z,w)$ be
$\tor$-invariant monomials of degree p such that $f(z,w)=m(z,w)+m(\g
3\cdot(z,w))$ and $h(z,w)=m'(z,w)$ meet the conditions of Proposition {\rm
III.1}. There exist functions $P_f$ and $P_h$ satisfying \mbox{\rm
(\ref{eq:irrcond1})-(\ref{eq:irrcond2})} such that
\beq
f(z,w)&=&\sum_{\vec{k_1}\in \tilde{A}(\k)}\sum_{\vec{k_2}\in \tilde{A}(\k)}
\ldots\sum_{\vec{k_p}\in\tilde{A}(\k)} a(\vec{k_1})\,a(\vec{k_2})\,\ldots
a(\vec{k_p})\,P_f(\vec{q_1},\vec{k_1},\ldots, \vec{k_p})\label{eq:pff}\\
&&\nonumber\\
h(z,w)&=&\sum_{\vec{k_1}\in\tilde{A}(\k)}\sum_{\vec{k_2}\in \tilde{A}(\k)}
\ldots\sum_{\vec{k_p}\in \tilde{A}(\k)} a(\vec{k_1})\,a(\vec{k_2})\,\ldots
a(\vec{k_p})\,P_h(\vec{p_1},\vec{k_1},\ldots, \vec{k_p})\label{eq:phh}
\eeq
and
\be
\sum_{\vec{k_1}\in \tilde{A}(\k)} \ldots\sum_{\vec{k_p}\in\tilde{A}(\k)}
a(\vec{k_1})\,\ldots a(\vec{k_p})\,P_f(\vec{p_1},\vec{k_1},\ldots,
\vec{k_p})=f(w_1,\overline{w}_4,0,z_1,z_2,0)\label{eq:pfh}
\ee
\beq
\lefteqn{\sum_{\vec{k_1}\in\tilde{A}(\k)} \ldots\sum_{\vec{k_p}\in
\tilde{A}(\k)} a(\vec{k_1})\,\ldots
a(\vec{k_p})\,P_h(\vec{q_1},\vec{k_1},\ldots, \vec{k_p})}\nonumber\\
&=&\left\{
\begin{array}{cc}
h(w_2,w_3,z_1,0,0,\overline{z}_2)&\mbox{\rm if}\;\;h(0,w)\neq0\\
h(w_2,w_3,z_1,0,0,\overline{z}_2) + h(w_1,w_4,z_1,0,0,{z}_2)& \mbox{\rm
if}\;\;h(0,w)=0
\end{array}
\right.\label{eq:phf}
\eeq
where $a(\vec{k})$ is given by \mbox{\rm (\ref{eq:restrict})}.
\end{lemma}
\noindent {\em {\bf Proof}.}\begin{quote}
The procedure for constructing $P_f$ and $P_h$ is essentially the same so we
present it in detail only for $P_h$.
\begin{enumerate}
\item Let $\{\vec{c_1},\vec{c_2},\ldots,\vec{c_n}\}$ denote the wave vector set
associated with the monomial $m'(z,w)$ used to define $h(z,w)$; note that
$n\leq p$ since $m'(z,w)$ has degree $p$. In order to allow for monomials
satisfying the special condition (\ref{eq:insect3}), define the numerical
factor
\be
\epsilon=\left\{\begin{array}{cc}
\frac{1}{2}&\mbox{\rm if}\;\;\{\vec{c_1},\ldots,\vec{c_n}\}\subseteq
\tilde{A}(\k)\cap[\gamma_{\vec{p_1}}\tilde{A}(\k)]\;\; \mbox{\rm and}\\
& \{\vec{c_1},\ldots,\vec{c_n}\}\neq
\gamma_{\vec{p_1}}\{\vec{c_1},\ldots,\vec{c_n}\}\\
&\\
1&\mbox{\rm otherwise};
\end{array}
\right.
\ee
the $\epsilon=1/2$ monomials are described in Corollary {\mbox\rm III.1}. First
we define $P_h$ when $n=p$, then $m'(z,w)$ is entirely specified by its wave
vector set:
\be
m'(z,w)=a(\vec{c_1})\,a(\vec{c_2})\,\ldots a(\vec{c_{n}})\hspace{0.5in}(n=p).
\label{eq:mprimea}
\ee
The inner product relations for $\{\vec{c_1},\vec{c_2},\ldots,\vec{c_n}\}$,
\beq
\vec{k}\cdot\vec{k_i}&=&\vec{p_1}\cdot\vec{c_i}\;\;\hspace{0.25in} \mbox{\rm
for }i=1,\ldots n\label{eq:ph1}\\
\vec{k_i}\cdot\vec{k_j}&=&\vec{c_i}\cdot\vec{c_j}\;\;\hspace{0.25in} \mbox{\rm
for }i,j=1,\ldots n,\label{eq:ph2}
\eeq
determine our choice for $P_h(\vec{k},\vec{k_1},\ldots, \vec{k_p})$:
\be
P_h(\vec{k},\vec{k_1},\ldots, \vec{k_p})=\epsilon\, \prod_{i=1}^{p}
\prod_{j=1}^{p}\,\delta_{\vec{k}\cdot\vec{k_i},\vec{p_1}\cdot\vec{c_i}}
\;\delta_{\vec{k_i}\cdot\vec{k_j},\vec{c_i}\cdot\vec{c_j}}\hspace{0.5in}(n=p).
\label{eq:phdef1}
\ee
The $\otwo$-invariance of $P_h$ required by (\ref{eq:irrcond2}) is assured by
the $\otwo$-invariance of the inner product. Translation symmetry required by
(\ref{eq:irrcond1}) is discussed below. By Proposition {\mbox\rm III.5}, if
$\epsilon=1/2$ then
\be
P_h(\vec{p_1},\vec{k_1},\ldots,\vec{k_p})=\left\{\begin{array}{cc}
\frac{1}{2}&\mbox{\rm if}\;\; \{\vec{k_1},\vec{k_2},\ldots,\vec{k_n}\}=
\{\vec{c_1},\vec{c_2},\ldots,\vec{c_n}\}\\
&\\
\frac{1}{2}&\mbox{\rm if}\;\; \{\vec{k_1},\vec{k_2},\ldots,\vec{k_n}\}=
\gamma_{\vec{p_1}} \cdot\{\vec{c_1},\vec{c_2},\ldots,\vec{c_n}\}\\
&\\
0&\mbox{\rm otherwise}
\end{array}\right.
\ee
and if $\epsilon=1$ then
\be
P_h(\vec{p_1},\vec{k_1},\ldots,\vec{k_p})=\left\{\begin{array}{cc}
{1}&\mbox{\rm if}\;\; \{\vec{k_1},\vec{k_2},\ldots,\vec{k_n}\}=
\{\vec{c_1},\vec{c_2},\ldots,\vec{c_n}\}\\
&\\
0&\mbox{\rm otherwise}.
\end{array}\right.\label{eq:php1a}
\ee
Thus for $\epsilon=1$ we find
\beq
\sum_{\vec{k_1}\in\tilde{A}(\k)} \ldots\sum_{\vec{k_p}\in \tilde{A}(\k)}
a(\vec{k_1})\,\ldots a(\vec{k_p})\,
P_h(\vec{p_1},\vec{k_1},\ldots, \vec{k_p})\nonumber\\
&=&a(\vec{c_1})\,a(\vec{c_2})\,\ldots a(\vec{c_n})\nonumber\\
&=& m'(z,w)\nonumber\\
&=& h(z,w),\label{eq:sumphp1a}
\eeq
and for $\epsilon=1/2$ the same sum can be evaluated using Corollary \mbox{\rm
III.1} to give the same result
\beq
\lefteqn{\sum_{\vec{k_1}\in\tilde{A}(\k)} \ldots\sum_{\vec{k_p}\in
\tilde{A}(\k)} a(\vec{k_1})\,\ldots a(\vec{k_p})\,
P_h(\vec{p_1},\vec{k_1},\ldots, \vec{k_p})}\nonumber\\
&=&\frac{1}{2}\left[a(\vec{c_1})\,a(\vec{c_2})\,\ldots a(\vec{c_n})+
a(\gamma_{\vec{p_1}}\cdot\vec{c_1})\,a(\gamma_{\vec{p_1}}\cdot\vec{c_2})\,\ldots a(\gamma_{\vec{p_1}}\cdot\vec{c_n})\right]\nonumber\\
&=& \frac{1}{2}\left[m'(0,0,w_1,0,0,w_4)+
m'(0,0,w_1,0,0,\overline{w}_4)\right]\nonumber\\
&=& \frac{1}{2}\left[m'(z,w)+m'(z,w)\right]\nonumber\\
&=& h(z,w).
\eeq
This verifies (\ref{eq:phh}) for the case $n=p$.
{}From Proposition {\mbox\rm III.5} we can evaluate
$P_h(\vec{q_1},\vec{k_1},\ldots, \vec{k_p})$ similarly.
\be
P_h(\vec{q_1},\vec{k_1},\ldots,\vec{k_p})=\left\{\begin{array}{cc}
\epsilon&\mbox{\rm if}\;\; \{\vec{k_1},\vec{k_2},\ldots,\vec{k_n}\}=
\rot{-\phi}\{\vec{c_1},\vec{c_2},\ldots,\vec{c_n}\}\;\;\mbox{\rm and}\\
&\{\vec{c_1},\vec{c_2},\ldots,\vec{c_n}\}\subseteq
\tilde{A}(\k)\cap[\rot\phi\tilde{A}(\k)]\\
&\\
\epsilon&\mbox{\rm if}\;\; \{\vec{k_1},\vec{k_2},\ldots,\vec{k_n}\}= \g 3
\rot{-\phi}\cdot\{\vec{c_1},\vec{c_2},\ldots,\vec{c_n}\}\;\;\mbox{\rm and}\\
&\{\vec{c_1},\vec{c_2},\ldots,\vec{c_n}\}\subseteq
\tilde{A}(\k)\cap[\rot\phi\tilde{A}(\k)]\\
&\\
0&\mbox{\rm otherwise}.
\end{array}\right.\label{eq:phfcn}
\ee
When
\be
\{\vec{c_1},\vec{c_2},\ldots,\vec{c_n}\}\not\subseteq
\tilde{A}(\k)\cap[\rot\phi\tilde{A}(\k)],\label{eq:pheq0}
\ee
then $P_h(\vec{q_1},\vec{k_1},\ldots, \vec{k_p})=0$ in (\ref{eq:phfcn}) and the
left hand side of (\ref{eq:phf}) is zero. The right hand side is also zero
since
(\ref{eq:pheq0}) implies $h(z_1,z_2,w_1,0,0,w_4)=0$. Now consider
(\ref{eq:phf}) when
\be
\{\vec{c_1},\vec{c_2},\ldots,\vec{c_n}\}\subseteq
\tilde{A}(\k)\cap[\rot\phi\tilde{A}(\k)].\label{eq:phf00}
\ee
If $h(0,w)=m'(0,w)=0$, then $\epsilon=1$ and
$\gamma_{\vec{p_1}}\{\vec{c_1},\vec{c_2},\ldots,\vec{c_n}\}\neq
\{\vec{c_1},\vec{c_2},\ldots,\vec{c_n}\}$; in this case we find
\beq
\lefteqn{\sum_{\vec{k_1}\in\tilde{A}(\k)} \ldots\sum_{\vec{k_p}\in
\tilde{A}(\k)} a(\vec{k_1})\,\ldots a(\vec{k_p})\,
P_h(\vec{q_1},\vec{k_1},\ldots, \vec{k_p})}\nonumber\\
&=&\left[a(\rot{-\phi}\cdot\vec{c_1})\,a(\rot{-\phi}\cdot\vec{c_2})\,\ldots
a(\rot{-\phi}\cdot\vec{c_n})+\nonumber\right.\\
&&\left.\hspace{0.5in} a(\g 3\rot{-\phi}\cdot\vec{c_1})\, a(\g
3\rot{-\phi}\cdot\vec{c_2})\,\ldots a(\g
3\rot{-\phi}\cdot\vec{c_n})\right]\nonumber\\
&=& \left[m'(w_2,w_3,z_1,0,0,\overline{z}_2)+
m'(w_1,w_4,z_1,0,0,{z}_2)\right]\nonumber\\
&=& \left[h(w_2,w_3,z_1,0,0,\overline{z}_2)+ h(w_1,w_4,z_1,0,0,{z}_2)\right].
\label{eq:phfhh}
\eeq
If $h(0,w)=m'(0,w)\neq0$, then $m'$ is described in Corollary III.1, and the
two
terms in (\ref{eq:phfhh}) are equal. If $\epsilon=1$, then the sets
$\{\vec{k_1},\vec{k_2},\ldots,\vec{k_n}\}$ in (\ref{eq:phfcn}) are identical
and we get only one term from the sum in (\ref{eq:phfhh}). If $\epsilon=1/2$
the sets $\{\vec{k_1},\vec{k_2},\ldots,\vec{k_n}\}$ in (\ref{eq:phfcn}) are
distinct and we get both terms in (\ref{eq:phfhh}). Either way the result is
the same:
\beq
\lefteqn{\sum_{\vec{k_1}\in\tilde{A}(\k)} \ldots\sum_{\vec{k_p}\in
\tilde{A}(\k)} a(\vec{k_1})\,\ldots a(\vec{k_p})\,
P_h(\vec{q_1},\vec{k_1},\ldots, \vec{k_p})}\nonumber\\
&=&\left\{\begin{array}{cc}
m'(w_2,w_3,z_1,0,0,\overline{z}_2)&\mbox{\rm if}\;\;\epsilon=1\\
&\\
\frac{1}{2}\left[m'(w_2,w_3,z_1,0,0,\overline{z}_2)+
m'(w_1,w_4,z_1,0,0,{z}_2)\right]&\mbox{\rm if}\;\;\epsilon=1/2
\end{array}\right.\nonumber\\
&=&m'(w_2,w_3,z_1,0,0,\overline{z}_2)\nonumber\\
&=&h(w_2,w_3,z_1,0,0,\overline{z}_2).
\label{eq:phcor}
\eeq
This verifies (\ref{eq:phf}) for the case $n=p$.

\item This construction of $P_h$ easily extends to the general case with $n\leq
p$. For these monomials the expression (\ref{eq:mprimea}) for $m'(z,w)$ becomes
\be
m'(z,w)=a(\vec{c_1})\,a(\vec{c_2})\ldots a(\vec{c_{n}})a(\vec{c}_{l_{n+1}})
\ldots a(\vec{c}_{l_{p}})\label{eq:mprimeb}
\ee
where the subscripts $\{l_{n+1},\ldots,l_{p}\}$ depend on the monomial $m'$
considered. Noting that the inner product relations
$\vec{k}_j\cdot\vec{c}_{l_{j}}=\k^2$ for $j=n+1,\ldots, p$
require $\vec{k}_j=\vec{c}_{l_{j}}$ we replace (\ref{eq:phdef1}) by
\be
P_h(\vec{k},\vec{k_1},\ldots, \vec{k_p})=\epsilon\,\left(\prod_{i=1}^{n}
\prod_{j=1}^{n}\,\delta_{\vec{k}\cdot\vec{k_i},\vec{p_1}\cdot\vec{c_i}}
\;\delta_{\vec{k_i}\cdot\vec{k_j},\vec{c_i}\cdot\vec{c_j}}\right)\,
\left(\prod_{j=n+1}^{p}\, \delta_{\vec{k_j}\cdot\vec{c}_{l_j},\k^2}\right).
\label{eq:phdef2}
\ee
Then the previous calculations generalize immediately by making the obvious
changes. For example for $\epsilon=1$ (\ref{eq:php1a}) becomes
\be
P_h(\vec{p_1},\vec{k_1},\ldots,\vec{k_p})=\left\{\begin{array}{cc}
{1}&\mbox{\rm if}\;\; \{\vec{k_1},\vec{k_2},\ldots,\vec{k_p}\}=
\{\vec{c_1},\vec{c_2},\ldots,\vec{c_n},\vec{c}_{l_{n+1}},\ldots,
\vec{c}_{l_{p}}\}\\
&\\
0&\mbox{\rm otherwise};
\end{array}\right.\label{eq:php1b}
\ee
so when we evaluate the sum in (\ref{eq:sumphp1a}) we still find
\beq
\lefteqn{\sum_{\vec{k_1}\in\tilde{A}(\k)} \ldots\sum_{\vec{k_p}\in
\tilde{A}(\k)} a(\vec{k_1})\,\ldots a(\vec{k_p})\,
P_h(\vec{p_1},\vec{k_1},\ldots, \vec{k_p})}\nonumber\\
&=&a(\vec{c_1})\,a(\vec{c_2})\,\ldots a(\vec{c_n})
\sum_{\vec{k}_{n+1}\in\tilde{A}(\k)}...\sum_{\vec{k_p}\in \tilde{A}(\k)}
a(\vec{k}_{n+1})\ldots a(\vec{k_p})\left(\prod_{j=n+1}^{p}\,
\delta_{\vec{k_j}\cdot\vec{c}_{l_j},\k^2}\right) \nonumber\\
&=& a(\vec{c_1})\,\ldots a(\vec{c_n}) a(\vec{c}_{l_{n+1}})\,\ldots
a(\vec{c}_{l_p})\nonumber\\
&=& m'(z,w)\nonumber\\
&=& h(z,w).\label{eq:sumphp1b}
\eeq

\item It remains to verify that $P_h(\vec{k},\vec{k_1},\ldots, \vec{k_p})$ in
(\ref{eq:phdef2}) satisfies the requirement of translation symmetry in
(\ref{eq:irrcond1}). It suffices to prove that
\be
\vec{k}=\vec{k_1}+\vec{k_2}+\cdots+\vec{k_p}\label{eq:sumvec}
\ee
holds whenever
$P_h(\vec{k},\vec{k_1},\ldots, \vec{k_p})\neq0.$
Note that since $|\vec{k}|=\k=|\vec{p_1}|$ in (\ref{eq:sumvec}) we can always
rotate the arguments of $P_h$ so that $\vec{k}=\vec{p_1}$ using the
$\otwo$-invariance of $P_h$:
\be
P_h(\vec{k},\vec{k_1},\ldots, \vec{k_p})=
P_h(\vec{p_1},\rot\theta\cdot\vec{k_1},\ldots, \rot\theta\cdot\vec{k_p}).
\ee
where $\vec{k}=\rot{-\theta}\cdot\vec{p_1}$. Thus we need only verify
(\ref{eq:irrcond1}) for the specific case $\vec{k}=\vec{p_1}$. By Proposition
{\mbox\rm III.5} $P_h(\vec{p_1},\vec{k_1},\ldots,\vec{k_p})\neq0$ can only
occur if
\be
\{\vec{k_1},\vec{k_2},\ldots,\vec{k_p}\}=
\{\vec{c_1},\vec{c_2},\ldots,\vec{c_n},\vec{c}_{l_{n+1}},\ldots,
\vec{c}_{l_{p}}\}\label{eq:pset1b}
\ee
or
\be
\{\vec{k_1},\vec{k_2},\ldots,\vec{k_p}\}=\gamma_{\vec{p_1}}
\cdot\{\vec{c_1},\vec{c_2},\ldots,\vec{c_n},\vec{c}_{l_{n+1}},\ldots,
\vec{c}_{l_{p}}\}.\label{eq:pset2b}
\ee
If we express $\overline{w}_1 m'(z,w)$ in the notation of Proposition {\mbox\rm
III.2}, then
\be
m'(z,w)=\frac{\sigma_1^{\mu_1'}\sigma_2^{\mu_2'}
\rho_1^{\nu_1'}\rho_2^{\nu_2'}\rho_3^{\nu_3'}\rho_4^{\nu_4'}
\Omega_1^{\mu_1}\Omega_2^{\mu_2}
\omega_1^{\nu_1}\omega_2^{\nu_2}\omega_3^{\nu_3}\omega_4^{\nu_4}}
{\overline{w}_1}
\ee
and from (\ref{eq:mprimeb}) and (\ref{eq:th1})-(\ref{eq:th2})
\beq
\vec{c_1}+\vec{c_2}+\cdots+\vec{c_n}+\vec{c}_{l_{n+1}}+\cdots+\vec{c}_{l_{p}}
&=& \mu_1\,\vec{q_1}+\mu_2\,\vec{q_2}+(\nu_1+1)\vec{p_1}+\nu_2\vec{p_2}
+\nu_3\vec{p_3}+\nu_4\vec{p_4}\nonumber\\
&=&\mu_1\,(l_1,0)+\mu_2\,(0,l_1)+(\nu_1+1)(l_2,n_2)+\nonumber\\
&&\nu_2(l_2,-n_2) +\nu_3(n_2,l_2)+\nu_4\,(n_2,-l_2)\nonumber\\
&=&(l_2,n_2)=\vec{p_1}.
\eeq
Since $\gamma_{\vec{p_1}} \cdot\vec{p_1}=\vec{p_1}$ this implies
\be
\gamma_{\vec{p_1}} \cdot(\vec{c_1}+\vec{c_2}+\cdots+\vec{c_n}+
\vec{c}_{l_{n+1}}+\cdots+\vec{c}_{l_{p}})=\vec{p_1}.
\ee
Thus for $\vec{k}=\vec{p_1}$, both possible solutions
(\ref{eq:pset1b})-(\ref{eq:pset2b}) satisfy the condition (\ref{eq:sumvec}).
This proves that $P_h(\vec{k},\vec{k_1},\ldots, \vec{k_p})$ satisfies the
condition (\ref{eq:irrcond1}).

\item  For $P_f(\vec{k},\vec{k_1},\ldots, \vec{k_p})$ we let
$\{\vec{c_1},\vec{c_2},\ldots,\vec{c_n}\}$ denote the wave vector set
associated with the monomial $m(z,w)$ used to define $f(z,w)=m(z,w)+m(\g
3\cdot(z,w))$. Since $f$ is defined to have two terms even when $m(z,w)$ and
$m(\g 3\cdot(z,w))$ are identical, no numerical factor $\epsilon$ is required
in this case. The monomial $m(z,w)$ can be written as
\be
m(z,w)=a(\vec{c_1})\,a(\vec{c_2})\ldots a(\vec{c_{n}})a(\vec{c}_{l_{n+1}})
\ldots a(\vec{c}_{l_{p}})\label{eq:mb}
\ee
in analogy with (\ref{eq:mprimeb}), and using the inner product relations,
\beq
\vec{k}\cdot\vec{k_i}&=&\vec{q_1}\cdot\vec{c_i}\;\;\hspace{0.25in} \mbox{\rm
for }i=1,\ldots n\label{eq:pf1}\\
\vec{k_i}\cdot\vec{k_j}&=&\vec{c_i}\cdot\vec{c_j}\;\;\hspace{0.25in} \mbox{\rm
for }i,j=1,\ldots n,\label{eq:pf2}
\eeq
we define $P_f$:
\be
P_f(\vec{k},\vec{k_1},\ldots, \vec{k_p})=\left(\prod_{i=1}^{n}
\prod_{j=1}^{n}\,\delta_{\vec{k}\cdot\vec{k_i},\vec{q_1}\cdot\vec{c_i}}
\;\delta_{\vec{k_i}\cdot\vec{k_j},\vec{c_i}\cdot\vec{c_j}}\right)\,
\left(\prod_{j=n+1}^{p}\, \delta_{\vec{k_j}\cdot\vec{c}_{l_j},\k^2}\right).
\label{eq:pfdef2}
\ee
It then follows from Proposition {\mbox\rm III.5} that for $\vec{k}=\vec{q_1}$
and $\vec{k}=\vec{p_1}$ we have
\be
P_f(\vec{q_1},\vec{k_1},\ldots,\vec{k_p})=\left\{\begin{array}{cc}
{1}&\mbox{\rm if}\;\; \{\vec{k_1},\vec{k_2},\ldots,\vec{k_p}\}=
\{\vec{c_1},\vec{c_2},\ldots,\vec{c_n},\vec{c}_{l_{n+1}},\ldots,
\vec{c}_{l_{p}}\}\\
&\\
{1}&\mbox{\rm if}\;\; \{\vec{k_1},\vec{k_2},\ldots,\vec{k_p}\}=
\gamma_{\vec{q_1}}\cdot\{\vec{c_1},\vec{c_2},\ldots,\vec{c_n},\vec{c}_{l_{n+1}},\ldots, \vec{c}_{l_{p}}\}\\
&\\
0&\mbox{\rm otherwise};
\end{array}\right.\label{eq:pfp1b}
\ee
and
\be
P_f(\vec{p_1},\vec{k_1},\ldots,\vec{k_p})=\left\{\begin{array}{cc}
{1}&\mbox{\rm if}\;\; \{\vec{c_1},\ldots,\vec{c_n}\}\subseteq
\tilde{A}(\k)\cap[\rot{-\phi}\tilde{A}(\k)]\;{\rm and}\\
&\{\vec{k_1},\vec{k_2},\ldots,\vec{k_p}\}=
\rot{\phi}\cdot\{\vec{c_1},\vec{c_2},\ldots,\vec{c_n},\vec{c}_{l_{n+1}},\ldots,
\vec{c}_{l_{p}}\}\\
&\\
{1}&\mbox{\rm if}\;\;\{\vec{c_1},\ldots,\vec{c_n}\}\subseteq
\tilde{A}(\k)\cap[\rot{\phi}\tilde{A}(\k)]\; {\rm and}\\
& \{\vec{k_1},\vec{k_2},\ldots,\vec{k_p}\}= \g 3\rot{-\phi}\cdot
\{\vec{c_1},\vec{c_2},\ldots,\vec{c_n},\vec{c}_{l_{n+1}},\ldots,
\vec{c}_{l_{p}}\}\\
&\\
0&\mbox{\rm otherwise};
\end{array}\right.\label{eq:pfpq1}
\ee
respectively. Note that {\em both} contributions in (\ref{eq:pfpq1}) arise for
the same monomial when
$\{\vec{c_1},\ldots,\vec{c_n}\}\subseteq\{\pm\vec{q_1},\pm\vec{q_2}\}$ or
equivalently $m(z,0)\neq0$. From (\ref{eq:pfp1b}) the relation in
(\ref{eq:pff}) follows immediately, and we evaluate the sum in (\ref{eq:pfh})
using (\ref{eq:pfpq1}) and Corollary {\mbox\rm III.2}:
\beq
\lefteqn{\sum_{\vec{k_1}\in \tilde{A}(\k)}
\ldots\sum_{\vec{k_p}\in\tilde{A}(\k)} a(\vec{k_1})\,\ldots
a(\vec{k_p})\,P_f(\vec{p_1},\vec{k_1},\ldots, \vec{k_p})}\nonumber\\
&=&
\left\{\begin{array}{cc}
m(w_1,\overline{w}_4,0,0,0,0)+&\mbox{\rm if}\;\;m(z,0)\neq0\\
\hspace{0.5in}m(w_1,{w}_4,0,0,0,0)&\\
&\\
m(w_1,\overline{w}_4,0,z_1,z_2,0)&
\mbox{\rm if}\;\;m(z,0)=0\;\;\mbox{\rm and}\\
& \{\vec{c_1},\ldots,\vec{c_n}\}\subseteq
\tilde{A}(\k)\cap[\rot{-\phi}\tilde{A}(\k)]\\
&\\
m(w_1,{w}_4,z_1,0,0,z_2)&
\mbox{\rm if}\;\;m(z,0)=0\;\;\mbox{\rm and}\\
&\{\vec{c_1},\ldots,\vec{c_n}\}\subseteq
\tilde{A}(\k)\cap[\rot{\phi}\tilde{A}(\k)]\\
&\\
0&\mbox{\rm otherwise}
\end{array}\right.\nonumber\\
&&\nonumber\\
&=&
\left\{\begin{array}{cc}
m(w_1,\overline{w}_4,0,0,0,0)+&\mbox{\rm if}\;\;m(z,0)\neq0\\
\hspace{0.5in}m(\g 3\cdot(w_1,\overline{w}_4,0,0,0,0))&\\
&\\
m(w_1,\overline{w}_4,0,z_1,z_2,0)&
\mbox{\rm if}\;\;m(z,0)=0\;\;\mbox{\rm and}\\
& \{\vec{c_1},\ldots,\vec{c_n}\}\subseteq
\tilde{A}(\k)\cap[\rot{-\phi}\tilde{A}(\k)]\\
&\\
m(\g 3\cdot(w_1,\overline{w}_4,0,z_1,z_2,0))&
\mbox{\rm if}\;\;m(z,0)=0\;\;\mbox{\rm and}\\
&\{\vec{c_1},\ldots,\vec{c_n}\}\subseteq
\tilde{A}(\k)\cap[\rot{\phi}\tilde{A}(\k)]\\
&\\
0&\mbox{\rm otherwise}
\end{array}\right.
\nonumber\\
&&\nonumber\\
&=&f(w_1,\overline{w}_4,0,z_1,z_2,0).
\eeq
Finally the translation symmetry (\ref{eq:irrcond1}) of $P_f$ can be proved as
in the case of $P_h$. {\bf $\Box$}
\end{enumerate}
\end{quote}

\begin{lemma}  Let $\overline{z}_1 m(z,w)$ and $\overline{w}_1 m'(z,w)$ be
$\tor$-invariant monomials of degree p such that $f(z,w)=m(z,w)+m(\g
3\cdot(z,w))$ and $h(z,w)=m'(z,w)$ meet the conditions of Proposition {\rm
III.1}. There exists a single function $P$ satisfying \mbox{\rm
(\ref{eq:irrcond1})-(\ref{eq:irrcond2})} such that
\beq
f(z,w)&=&\sum_{\vec{k_1}\in \tilde{A}(\k)}\sum_{\vec{k_2}\in \tilde{A}(\k)}
\ldots\sum_{\vec{k_p}\in\tilde{A}(\k)} a(\vec{k_1})\,a(\vec{k_2})\,\ldots
a(\vec{k_p})\,P(\vec{q_1},\vec{k_1},\ldots, \vec{k_p})\label{eq:pf}\\
&&\nonumber\\
h(z,w)&=&\sum_{\vec{k_1}\in\tilde{A}(\k)}\sum_{\vec{k_2}\in \tilde{A}(\k)}
\ldots\sum_{\vec{k_p}\in \tilde{A}(\k)} a(\vec{k_1})\,a(\vec{k_2})\,\ldots
a(\vec{k_p})\,P(\vec{p_1},\vec{k_1},\ldots, \vec{k_p})\label{eq:ph}
\eeq
if and only if
\be
f(z_1,z_2,0,w_2,w_3,0)=h(w_2,w_3,z_1,0,0,\overline{z}_2).\label{eq:rotconda}
\ee
\end{lemma}
\noindent {\em {\bf Proof}.}\begin{quote} The necessity of (\ref{eq:rotconda})
is established by Proposition {\rm III.4} and the proof of necessity in Theorem
{\rm III.1}. Assume that (\ref{eq:rotconda}) holds and let
$P_f(\vec{k},\vec{k_1},\ldots, \vec{k_p})$ and $P_h(\vec{k},\vec{k_1},\ldots,
\vec{k_p})$ be the functions of Lemma {\rm III.1} for $f$ and $h$ respectively.
\begin{enumerate}
\item If $f(z_1,z_2,0,w_2,w_3,0)=0$ then $h(w_2,w_3,z_1,0,0,\overline{z}_2)=0$
and
\beq
0&=&\sum_{\vec{k_1}\in \tilde{A}(\k)} \ldots\sum_{\vec{k_p}\in\tilde{A}(\k)}
a(\vec{k_1})\,\ldots a(\vec{k_p})\,P_f(\vec{p_1},\vec{k_1},\ldots, \vec{k_p})
\label{eq:pfh0}\\
&&\nonumber\\
0&=&\sum_{\vec{k_1}\in\tilde{A}(\k)} \ldots\sum_{\vec{k_p}\in \tilde{A}(\k)}
a(\vec{k_1})\,\ldots a(\vec{k_p})\,P_h(\vec{q_1},\vec{k_1},\ldots, \vec{k_p})
\label{eq:phf0}
\eeq
from (\ref{eq:pfh})-(\ref{eq:phf}). In this case we can define
\be
P(\vec{k},\vec{k_1},\ldots, \vec{k_p})\equiv P_f(\vec{k},\vec{k_1},\ldots,
\vec{k_p}) + P_h(\vec{k},\vec{k_1},\ldots, \vec{k_p});
\ee
the equations (\ref{eq:pf})-(\ref{eq:ph}) then follow from
(\ref{eq:pff})-(\ref{eq:phh}) and (\ref{eq:pfh0})-(\ref{eq:phf0}).

\item If instead $f(z_1,z_2,0,w_2,w_3,0)\neq 0$ then
$h(w_2,w_3,z_1,0,0,\overline{z}_2)\neq 0$ and the monomial $m'(z,w)$ must be
independent of $(w_2,w_3,\overline{w}_2,\overline{w}_3)$:
\be
m'(z_1,z_2,w_1,w_2,w_3,w_4)=m'(z_1,z_2,w_1,0,0,w_4)\neq 0.
\ee
In this case we define
$P(\vec{k},\vec{k_1},\ldots, \vec{k_p})\equiv P_f(\vec{k},\vec{k_1},\ldots,
\vec{k_p});$
this gives (\ref{eq:pf}) by construction, and the right hand side of
(\ref{eq:ph}) can be evaluated using (\ref{eq:pfh}):
\beq
\sum_{\vec{k_1}\in\tilde{A}(\k)} \ldots\sum_{\vec{k_p}\in \tilde{A}(\k)}
a(\vec{k_1})\ldots a(\vec{k_p})\,P(\vec{p_1},\vec{k_1},\ldots, \vec{k_p})&=&
f(w_1,\overline{w}_4,0,z_1,z_2,0)\nonumber\\
&=&h(z_1,z_2,w_1,0,0,w_4)\nonumber\\
&=&h(z,w).
\eeq
This verifies (\ref{eq:ph}). {\bf $\Box$}\end{enumerate}
\end{quote}

\begin{lemma}
For a homogeneous vector field $\Ft$, equivariant with respect to the
representation of $\semi$ in \mbox{\rm (\ref{eq:48repa})-(\ref{eq:48repc})},
there exists a $\euc$-symmetric vector field $\strobe$ on $E^c(\rtwo)$ such
that \be
\Ft=\strobe|_{E^c(\ext)}\label{eq:48Ftrestrict}
\ee
if and only if for all $\rot {\theta}\in\euc$
\be
\rot {\theta}\cdot\Ft(\Phi)=\Ft(\rot {\theta}\cdot\Phi)\label{eq:48iff}
\ee
for $\Phi\in E^c(\ext)\,\cap\,[\rot{-\theta}\,E^c(\ext)]$.
\end{lemma}
\noindent {\em {\bf Proof}.}\begin{quote} The proof that (\ref{eq:48iff}) is a
necessary condition was given after Theorem III.1. The sufficiency of
(\ref{eq:48iff}) follows from the observation that $\Ft$ is a sum of the
elementary vector fields $[f,h]$
\beq
f(z,w)&=&m(z,w)+m(\g 3\cdot(z,w))\\
h(z,w)&=&m'(z,w)
\eeq
considered in Lemma {\rm III.2}. The general vector field $\Ft$ can be extended
to $\strobe$ if each of the elementary vector fields extends as in Lemma {\rm
III.2}. By Proposition {\rm III.4} the necessary and sufficient condition of
Lemma {\rm III.2} for these extensions is given by (\ref{eq:48iff}) with
$\theta=\phi$.
{\bf $\Box$}\end{quote}

\subsection{\hspace{0.125in}The $[8,8]$-mode interaction}

For the binary mode interaction between two eight-dimensional representations
we have for $(l_1,n_1)$
\be
\irreds {1}=\{(z_1\gef{\vec{q_1}}(\vec{r}) + z_2\gef{\vec{q_2}}(\vec{r}) +
z_3\gef{\vec{q_3}}(\vec{r}) + z_4\gef{\vec{q_4}}(\vec{r}) +
cc)|\;(z_1,z_2,z_3,z_4)\in {\bf C}^4\},\label{eq:irredl1n1}
\ee
with wave vectors $\vec{q_1}=(l_1,n_1)$, $\vec{q_2}=(l_1,-n_1)$,
$\vec{q_3}=(n_1,l_1)$ and $\vec{q_4}=(n_1,-l_1)$, and for $(l_2,n_2)$
\be
\irreds {2}=\{(w_1\gef{\vec{p_1}}(\vec{r}) + w_2\gef{\vec{p_2}}(\vec{r}) +
w_3\gef{\vec{p_3}}(\vec{r}) + w_4\gef{\vec{p_4}}(\vec{r}) +
cc)|\;(w_1,w_2,w_3,w_4)\in {\bf C}^4\}\label{eq:irredl2n2}
\ee
with wave vectors $\vec{p_1}=(l_2,n_2)$, $\vec{p_2}=(l_2,-n_2)$,
$\vec{p_3}=(n_2,l_2)$ and $\vec{p_4}=(n_2,-l_2)$. The reducible representation
of $\semi$ on the sixteen-dimensional space $\irreds {1}\oplus\irreds {2}$ is
generated by
\beq
\g{1}\cdot(z,w)&=&(\overline{z}_2,\overline{z}_1,\overline{z}_4,\overline{z}_3,
\overline{w}_2,\overline{w}_1,\overline{w}_4,\overline{w}_3)\label{eq:88repa}\\
\g{2}\cdot(z,w)&=&(z_3,\overline{z}_4,z_1,\overline{z}_2,w_3,\overline{w}_4,w_1,\overline{w}_2)\label{eq:88repb}\\
\tran{a}{b}\cdot(z,w)&=&
(e^{-i(l_1a+n_1b)}z_1,e^{-i(l_1a-n_1b)}z_2,
e^{-i(n_1a+l_1b)}z_3,e^{-i(n_1a-l_1b)}z_4,\nonumber\\
&&\hspace{0.5in}e^{-i(l_2a+n_2b)}w_1,e^{-i(l_2a-n_2b)}w_2,
e^{-i(n_2a+l_2b)}w_3,e^{-i(n_2a-l_2b)}w_4) \label{eq:88repc}
\eeq
where $z\equiv(z_1,z_2,z_3,z_4)$ and $w\equiv(w_1,w_2,w_3,w_4)$.

We adopt the following notation for vector fields that commute with this
representation.
\begin{prop}
$\Ft$ is $\semi$-equivariant if and only if it has the form
\be
\Ft(z,w)=\left(
\begin{array}{c}
f(z,w)\\
\\
f(\g 3\cdot(z,w))\\
\\
f(\g 2\cdot(z,w))\\
\\
f(\g 2\g 1\cdot(z,w))\\
\\
h(z,w)\\
\\
h(\g 3\cdot(z,w))\\
\\
h(\g 2\cdot(z,w))\\
\\
h(\g 2\g 1\cdot(z,w))
\end{array}
\right)=\left(
\begin{array}{c}
f(z_1,z_2,z_3,z_4,w_1,w_2,w_3,w_4)\\
\\
f(z_2,z_1,z_4,z_3,w_2,w_1,w_4,w_3)\\
\\
f(z_3,\overline{z}_4,z_1,\overline{z}_2,w_3,\overline{w}_4,w_1,\overline{w}_2)\\
\\
f(\overline{z}_4,z_3,\overline{z}_2,z_1,\overline{w}_4,w_3,\overline{w}_2,w_1)\\
\\
h(z_1,z_2,z_3,z_4,w_1,w_2,w_3,w_4)\\
\\
h(z_2,z_1,z_4,z_3,w_2,w_1,w_4,w_3)\\
\\
h(z_3,\overline{z}_4,z_1,\overline{z}_2,w_3,\overline{w}_4,w_1,\overline{w}_2)\\
\\
h(\overline{z}_4,z_3,\overline{z}_2,z_1,\overline{w}_4,w_3,\overline{w}_2,w_1)
\end{array}
\right)\label{eq:88map}
\ee
where $f(z,w)$ and $h(z,w)$ are complex-valued functions satisfying following
conditions:
\begin{enumerate}
\item $f(\overline{z},\overline{w})=\overline{f(z,w)}$ and
$h(\overline{z},\overline{w})=\overline{h(z,w)}$
\item $\overline{z}_1\,f(z,w)$ and $\overline{w}_1\,h(z,w)$ are invariant under
the translations $\tor$.
\end{enumerate}
\end{prop}
\noindent {\em {\bf Proof}.}\begin{quote}The proof follows that for Proposition
III.1. {\bf $\Box$}\end{quote}
We continue to use the notation
$\Ft=[f,h]$
to indicate the vector field (\ref{eq:88map}).

For $\semi$-symmetric vector fields $\Ft=[f,h]$ such that $f$ and $h$ are
homogeneous functions, we can reduce to the case where each function is a
single monomial
\beq
f(z,w)&=&m(z,w)\label{eq:88fmon}\\
h(z,w)&=&m'(z,w)\label{eq:88hmon}
\eeq
where $\overline{z}_1 m(z,w)$ and $\overline{w}_1 m'(z,w)$ are
$\tor$-invariant.
With the previous notation for the quadratic invariants
(\ref{eq:zelinv})-(\ref{eq:welinv}), these monomials are easily described.
\begin{prop}
A $\tor$-invariant monomial $M(z,w)$ can always be written in the form
\be
M(z,w)=\left(\prod_{i=1}^4\,\sigma_i^{\mu_i'}\rho_i^{\nu_i'}\right)
\Omega_1^{\mu_1}\ldots\Omega_4^{\mu_4}
\omega_1^{\nu_1}\ldots\omega_4^{\nu_4}\label{eq:88mon}
\ee
where $(\mu_1,\ldots,\mu_4,\nu_1,\ldots,\nu_4)$ are integers satisfying
\beq
l_1(\mu_1+\mu_2)+n_1(\mu_3+\mu_4)+l_2(\nu_1+\nu_2)+n_2(\nu_3+\nu_4)
&=&0\label{eq:88th1}\\
n_1(\mu_1-\mu_2)+l_1(\mu_3-\mu_4)+n_2(\nu_1-\nu_2)+l_2(\nu_3-\nu_4)
&=&0.\label{eq:88th2}
\eeq
\end{prop}

The geometry of the wave vector set (\ref{eq:akint})
\be
\tilde{A}(\k)=\{\pm\vec{q_1},\pm\vec{q_2},\pm\vec{q_3},\pm\vec{q_4},\pm\vec{p_1},\pm\vec{p_2},\pm\vec{p_3},\pm\vec{p_4}\}\label{eq:88wavev}
\ee
is shown in Fig. 2. For this mode interaction there are two independent
rotations $\rot{\phi}$ and $\rot{\alpha+\phi}$ that must be taken into account.
One can verify that $\phi\neq\alpha,\,2\theta_1$ and $\alpha\neq
2\theta_1,\,2\theta_1+\phi,\,2\theta_1-\phi$,
and this implies that the set $\tilde{A}(\k)$ satisfies
\beq
\tilde{A}(\k)\cap[\rot{\phi}\tilde{A}(\k)]&=&\{\pm\vec{q_2},\pm\vec{q_3},\pm\vec{p_1},\pm\vec{p_4}\}\label{eq:88wavevint1}\\
\tilde{A}(\k)\cap[\rot{-\phi}\tilde{A}(\k)]&=&\{\pm\vec{q_1},\pm\vec{q_4},\pm\vec{p_2},\pm\vec{p_3}\}\label{eq:88wavevint2}\\
&&\nonumber\\
\tilde{A}(\k)\cap[\rot{\alpha+\phi}\tilde{A}(\k)]&=&\{\pm\vec{q_2},\pm\vec{q_3},\pm\vec{p_2},\pm\vec{p_3}\}\label{eq:88wavevint3}\\
\tilde{A}(\k)\cap[\rot{-\alpha-\phi}\tilde{A}(\k)]&=&\{\pm\vec{q_1},\pm\vec{q_4},\pm\vec{p_1},\pm\vec{p_4}\}.\label{eq:88wavevint4}
\eeq
These non-zero intersections show that $\rot{\pm\phi}E^c$ and
$\rot{\pm(\alpha+\phi)}E^c$ intersect $E^c$ along eight-dimensional subspaces,
\beq
E^c(\ext)\,\cap\,[\rot{\phi}\,E^c(\ext)]&=&(0,z_2,z_3,0,w_1,0,0,w_4)
\label{eq:88sub1}\\
E^c(\ext)\,\cap\,[\rot{-\phi}\,E^c(\ext)]&=&(z_1,0,0,z_4,0,w_2,w_3,0)
\label{eq:88sub2}\\
&&\nonumber\\
E^c(\ext)\,\cap\,[\rot{\alpha+\phi}\,E^c(\ext)]&=&(0,z_2,z_3,0,w_2,w_3,0)
\label{eq:88sub3}\\
E^c(\ext)\,\cap\,[\rot{-\alpha-\phi}\,E^c(\ext)]&=&(z_1,0,0,z_4,w_1,0,0,w_4),
\label{eq:88sub4}
\eeq
that are related by reflection symmetry:
\beq
\g 3:E^c(\ext)\,\cap\,[\rot{\phi}\,E^c(\ext)]&\rightarrow&
E^c(\ext)\,\cap\,[\rot{-\phi}\,E^c(\ext)]\\
\g 3:E^c(\ext)\,\cap\,[\rot{\alpha+\phi}\,E^c(\ext)]&\rightarrow&
E^c(\ext)\,\cap\,[\rot{-\alpha-\phi}\,E^c(\ext)].
\eeq
\begin{prop}
For the representation \mbox{\rm (\ref{eq:88repa})-(\ref{eq:88repc})}, the
subspaces \mbox{\rm (\ref{eq:88sub1})-(\ref{eq:88sub4})} are fixed point
subspaces for isotropy subgroups of $\semi$; hence they are invariant under any
$\semi$ symmetric vector field $\Ft=[f,h]$. This invariance implies
\beq
f(0,z_2,z_3,0,w_1,0,0,w_4)&=&f(0,z_2,z_3,0,w_2,w_3,0)=0\label{eq:88fpspace1}\\
h(z_1,0,0,z_4,0,w_2,w_3,0)&=&h(0,z_2,z_3,0,w_2,w_3,0)=0.\label{eq:88fpspace2}
\eeq
\end{prop}
\noindent {\em {\bf Proof}.}\begin{quote} The proof follows that of Proposition
III.3. The isotropy subgroup $\Sigma$ for $(0,z_2,z_3,0,w_1,0,0,w_4)$ contains
the translation $\tran{a}{b}$ where
\be
(a,b)=\frac{2\pi}{\k^2}(l_1\,l_2\,+\,n_1\,n_2,l_1 n_2\,-\,n_1 l_2).
\label{eq:88tran1}
\ee
Thus $\tran ab\cdot(z,w)=(z,w)$ is a necessary condition for $(z,w)\in\mbox{\rm
Fix}(\Sigma)$; this condition implies
\beq
e^{-i(l_1a+n_1b)}z_1&=&z_1\;\;\;\;\;\;\;e^{-i(n_1a-l_1b)}z_4=z_4\\
e^{-i(l_2a-n_2b)}w_2&=&w_2\;\;\;\;\;\;\;e^{-i(n_2a+l_2b)}w_3=w_3.
\eeq
If $z_1\neq 0$, then the  $z_1$ equation requires $(l_1a+n_1b)=2\pi\,(\mbox{\rm
integer})$ or equivalently
\be
\frac{l_2}{\k}\,\frac{(l_1^2-n_1^2)}{\k}-\frac{2l_1n_1}{\k}\,\frac{n_2}{\k}
=\mbox{\rm integer}.
\ee
However since $\cos\theta_2={l_2}/{\k}$, $\cos 2\theta_1=(l_1^2-n_1^2)/{\k}$,
$\sin \theta_2={n_2}/{\k}$, and $\sin 2\theta_1={2l_1n_1}/{\k}$, this becomes
$\cos(\theta_2+2\theta_1)=\mbox{\rm integer}$
which is impossible. Therefore we must set $z_1=0$. In a similar fashion the
remaining conditions require $z_4=w_2=w_3=0$ which proves that
\be
\mbox{\rm Fix}(\Sigma)=(0,z_2,z_3,0,w_1,0,0,w_4).
\ee
The proof for $(0,z_2,z_3,0,w_2,w_3,0)$ is the same with
\be
(a,b)=\frac{2\pi}{\k^2}(l_1 n_2\,+\,n_1,l_2,\;l_1\,l_2\,-\,n_1\,n_2)
\label{eq:88tran2}
\ee
replacing (\ref{eq:88tran1}).
{\bf $\Box$} \end{quote}

When $\Ft$ is obtained by restriction from $\strobe$ there are additional
conditions on $f$ and $h$. For this mode interaction one finds two independent
conditions arising from $\rot{\phi}$ on
$E^c(\ext)\,\cap\,[\rot{-\phi}\,E^c(\ext)]$ and $\rot{(\alpha+\phi)}$ on
$E^c(\ext)\,\cap\,[\rot{\alpha+\phi}\,E^c(\ext)]$.
\begin{prop}
For $\theta=\phi$ and $\theta=\alpha+\phi$, the condition
\be
\rot {\theta}\cdot\Ft(\Phi)=\Ft(\rot {\theta}\cdot\Phi)\label{eq:88rotcond0}
\ee
for all $\Phi\in E^c(\ext)\,\cap\,[\rot{-\theta}\,E^c(\ext)]$ is satisfied if
and only if
\beq
f(z_1,0,0,z_4,0,w_2,w_3,0)&=&
h(0,w_2,w_3,0,z_1,0,0,{z}_4)\label{eq:88rotcond1}\\
f(z_1,0,0,z_4,w_1,0,0,w_4)&=&
h(w_1,0,0,\overline{w}_4,z_1,0,0,\overline{z}_4),\label{eq:88rotcond2}
\eeq
or equivalently
\beq
h(0,z_2,z_3,0,w_1,0,0,w_4)&=&
f(w_1,0,0,w_4,0,z_2,z_3,0)\label{eq:88rotcondeqv1}\\
h(z_1,0,0,z_4,w_1,0,0,w_4)&=&
f(w_1,0,0,\overline{w}_4,z_1,0,0,\overline{z}_4).\label{eq:88rotcondeqv2}
\eeq
\end{prop}
\noindent {\em {\bf Proof}.}\begin{quote} The proof follows that for
Proposition III.4. {\bf $\Box$}\end{quote}

The restriction in (\ref{eq:Ftrestrict}) is accomplished by choosing
$a(\vec{k})$ in (\ref{eq:genef2}) so that $\Phi\in E^c(\ext)$:
\beq
a(\vec{k})&=&\sum_{i=1}^4\,\left[
z_i\;\delta_{\vec{k},\vec{q_i}} + \overline{z}_i\;\delta_{\vec{k},\vec{-q_i}}
+ w_i\;\delta_{\vec{k},\vec{p_i}} + \overline{w}_i\;\delta_{\vec{k},\vec{-p_i}}
\right].\label{eq:88restrict}
\eeq
This determines a $\semi$-equivariant vector field $\Ft=[f,h]$ on $E^c(\ext)$
such
that $f$ and $h$ are homogeneous functions of degree p; when $f$ and $h$ are
determined by monomials as in (\ref{eq:88fmon}) and (\ref{eq:88hmon}) then
\beq
m(z,w)&=&\sum_{\vec{k_1}\in \tilde{A}(\k)}
\ldots\sum_{\vec{k_p}\in\tilde{A}(\k)} a(\vec{k_1})\,\ldots
a(\vec{k_p})\,P(\vec{q_1},\vec{k_1},\ldots, \vec{k_p})\label{eq:88fmona}\\
&&\nonumber\\
m'(z,w)&=&\sum_{\vec{k_1}\in\tilde{A}(\k)} \ldots\sum_{\vec{k_p}\in
\tilde{A}(\k)} a(\vec{k_1})\,\ldots a(\vec{k_p})\,P(\vec{p_1},\vec{k_1},\ldots,
\vec{k_p}).\label{eq:88hmonb}
\eeq
Given monomials $m$ and $m'$ in (\ref{eq:88fmon}) and (\ref{eq:88hmon}), the
essential question is whether there exists a function $P$, satisfying
conditions (\ref{eq:irrcond1})-(\ref{eq:irrcond2}), such that
(\ref{eq:88fmona}) and (\ref{eq:88hmonb}) hold. We answer this as before by
constructing $P$; the procedure is the same as in the [4,8] mode interaction.

The inner products amongst the vectors in $\tilde{A}(\k)$ are used to specify
subsets of $\tilde{A}(\k)$. Given a vector $\vec{c_i}\in\tilde{A}(\k)$ denote
the reflection in $\euc$ that fixes $\vec{c_i}$ by $\gamma_{\vec{c_i}}$, i.e.
$\gamma_{\vec{c_i}}\cdot\vec{c_i}=\vec{c_i}.$
For example, $\gamma_{\vec{q_1}}=\rot{\theta_1}\g 3\rot{-\theta_1}$. One can
check that:
\beq
\tilde{A}(\k)\cap[\gamma_{\vec{q_1}}\tilde{A}(\k)]&=&
\tilde{A}(\k)\cap[\gamma_{\vec{q_4}}\tilde{A}(\k)]=
\{\pm\vec{q_1},\pm\vec{q_4}\}\label{eq:88insect1}\\
\tilde{A}(\k)\cap[\gamma_{\vec{q_2}}\tilde{A}(\k)]&=&
\tilde{A}(\k)\cap[\gamma_{\vec{q_3}}\tilde{A}(\k)]=
\{\pm\vec{q_2},\pm\vec{q_3}\},\label{eq:88insect2}
\eeq
and
\beq
\tilde{A}(\k)\cap[\gamma_{\vec{p_1}}\tilde{A}(\k)]&=&
\tilde{A}(\k)\cap[\gamma_{\vec{p_4}}\tilde{A}(\k)]=
\{\pm\vec{p_1},\pm\vec{p_4}\}\label{eq:88insect3}\\
\tilde{A}(\k)\cap[\gamma_{\vec{p_2}}\tilde{A}(\k)]&=&
\tilde{A}(\k)\cap[\gamma_{\vec{p_3}}\tilde{A}(\k)]=
\{\pm\vec{p_2},\pm\vec{p_3}\}.\label{eq:88insect4}
\eeq
\begin{prop} Let $\{\vec{c_1},\vec{c_2},\ldots,\vec{c_n}\}$ denote a fixed
subset of $\tilde{A}(\k)$. Given $(\vec{k},\vec{k'})$, let
$\{\vec{k_1},\vec{k_2},\ldots,\vec{k_n}\}$ denote a second set of $n$ vectors
in $\tilde{A}(\k)$ required to satisfy the conditions:
\beq
\vec{k}\cdot\vec{k_i}&=&\vec{k'}\cdot\vec{c_i}\;\;\hspace{0.25in} \mbox{\rm for
}i=1,\ldots n\label{eq:88cinner1}\\
\vec{k_i}\cdot\vec{k_j}&=&\vec{c_i}\cdot\vec{c_j}\;\;\hspace{0.25in} \mbox{\rm
for }i,j=1,\ldots n.\label{eq:88cinner2}
\eeq
\begin{enumerate}
\item If $(\vec{k},\vec{k'})=(\vec{p_1},\vec{p_1})$, then the conditions
\mbox{\rm (\ref{eq:88cinner1})-(\ref{eq:88cinner2})}
can be satisfied in only one way:
\be
\{\vec{k_1},\vec{k_2},\ldots,\vec{k_n}\}=
\{\vec{c_1},\vec{c_2},\ldots,\vec{c_n}\}\label{eq:88pset1}
\ee
unless
\be
\{\vec{c_1},\vec{c_2},\ldots,\vec{c_n}\}\subseteq
\tilde{A}(\k)\cap[\gamma_{\vec{p_1}}\tilde{A}(\k)] \label{eq:88insect5}
\ee
in which case there is a second (not necessarily distinct) solution:
\be
\{\vec{k_1},\vec{k_2},\ldots,\vec{k_n}\}= \gamma_{\vec{p_1}}
\cdot\{\vec{c_1},\vec{c_2},\ldots,\vec{c_n}\}.\label{eq:88pset2}
\ee
\item If $(\vec{k},\vec{k'})=(\vec{q_1},\vec{q_1})$, then the conditions
\mbox{\rm (\ref{eq:88cinner1})-(\ref{eq:88cinner2})} can be satisfied in only
one way:
\be
\{\vec{k_1},\vec{k_2},\ldots,\vec{k_n}\}=
\{\vec{c_1},\vec{c_2},\ldots,\vec{c_n}\}\label{eq:88qset1}
\ee
unless
\be
\{\vec{c_1},\vec{c_2},\ldots,\vec{c_n}\}\subseteq
\tilde{A}(\k)\cap[\gamma_{\vec{q_1}}\tilde{A}(\k)] \label{eq:88qinsect3}
\ee
in which case there is a second (not necessarily distinct) solution:
\be
\{\vec{k_1},\vec{k_2},\ldots,\vec{k_n}\}= \gamma_{\vec{q_1}}
\cdot\{\vec{c_1},\vec{c_2},\ldots,\vec{c_n}\}.\label{eq:88qset2}
\ee
\item If $(\vec{k},\vec{k'})=(\vec{q_1},\vec{p_1})$, then the conditions
\mbox{\rm (\ref{eq:88cinner1})-(\ref{eq:88cinner2})} have no solutions unless
\be
\{\vec{c_1},\vec{c_2},\ldots,\vec{c_n}\}\subseteq
\tilde{A}(\k)\cap[\rot{\phi}\tilde{A}(\k)],\label{eq:88qpcond1}
\ee
or
\be
\{\vec{c_1},\vec{c_2},\ldots,\vec{c_n}\}\subseteq
\tilde{A}(\k)\cap[\rot{\phi}\gamma_{\vec{q_1}}\tilde{A}(\k)].
\label{eq:88qpcond2}
\ee
When \mbox{\rm (\ref{eq:88qpcond1})} holds then
\be
\{\vec{k_1},\vec{k_2},\ldots,\vec{k_n}\}=
\rot{-\phi}\{\vec{c_1},\vec{c_2},\ldots,\vec{c_n}\}\label{eq:88qpset1}
\ee
and when \mbox{\rm (\ref{eq:88qpcond2})} holds then
\be
\{\vec{k_1},\vec{k_2},\ldots,\vec{k_n}\}= \rot{-\phi}\gamma_{\vec{p_1}} \cdot
\{\vec{c_1},\vec{c_2},\ldots,\vec{c_n}\}=\gamma_{\vec{q_1}} \rot{-\phi} \cdot
\{\vec{c_1},\vec{c_2},\ldots,\vec{c_n}\}.\label{eq:88qpset2}
\ee
\item If $(\vec{k},\vec{k'})=(\vec{p_1},\vec{q_1})$, then the conditions
\mbox{\rm (\ref{eq:88cinner1})-(\ref{eq:88cinner2})} have no solutions unless
\be
\{\vec{c_1},\vec{c_2},\ldots,\vec{c_n}\}\subseteq
\tilde{A}(\k)\cap[\rot{-\phi}\tilde{A}(\k)],\label{eq:88qpcond3}
\ee
or
\be
\{\vec{c_1},\vec{c_2},\ldots,\vec{c_n}\}\subseteq
\tilde{A}(\k)\cap[\gamma_{\vec{q_1}}\rot{-\phi}\tilde{A}(\k)].
\label{eq:88qpcond4}
\ee
When \mbox{\rm (\ref{eq:88qpcond3})} holds then
\be
\{\vec{k_1},\vec{k_2},\ldots,\vec{k_n}\}=
\rot{\phi}\{\vec{c_1},\vec{c_2},\ldots,\vec{c_n}\}\label{eq:88qpset3}
\ee
and when \mbox{\rm (\ref{eq:88qpcond4})} holds then
\be
\{\vec{k_1},\vec{k_2},\ldots,\vec{k_n}\}= \rot{\phi}\gamma_{\vec{q_1}} \cdot
\{\vec{c_1},\vec{c_2},\ldots,\vec{c_n}\}.\label{eq:88qpset4}
\ee
\end{enumerate}
\end{prop}
\noindent {\em {\bf Proof}.}\begin{quote}The proof follows that for Proposition
III.5. {\bf $\Box$}
\end{quote}

\begin{lemma}  Let $\overline{z}_1 m(z,w)$ and $\overline{w}_1 m'(z,w)$ be
$\tor$-invariant monomials of degree p such that $f(z,w)=m(z,w)$ and
$h(z,w)=m'(z,w)$ meet the conditions of Proposition {\rm III.6}. There exist
functions $P_f$ and $P_h$ satisfying \mbox{\rm
(\ref{eq:irrcond1})-(\ref{eq:irrcond2})} such that
\beq
f(z,w)&=&\sum_{\vec{k_1}\in \tilde{A}(\k)}\sum_{\vec{k_2}\in \tilde{A}(\k)}
\ldots\sum_{\vec{k_p}\in\tilde{A}(\k)} a(\vec{k_1})\,a(\vec{k_2})\,\ldots
a(\vec{k_p})\,P_f(\vec{q_1},\vec{k_1},\ldots, \vec{k_p})\label{eq:88pff}\\
&&\nonumber\\
h(z,w)&=&\sum_{\vec{k_1}\in\tilde{A}(\k)}\sum_{\vec{k_2}\in \tilde{A}(\k)}
\ldots\sum_{\vec{k_p}\in \tilde{A}(\k)} a(\vec{k_1})\,a(\vec{k_2})\,\ldots
a(\vec{k_p})\,P_h(\vec{p_1},\vec{k_1},\ldots, \vec{k_p})\label{eq:88phh}
\eeq
and
\beq
\lefteqn{\sum_{\vec{k_1}\in\tilde{A}(\k)} \ldots\sum_{\vec{k_p}\in
\tilde{A}(\k)} a(\vec{k_1})\,\ldots
a(\vec{k_p})\,P_f(\vec{p_1},\vec{k_1},\ldots, \vec{k_p})}\nonumber\\
&=&\left\{
\begin{array}{cc}
f(w_1,0,0,{w}_4,0,z_2,z_3,0)&\mbox{\rm if}\;\;
f(w_1,0,0,{w}_4,0,z_2,z_3,0)\neq0\\
&\\
f(w_1,0,0,\overline{w}_4,z_1,0,0,\overline{z}_4)&\mbox{\rm if}\;\;
f(w_1,0,0,{w}_4,z_1,0,0,{z}_4)\neq0\\
&\\
0&\mbox{\rm otherwise}
\end{array}
\right.\label{eq:88pfh}
\eeq
\beq
\lefteqn{\sum_{\vec{k_1}\in\tilde{A}(\k)} \ldots\sum_{\vec{k_p}\in
\tilde{A}(\k)} a(\vec{k_1})\,\ldots
a(\vec{k_p})\,P_h(\vec{q_1},\vec{k_1},\ldots, \vec{k_p})}\nonumber\\
&=&\left\{
\begin{array}{cc}
h(0,w_2,w_3,0,z_1,0,0,{z}_4)&\mbox{\rm if}\;\;
h(0,w_2,w_3,0,z_1,0,0,{z}_4)\neq0\\
&\\
h(w_1,0,0,\overline{w}_4,z_1,0,0,\overline{z}_4)&\mbox{\rm if}\;\;
h(w_1,0,0,{w}_4,z_1,0,0,{z}_4)\neq0\\
&\\
0&\mbox{\rm otherwise.}
\end{array}
\right.\label{eq:88phf}
\eeq
where $a(\vec{k})$ is given by \mbox{\rm (\ref{eq:88restrict})}.
\end{lemma}
\noindent {\em {\bf Proof}}\begin{quote}The proof follows that for Lemma III.1.
\begin{enumerate}
\item Let $\{\vec{c_1},\vec{c_2},\ldots,\vec{c_n}\}$ denote the wave vector set
associated with the monomial $m(z,w)$ used to define $f(z,w)$. The monomial
$m(z,w)$ can be written as
\be
m(z,w)=a(\vec{c_1})\,a(\vec{c_2})\ldots a(\vec{c_{n}})a(\vec{c}_{l_{n+1}})
\ldots a(\vec{c}_{l_{p}})\label{eq:88mb}
\ee
in analogy with (\ref{eq:mb}).
For $P_f$ we introduce the numerical factor
\be
\epsilon=\left\{\begin{array}{cc}
\frac{1}{2}&\mbox{\rm if}\;\;\{\vec{c_1},\ldots,\vec{c_n}\}\subseteq
\tilde{A}(\k)\cap[\gamma_{\vec{q_1}}\tilde{A}(\k)]\;\; \mbox{\rm and}\\
& \{\vec{c_1},\ldots,\vec{c_n}\}\neq
\gamma_{\vec{q_1}}\{\vec{c_1},\ldots,\vec{c_n}\}\\
&\\
1&\mbox{\rm otherwise},
\end{array}
\right.
\ee
and define $P_f$ as
\be
P_f(\vec{k},\vec{k_1},\ldots, \vec{k_p})=\epsilon\left(\prod_{i=1}^{n}
\prod_{j=1}^{n}\,\delta_{\vec{k}\cdot\vec{k_i},\vec{q_1}\cdot\vec{c_i}}
\;\delta_{\vec{k_i}\cdot\vec{k_j},\vec{c_i}\cdot\vec{c_j}}\right)\,
\left(\prod_{j=n+1}^{p}\, \delta_{\vec{k_j}\cdot\vec{c}_{l_j},\k^2}\right).
\label{eq:88pfdef1}
\ee
{}From Proposition III.10 we then have
\be
P_f(\vec{q_1},\vec{k_1},\ldots,\vec{k_p})=\left\{\begin{array}{cc}
{\epsilon}&\mbox{\rm if}\;\; \{\vec{k_1},\ldots,\vec{k_p}\}=
\{\vec{c_1},\ldots,\vec{c_n},\vec{c}_{l_{n+1}},\ldots, \vec{c}_{l_{p}}\}\\
&\\
{\epsilon}&\mbox{\rm if}\;\;\{\vec{c_1},\ldots,\vec{c_n}\}\subseteq
\tilde{A}(\k)\cap[\gamma_{\vec{q_1}}\tilde{A}(\k)] \;\;\mbox{\rm and}\\
&\{\vec{k_1},\ldots,\vec{k_p}\}=
\gamma_{\vec{q_1}}\cdot\{\vec{c_1},\ldots,\vec{c_n},\vec{c}_{l_{n+1}},\ldots,
\vec{c}_{l_{p}}\}\\
&\\
0&\mbox{\rm otherwise};
\end{array}\right.\label{eq:88pfp1b}
\ee
and
\be
P_f(\vec{p_1},\vec{k_1},\ldots,\vec{k_p})=\left\{\begin{array}{cc}
{\epsilon}&\mbox{\rm if}\;\; \{\vec{c_1},\ldots,\vec{c_n}\}\subseteq
\tilde{A}(\k)\cap[\rot{-\phi}\tilde{A}(\k)]\;{\rm and}\\
&\{\vec{k_1},\ldots,\vec{k_p}\}= \rot{\phi}
\cdot\{\vec{c_1},\ldots,\vec{c_n},\vec{c}_{l_{n+1}},\ldots, \vec{c}_{l_{p}}\}\\
&\\
{\epsilon}&\mbox{\rm if}\;\;\{\vec{c_1},\ldots,\vec{c_n}\}\subseteq
\tilde{A}(\k)\cap[\gamma_{\vec{q_1}}\rot{-\phi}\tilde{A}(\k)]\; {\rm and}\\
& \{\vec{k_1},\ldots,\vec{k_p}\}= \rot{\phi}\gamma_{\vec{q_1}}\cdot
\{\vec{c_1},\ldots,\vec{c_n},\vec{c}_{l_{n+1}},\ldots, \vec{c}_{l_{p}}\}\\
&\\
0&\mbox{\rm otherwise};
\end{array}\right.\label{eq:88pfpq1}
\ee
The calculations to verify (\ref{eq:88pff}) and (\ref{eq:88pfh}) are now
straightforward.

\item Let $\{\vec{c_1},\vec{c_2},\ldots,\vec{c_n}\}$ denote the wave vector set
associated with the monomial $m'(z,w)$ used to define $h(z,w)$. The monomial
$m'(z,w)$ can be written as
\be
m'(z,w)=a(\vec{c_1})\,a(\vec{c_2})\ldots a(\vec{c_{n}})a(\vec{c}_{l_{n+1}})
\ldots a(\vec{c}_{l_{p}}).\label{eq:88mprimeb}
\ee
For $P_h$ we require the numerical factor
\be
\epsilon=\left\{\begin{array}{cc}
\frac{1}{2}&\mbox{\rm if}\;\;\{\vec{c_1},\ldots,\vec{c_n}\}\subseteq
\tilde{A}(\k)\cap[\gamma_{\vec{p_1}}\tilde{A}(\k)]\;\; \mbox{\rm and}\\
& \{\vec{c_1},\ldots,\vec{c_n}\}\neq
\gamma_{\vec{p_1}}\{\vec{c_1},\ldots,\vec{c_n}\}\\
&\\
1&\mbox{\rm otherwise},
\end{array}
\right.
\ee
and $P_h$ is defined as
\be
P_h(\vec{k},\vec{k_1},\ldots, \vec{k_p})=\epsilon\left(\prod_{i=1}^{n}
\prod_{j=1}^{n}\,\delta_{\vec{k}\cdot\vec{k_i},\vec{p_1}\cdot\vec{c_i}}
\;\delta_{\vec{k_i}\cdot\vec{k_j},\vec{c_i}\cdot\vec{c_j}}\right)\,
\left(\prod_{j=n+1}^{p}\, \delta_{\vec{k_j}\cdot\vec{c}_{l_j},\k^2}\right).
\label{eq:88phdef1}
\ee
The verification of (\ref{eq:88phh}) and (\ref{eq:88phf}) proceeds as for
$P_f$.

\end{enumerate}

 {\bf $\Box$} \end{quote}

\begin{lemma}  Let $\overline{z}_1 m(z,w)$ and $\overline{w}_1 m'(z,w)$ be
$\tor$-invariant monomials of degree p such that $f(z,w)=m(z,w)$ and
$h(z,w)=m'(z,w)$ meet the conditions of Proposition {\rm III.6}. There exists a
single function $P$ satisfying \mbox{\rm
(\ref{eq:irrcond1})-(\ref{eq:irrcond2})} such that
\beq
f(z,w)&=&\sum_{\vec{k_1}\in \tilde{A}(\k)}\sum_{\vec{k_2}\in \tilde{A}(\k)}
\ldots\sum_{\vec{k_p}\in\tilde{A}(\k)} a(\vec{k_1})\,a(\vec{k_2})\,\ldots
a(\vec{k_p})\,P(\vec{q_1},\vec{k_1},\ldots, \vec{k_p})\label{eq:88pf}\\
&&\nonumber\\
h(z,w)&=&\sum_{\vec{k_1}\in\tilde{A}(\k)}\sum_{\vec{k_2}\in \tilde{A}(\k)}
\ldots\sum_{\vec{k_p}\in \tilde{A}(\k)} a(\vec{k_1})\,a(\vec{k_2})\,\ldots
a(\vec{k_p})\,P(\vec{p_1},\vec{k_1},\ldots, \vec{k_p})\label{eq:88ph}
\eeq
if and only if
\beq
f(z_1,0,0,z_4,0,w_2,w_3,0)&=&
h(0,w_2,w_3,0,z_1,0,0,{z}_4)\label{eq:88rotconda}\\
f(z_1,0,0,z_4,w_1,0,0,w_4)&=&
h(w_1,0,0,\overline{w}_4,z_1,0,0,\overline{z}_4).\label{eq:88rotcondb}
\eeq
\end{lemma}
\noindent {\em {\bf Proof}.}\begin{quote} The proof follows that for Lemma
II.2.
The necessity of (\ref{eq:88rotconda}) and (\ref{eq:88rotcondb}) is established
by Proposition {\rm III.9} and the proof of necessity in Theorem {\rm III.1}.
Assume that (\ref{eq:88rotconda}) and (\ref{eq:88rotcondb}) hold and let
$P_f(\vec{k},\vec{k_1},\ldots, \vec{k_p})$ and $P_h(\vec{k},\vec{k_1},\ldots,
\vec{k_p})$ be the functions of Lemma {\rm III.4} for $f$ and $h$ respectively.
\begin{enumerate}
\item If $f(z_1,0,0,z_4,0,w_2,w_3,0)=f(z_1,0,0,z_4,w_1,0,0,w_4)=0$ then define
\be
P(\vec{k},\vec{k_1},\ldots, \vec{k_p})\equiv P_f(\vec{k},\vec{k_1},\ldots,
\vec{k_p}) + P_h(\vec{k},\vec{k_1},\ldots, \vec{k_p}).
\ee
The equations (\ref{eq:88pf}) and (\ref{eq:88ph}) follow from the previous
Lemma.

\item If $f(z_1,0,0,z_4,0,w_2,w_3,0)\neq0$ or $f(z_1,0,0,z_4,w_1,0,0,w_4)\neq0$
then define
\be
P(\vec{k},\vec{k_1},\ldots, \vec{k_p})\equiv P_f(\vec{k},\vec{k_1},\ldots,
\vec{k_p}).
\ee
Again (\ref{eq:88pf}) and (\ref{eq:88ph}) can be verified using  the previous
Lemma.{\bf $\Box$}
\end{enumerate}\end{quote}

These results imply the analogue of Lemma III.3 for this mode interaction.
\begin{lemma}
For a homogeneous vector field $\Ft$, equivariant with respect to the
representation of $\semi$ in \mbox{\rm (\ref{eq:88repa})-(\ref{eq:88repc})},
there exists a $\euc$-symmetric vector field $\strobe$ on $E^c(\rtwo)$ such
that \be
\Ft=\strobe|_{E^c(\ext)}\label{eq:88Ftrestrict}
\ee
if and only if for all $\rot\theta\in\euc$,
\be
\rot {\theta}\cdot\Ft(\Phi)=\Ft(\rot {\theta}\cdot\Phi)\label{eq:88iff}
\ee
for all $\Phi\in E^c(\ext)\,\cap\,[\rot{-\theta}\,E^c(\ext)]$.
\end{lemma}

\subsection{\hspace{0.125in}The $[8,4]$-mode interaction}

In the notation of section II.B, the eight-dimensional representation
$(l_1,n_1)$ is
\be
\irreds {1}=\{(z_1\gef{\vec{q_1}}(\vec{r}) + z_2\gef{\vec{q_2}}(\vec{r}) +
z_3\gef{\vec{q_3}}(\vec{r}) + z_4\gef{\vec{q_4}}(\vec{r}) +
cc)|\;(z_1,z_2,z_3,z_4)\in {\bf C}^4\},\label{eq:irredl1n1b}
\ee
with wave vectors $\vec{q_1}=(l_1,n_1)$, $\vec{q_2}=(l_1,-n_1)$,
$\vec{q_3}=(n_1,l_1)$ and $\vec{q_4}=(n_1,-l_1)$ and the four-dimensional
representation $(l_2,l_2)$ is
\be
\irreds {2}=\{w_1\gef{\vec{p_1}}(\vec{r}) + w_2\gef{\vec{p_1}}(\vec{r}) +
cc)|\;(w_1,w_2)\in {\bf C}^2\}\label{eq:irredll}
\ee
has wave vectors $\vec{p_1}=(l_2,l_2)$ and $\vec{p_2}=(l_2,-l_2)$. The
reducible representation of $\semi$ on the twelve-dimensional space $\irreds
{1}\oplus\irreds {2}$ is generated by
\beq
\g{1}\cdot(z,w)&=&(\overline{z}_2,\overline{z}_1,\overline{z}_4,\overline{z}_3,
\overline{w}_2,\overline{w}_1)\label{eq:84repa}\\
\g{2}\cdot(z,w)&=&(z_3,\overline{z}_4,z_1,\overline{z}_2,w_1,\overline{w}_2)
\label{eq:84repb}\\
\tran{a}{b}\cdot(z,w)&=&
(e^{-i(l_1a+n_1b)}z_1,e^{-i(l_1a-n_1b)}z_2,
e^{-i(n_1a+l_1b)}z_3,e^{-i(n_1a-l_1b)}z_4,\nonumber\\
&&\hspace{0.5in}e^{-il_2(a+b)}w_1,e^{-il_2(a-b)}w_2) \label{eq:84repc}
\eeq
where $z\equiv(z_1,z_2,z_3,z_4)$ and $w\equiv(w_1,w_2)$.

We adopt the following notation for vector fields that commute with this
representation.
\begin{prop}
$\Ft$ is $\semi$-equivariant if and only if it has the form
\be
\Ft(z,w)=\left(
\begin{array}{c}
f(z,w)\\
\\
f(\g 3\cdot(z,w))\\
\\
f(\g 2\cdot(z,w))\\
\\
f(\g 2\g 1\cdot(z,w))\\
\\
h(z,w)\\
\\
h(\g 3\cdot(z,w))
\end{array}
\right)=\left(
\begin{array}{c}
f(z_1,z_2,z_3,z_4,w_1,w_2)\\
\\
f(z_2,z_1,z_4,z_3,w_2,w_1)\\
\\
f(z_3,\overline{z}_4,z_1,\overline{z}_2,w_1,\overline{w}_2)\\
\\
f(z_4,\overline{z}_3,z_2,\overline{z}_1,w_2,\overline{w}_1)\\
\\
h(z_1,z_2,z_3,z_4,w_1,w_2)\\
\\
h(z_2,z_1,z_4,z_3,w_2,w_1)
\end{array}
\right)\label{eq:84map}
\ee
where $f(z,w)$ and $h(z,w)$ are complex-valued functions satisfying following
conditions:
\begin{enumerate}
\item $f(\overline{z},\overline{w})=\overline{f(z,w)}$ and
$h(\overline{z},\overline{w})=\overline{h(z,w)}$
\item $h(z,w)$ is $\g 2$-invariant:
\be
h(z_1,z_2,z_3,z_4,w_1,w_2)=
h(z_3,\overline{z}_4,z_1,\overline{z}_2,w_1,\overline{w}_2)\label{eq:84fg}
\ee
\item $\overline{z}_1\,f(z,w)$ and $\overline{w}_1\,h(z,w)$ are invariant under
the translations $\tor$.
\end{enumerate}
\end{prop}
\noindent {\em {\bf Proof}.}\begin{quote}The proof follows that for Proposition
 {\rm III.1}  {\bf $\Box$}\end{quote}
We write
$\Ft=[f,h]$
to indicate the map (\ref{eq:84map}) corresponding to $f$ and $h$.

For $\semi$-symmetric vector fields $\Ft=[f,h]$, we assume $f$ and $h$ are
homogeneous functions and reduce to the case with each function determined by a
single monomial
\beq
f(z,w)&=&m(z,w)\label{eq:84fmon}\\
h(z,w)&=&m'(z,w)+m'(\g 2\cdot(z,w))\label{eq:84hmon}
\eeq
where $\overline{z}_1 m(z,w)$ and $\overline{w}_1 m'(z,w)$ are
$\tor$-invariant.
With the same notation for the quadratic invariants
(\ref{eq:zelinv})-(\ref{eq:welinv}), these monomials are readily characterized.
\begin{prop}
A $\tor$-invariant monomial $M(z,w)$ can always be written in the form
\be
M(z,w)=(\sigma_1^{\mu_1'}\sigma_2^{\mu_2'}\sigma_3^{\mu_3'}\sigma_4^{\mu_4'}
\rho_1^{\nu_1'}\rho_2^{\nu_2'})\,
\Omega_1^{\mu_1}\Omega_2^{\mu_2}\Omega_3^{\mu_3}\Omega_4^{\mu_4}
\omega_1^{\nu_1}\omega_2^{\nu_2}\label{eq:84mon}
\ee
where $(\mu_1,\mu_2,\mu_3,\mu_4,\nu_1,\nu_2)$ are integers satisfying
\beq
l_1(\mu_1+\mu_2)+n_1(\mu_3+\mu_4)+l_2(\nu_1+\nu_2)&=&0\label{eq:84th1}\\
n_1(\mu_1-\mu_2)+l_1(\mu_3-\mu_4)+n_2(\nu_1-\nu_2)&=&0.\label{eq:84th2}
\eeq
\end{prop}

The geometry of the wave vector set (\ref{eq:akint})
\be
\tilde{A}(\k)=\{\pm\vec{q_1},\pm\vec{q_2},\pm\vec{q_3},\pm\vec{q_4},\pm\vec{p_1},\pm\vec{p_2}\}\label{eq:84wavev}
\ee
is shown in Fig. 3. For this mode interaction, as for the [4,8] interaction, it
is sufficient to consider a single rotation $\rot{\phi}$. One can verify that
$\phi\neq2\theta_1,$
and this implies that the set $\tilde{A}(\k)$ satisfies
\beq
\tilde{A}(\k)\cap[\rot{\phi}\tilde{A}(\k)]&=&\{\pm\vec{q_2},\pm\vec{q_3},\pm\vec{p_1},\pm\vec{p_2}\}\label{eq:84wavevint1}\\
\tilde{A}(\k)\cap[\rot{-\phi}\tilde{A}(\k)]&=&\{\pm\vec{q_1},\pm\vec{q_4},\pm\vec{p_1},\pm\vec{p_2}\}.\label{eq:84wavevint2}
\eeq
These non-zero intersections show that $\rot{\phi}E^c$ and $\rot{-\phi}E^c$
intersect $E^c$ along eight-dimensional subspaces,
\beq
E^c(\ext)\,\cap\,[\rot{\phi}\,E^c(\ext)]&=&(0,z_2,z_3,0,w_1,w_2)
\label{eq:84sub1}\\
E^c(\ext)\,\cap\,[\rot{-\phi}\,E^c(\ext)]&=&(z_1,0,0,z_4,w_1,w_2),
\label{eq:84sub2}
\eeq
that are related by reflection symmetry:
\be
\g 2:E^c(\ext)\,\cap\,[\rot{\phi}\,E^c(\ext)]\rightarrow
E^c(\ext)\,\cap\,[\rot{-\phi}\,E^c(\ext)].
\ee
\begin{prop}
For the representation \mbox{\rm (\ref{eq:84repa})-(\ref{eq:84repc})}, the
subspaces \mbox{\rm (\ref{eq:84sub1})-(\ref{eq:84sub2})} are fixed point
subspaces for isotropy subgroups of $\semi$; hence they are invariant under any
$\semi$ symmetric vector field $\Ft=[f,h]$. This invariance implies
\be
f(0,z_2,z_3,0,w_1,w_2)=0.\label{eq:84fpspace1}
\ee
\end{prop}
\noindent {\em {\bf Proof}.}\begin{quote}
The proof follows that of Proposition III.3. The isotropy subgroup $\Sigma$ for
$(0,z_2,z_3,0,w_1,w_2)$ contains the translations $\tran{a}{b}$ where
\be
(a,b)=\frac{2\pi}{\k^2}(l_2\,(l_1\,+\,n_1),\,l_2 (l_1\,-\,n_1))
={2\pi}(\cos\phi\,,\sin\phi)\label{eq:84tran1}
\ee
and
\be
(a,b)=\frac{2\pi}{\k^2}(-l_2 (l_1\,-\,n_1),\,l_2\,(l_1\,+\,n_1))
={2\pi}(-\sin\phi,\,\cos\phi).
\label{eq:84tran2}
\ee
Thus $\tran ab\cdot(z,w)=(z,w)$ is a necessary condition for $(z,w)\in\mbox{\rm
Fix}(\Sigma)$, and this condition implies
\beq
e^{-i(l_1a+n_1b)}z_1&=&z_1\label{eq:84fix1}\\
e^{-i(n_1a-l_1b)}z_4&=&z_4.\label{eq:84fix2}
\eeq
If $z_1\neq 0$, then for (\ref{eq:84tran1}) - (\ref{eq:84tran2}) the  $z_1$
equation requires
\beq
l_1\cos\phi+n_1\sin\phi& =&\mbox{\rm integer}\\
-l_1\sin\phi+n_1\cos\phi& =&\mbox{\rm integer},
\eeq
or equivalently $\rot{-\phi}\cdot\vec{q_1}\in\tilde{A}(\k)$
which is impossible since $\phi\neq 2\theta_1$. Hence $(z,w)\in\mbox{\rm
Fix}(\Sigma)$ requires $z_1=0$. A similar discussion of (\ref{eq:84fix2}) shows
that $z_4=0$. {\bf $\Box$} \end{quote}

When $\Ft$ is obtained by restriction from $\strobe$ there are additional
conditions on $f$ and $h$.
\begin{prop}
For a $\semi$-equivariant vector field $\Ft=[f,h]$, the condition
\be
\rot {\phi}\cdot\Ft(\Phi)=\Ft(\rot {\phi}\cdot\Phi)\label{eq:84rotcond0}
\ee
for all $\Phi\in E^c(\ext)\,\cap\,[\rot{-\phi}\,E^c(\ext)]$ is satisfied if and
only if
\be
f(z_1,0,0,z_4,w_1,w_2)=h(0,\overline{w}_2,w_1,0,z_1,\overline{z}_4),
\label{eq:84rotcond}
\ee
or equivalently
\be
h(0,z_2,z_3,0,w_1,w_2)=f(w_1,0,0,\overline{w}_2,z_3,\overline{z}_2).
\label{eq:84rotcondeqv}
\ee
\end{prop}
\noindent {\em {\bf Proof}.}\begin{quote}The proof follows that of Proposition
III.4. {\bf $\Box$}\end{quote}

The restriction in (\ref{eq:Ftrestrict}) is accomplished by choosing
$a(\vec{k})$ in (\ref{eq:genef2}) so that $\Phi\in E^c(\ext)$:
\beq
a(\vec{k})&=&z_1\;\delta_{\vec{k},\vec{q_1}} +
\overline{z}_1\;\delta_{\vec{k},\vec{-q_1}} + z_2\;\delta_{\vec{k},\vec{q_2}} +
\overline{z}_2\;\delta_{\vec{k},\vec{-q_2}} + z_3\;\delta_{\vec{k},\vec{q_3}} +
\overline{z}_3\;\delta_{\vec{k},\vec{-q_3}} + z_4\;\delta_{\vec{k},\vec{q_4}} +
\overline{z}_4\;\delta_{\vec{k},\vec{-q_4}}\nonumber\\
&&\hspace{0.0in} + w_1\;\delta_{\vec{k},\vec{p_1}} +
\overline{w}_1\;\delta_{\vec{k},\vec{-p_1}}  + w_2\;\delta_{\vec{k},\vec{p_2}}
+ \overline{w}_2\;\delta_{\vec{k},\vec{-p_2}}.\label{eq:84restrict}
\eeq
This determines a $\semi$-equivariant vector field $\Ft=[f,h]$ on $E^c(\ext)$
such
that $f$ and $h$ are homogeneous functions of degree p; when $f$ and $h$ are
determined by monomials as in (\ref{eq:84fmon}) and (\ref{eq:84hmon}) then
\beq
m(z,w)&=&\sum_{\vec{k_1}\in \tilde{A}(\k)}
\ldots\sum_{\vec{k_p}\in\tilde{A}(\k)} a(\vec{k_1})\,\ldots
a(\vec{k_p})\,P(\vec{q_1},\vec{k_1},\ldots, \vec{k_p})\label{eq:84fmona}\\
&&\nonumber\\
m'(z,w)+m'(\g 2\cdot(z,w))&=&\sum_{\vec{k_1}\in\tilde{A}(\k)}
\ldots\sum_{\vec{k_p}\in \tilde{A}(\k)} a(\vec{k_1})\,\ldots
a(\vec{k_p})\,P(\vec{p_1},\vec{k_1},\ldots, \vec{k_p}).\label{eq:84hmonb}
\eeq
Given monomials $m$ and $m'$ in (\ref{eq:84fmon}) and (\ref{eq:84hmon}), the
essential question is whether there exists a function $P$, satisfying
conditions (\ref{eq:irrcond1})-(\ref{eq:irrcond2}), such that
(\ref{eq:84fmona}) and (\ref{eq:84hmonb}) hold. We answer this by constructing
$P$; the procedure is the same as in the [4,8] mode interaction.

The inner products amongst the vectors in $\tilde{A}(\k)$ are used to specify
subsets of $\tilde{A}(\k)$. Given a vector $\vec{c_i}\in\tilde{A}(\k)$ denote
the reflection in $\euc$ that fixes $\vec{c_i}$ by $\gamma_{\vec{c_i}}$, i.e.
$\gamma_{\vec{c_i}}\cdot\vec{c_i}=\vec{c_i}.$

For example, $\gamma_{\vec{p_1}}=\g 2$, $\gamma_{\vec{p_2}}=\g 2\gt 1$, and
$\gamma_{\vec{q_1}}=\rot{-\phi}\g 2\rot{\phi}.$ Thus $\gamma_{\vec{p_1}}$ and
$\gamma_{\vec{p_2}}$ leave $\tilde{A}(\k)$ invariant, but for ${\vec{q_i}},
\;i=1,2,3,4$ we have:
\beq
\tilde{A}(\k)\cap[\gamma_{\vec{q_1}}\tilde{A}(\k)]&=&
\tilde{A}(\k)\cap[\gamma_{\vec{q_4}}\tilde{A}(\k)]=
\{\pm\vec{q_1},\pm\vec{q_4}\}\label{eq:84insect1}\\
\tilde{A}(\k)\cap[\gamma_{\vec{q_2}}\tilde{A}(\k)]&=&
\tilde{A}(\k)\cap[\gamma_{\vec{q_3}}\tilde{A}(\k)]=
\{\pm\vec{q_2},\pm\vec{q_3}\}.\label{eq:84insect2}
\eeq
\begin{prop} Let $\{\vec{c_1},\vec{c_2},\ldots,\vec{c_n}\}$ denote a fixed
subset of $\tilde{A}(\k)$. Given $(\vec{k},\vec{k'})$, let
$\{\vec{k_1},\vec{k_2},\ldots,\vec{k_n}\}$ denote a second set of $n$ vectors
in $\tilde{A}(\k)$ required to satisfy the conditions:
\beq
\vec{k}\cdot\vec{k_i}&=&\vec{k'}\cdot\vec{c_i}\;\;\hspace{0.25in} \mbox{\rm for
}i=1,\ldots n\label{eq:84cinner1}\\
\vec{k_i}\cdot\vec{k_j}&=&\vec{c_i}\cdot\vec{c_j}\;\;\hspace{0.25in} \mbox{\rm
for }i,j=1,\ldots n.\label{eq:84cinner2}
\eeq
\begin{enumerate}
\item If $(\vec{k},\vec{k'})=(\vec{p_1},\vec{p_1})$, then the conditions
\mbox{\rm (\ref{eq:84cinner1})-(\ref{eq:84cinner2})} have only two (not
necessarily distinct) solutions:
\be
\{\vec{k_1},\vec{k_2},\ldots,\vec{k_n}\}=
\{\vec{c_1},\vec{c_2},\ldots,\vec{c_n}\}\label{eq:84ppset1}
\ee
and
\be
\{\vec{k_1},\vec{k_2},\ldots,\vec{k_n}\}= \gamma_{\vec{p_1}}
\cdot\{\vec{c_1},\vec{c_2},\ldots,\vec{c_n}\}.\label{eq:84ppset2}
\ee
\item If $(\vec{k},\vec{k'})=(\vec{q_1},\vec{q_1})$, then the conditions
\mbox{\rm (\ref{eq:84cinner1})-(\ref{eq:84cinner2})} can be satisfied in only
one way:
\be
\{\vec{k_1},\vec{k_2},\ldots,\vec{k_n}\}=
\{\vec{c_1},\vec{c_2},\ldots,\vec{c_n}\}\label{eq:84qqset1}
\ee
unless
\be
\{\vec{c_1},\vec{c_2},\ldots,\vec{c_n}\}\subseteq
\tilde{A}(\k)\cap[\gamma_{\vec{q_1}}\tilde{A}(\k)] \label{eq:84insect3}
\ee
in which case there is a second (not necessarily distinct) solution:
\be
\{\vec{k_1},\vec{k_2},\ldots,\vec{k_n}\}= \gamma_{\vec{q_1}}
\cdot\{\vec{c_1},\vec{c_2},\ldots,\vec{c_n}\}.\label{eq:84qqset2}
\ee
\item If $(\vec{k},\vec{k'})=(\vec{q_1},\vec{p_1})$, then the conditions
\mbox{\rm (\ref{eq:84cinner1})-(\ref{eq:84cinner2})} have no solutions unless
\be
\{\vec{c_1},\vec{c_2},\ldots,\vec{c_n}\}\subseteq
\tilde{A}(\k)\cap[\rot{\phi}\tilde{A}(\k)],\label{eq:84qpcond1}
\ee
or
\be
\{\vec{c_1},\vec{c_2},\ldots,\vec{c_n}\}\subseteq
\tilde{A}(\k)\cap[\gamma_{\vec{p_1}}\rot{\phi}\tilde{A}(\k)].
\label{eq:84qpcond2}
\ee
When \mbox{\rm (\ref{eq:84qpcond1})} holds then
\be
\{\vec{k_1},\vec{k_2},\ldots,\vec{k_n}\}=
\rot{-\phi}\{\vec{c_1},\vec{c_2},\ldots,\vec{c_n}\}\label{eq:84qpset1}
\ee
and when \mbox{\rm (\ref{eq:84qpcond2})} holds then
\be
\{\vec{k_1},\vec{k_2},\ldots,\vec{k_n}\}= \rot{-\phi}\gamma_{\vec{p_1}} \cdot
\{\vec{c_1},\vec{c_2},\ldots,\vec{c_n}\}.\label{eq:84qpset2}
\ee
\item If $(\vec{k},\vec{k'})=(\vec{p_1},\vec{q_1})$, then the conditions
\mbox{\rm (\ref{eq:88cinner1})-(\ref{eq:88cinner2})} have no solutions unless
\be
\{\vec{c_1},\vec{c_2},\ldots,\vec{c_n}\}\subseteq
\tilde{A}(\k)\cap[\rot{-\phi}\tilde{A}(\k)],\label{eq:84qpcond3}
\ee
or
\be
\{\vec{c_1},\vec{c_2},\ldots,\vec{c_n}\}\subseteq
\tilde{A}(\k)\cap[\gamma_{\vec{q_1}}\rot{-\phi}\tilde{A}(\k)].
\label{eq:84qpcond4}
\ee
When \mbox{\rm (\ref{eq:84qpcond3})} holds then
\be
\{\vec{k_1},\vec{k_2},\ldots,\vec{k_n}\}=
\rot{\phi}\{\vec{c_1},\vec{c_2},\ldots,\vec{c_n}\}\label{eq:84qpset3}
\ee
and when \mbox{\rm (\ref{eq:84qpcond4})} holds then
\be
\{\vec{k_1},\vec{k_2},\ldots,\vec{k_n}\}= \rot{\phi}\gamma_{\vec{q_1}} \cdot
\{\vec{c_1},\vec{c_2},\ldots,\vec{c_n}\}.\label{eq:84qpset4}
\ee
\end{enumerate}
\end{prop}
\noindent {\em {\bf Proof}.}\begin{quote}The proof follows that of Proposition
III.5.  {\bf $\Box$}
\end{quote}

\begin{lemma}  Let $\overline{z}_1 m(z,w)$ and $\overline{w}_1 m'(z,w)$ be
$\tor$-invariant monomials of degree p such that $f(z,w)=m(z,w)$ and
$h(z,w)=m'(z,w)+m'(\g 2\cdot(z,w))$ meet the conditions of Proposition {\rm
III.11}. There exist functions $P_f$ and $P_h$ satisfying \mbox{\rm
(\ref{eq:irrcond1})-(\ref{eq:irrcond2})} such that
\beq
f(z,w)&=&\sum_{\vec{k_1}\in \tilde{A}(\k)}\sum_{\vec{k_2}\in \tilde{A}(\k)}
\ldots\sum_{\vec{k_p}\in\tilde{A}(\k)} a(\vec{k_1})\,a(\vec{k_2})\,\ldots
a(\vec{k_p})\,P_f(\vec{q_1},\vec{k_1},\ldots, \vec{k_p})\label{eq:84pff}\\
&&\nonumber\\
h(z,w)&=&\sum_{\vec{k_1}\in\tilde{A}(\k)}\sum_{\vec{k_2}\in \tilde{A}(\k)}
\ldots\sum_{\vec{k_p}\in \tilde{A}(\k)} a(\vec{k_1})\,a(\vec{k_2})\,\ldots
a(\vec{k_p})\,P_h(\vec{p_1},\vec{k_1},\ldots, \vec{k_p})\label{eq:84phh}
\eeq
and
\beq
\lefteqn{\sum_{\vec{k_1}\in\tilde{A}(\k)} \ldots\sum_{\vec{k_p}\in
\tilde{A}(\k)} a(\vec{k_1})\,\ldots
a(\vec{k_p})\,P_f(\vec{p_1},\vec{k_1},\ldots, \vec{k_p})}\nonumber\\
&=&\left\{
\begin{array}{cc}
f(w_1,0,0,{w}_2,0,0)&\mbox{\rm if}\;\; f(z_1,0,0,{z}_4,0,0)\neq0\\
&\\
f(w_1,0,0,\overline{w}_2,z_3,\overline{z}_2)&\mbox{\rm
if}\;\;f(z_1,0,0,{z}_4,0,0)=0\;\;
\mbox{\rm and}\\
\hspace{0.4in}  + f(w_1,0,0,{w}_2,z_1,\overline{z}_4)&
f(z_1,0,0,{z}_4,w_1,{w}_2)\neq0\\
&\\
0&\mbox{\rm otherwise}
\end{array}
\right.\label{eq:84pfh}
\eeq
\be
\sum_{\vec{k_1}\in\tilde{A}(\k)} \ldots\sum_{\vec{k_p}\in \tilde{A}(\k)}
a(\vec{k_1})\,\ldots a(\vec{k_p})\,P_h(\vec{q_1},\vec{k_1},\ldots, \vec{k_p})
=h(w_1,0,0,{w}_2,z_1,{z}_4)\label{eq:84phf}
\ee
where $a(\vec{k})$ is given by \mbox{\rm (\ref{eq:84restrict})}.
\end{lemma}
\noindent {\em {\bf Proof}}\begin{quote}The proof follows that of Lemma III.1.
\begin{enumerate}
\item Let $\{\vec{c_1},\vec{c_2},\ldots,\vec{c_n}\}$ denote the wave vector set
associated with the monomial $m(z,w)$ used to define $f(z,w)$. Then $m(z,w)$
can be written as
\be
m(z,w)=a(\vec{c_1})\,a(\vec{c_2})\ldots a(\vec{c_{n}})a(\vec{c}_{l_{n+1}})
\ldots a(\vec{c}_{l_{p}})\label{eq:84mb}
\ee
in analogy with (\ref{eq:mb}).
For $P_f$ we introduce the numerical factor
\be
\epsilon=\left\{\begin{array}{cc}
\frac{1}{2}&\mbox{\rm if}\;\;\{\vec{c_1},\ldots,\vec{c_n}\}\subseteq
\tilde{A}(\k)\cap[\gamma_{\vec{q_1}}\tilde{A}(\k)]\;\; \mbox{\rm and}\\
& \{\vec{c_1},\ldots,\vec{c_n}\}\neq
\gamma_{\vec{q_1}}\{\vec{c_1},\ldots,\vec{c_n}\}\\
&\\
1&\mbox{\rm otherwise},
\end{array}
\right.
\ee
and define $P_f$ as
\be
P_f(\vec{k},\vec{k_1},\ldots, \vec{k_p})=\epsilon\left(\prod_{i=1}^{n}
\prod_{j=1}^{n}\,\delta_{\vec{k}\cdot\vec{k_i},\vec{q_1}\cdot\vec{c_i}}
\;\delta_{\vec{k_i}\cdot\vec{k_j},\vec{c_i}\cdot\vec{c_j}}\right)\,
\left(\prod_{j=n+1}^{p}\, \delta_{\vec{k_j}\cdot\vec{c}_{l_j},\k^2}\right).
\label{eq:84pfdef1}
\ee
{}From Proposition III.15 we then have
\be
P_f(\vec{q_1},\vec{k_1},\ldots,\vec{k_p})=\left\{\begin{array}{cc}
{\epsilon}&\mbox{\rm if}\;\; \{\vec{k_1},\ldots,\vec{k_p}\}=
\{\vec{c_1},\ldots,\vec{c_n},\vec{c}_{l_{n+1}},\ldots, \vec{c}_{l_{p}}\}\\
&\\
{\epsilon}&\mbox{\rm if}\;\;\{\vec{c_1},\ldots,\vec{c_n}\}\subseteq
\tilde{A}(\k)\cap[\gamma_{\vec{q_1}}\tilde{A}(\k)] \;\;\mbox{\rm and}\\
&\{\vec{k_1},\ldots,\vec{k_p}\}=
\gamma_{\vec{q_1}}\cdot\{\vec{c_1},\ldots,\vec{c_n},\vec{c}_{l_{n+1}},\ldots,
\vec{c}_{l_{p}}\}\\
&\\
0&\mbox{\rm otherwise};
\end{array}\right.\label{eq:84pfp1b}
\ee
and
\be
P_f(\vec{p_1},\vec{k_1},\ldots,\vec{k_p})=\left\{\begin{array}{cc}
{\epsilon}&\mbox{\rm if}\;\; \{\vec{c_1},\ldots,\vec{c_n}\}\subseteq
\tilde{A}(\k)\cap[\rot{-\phi}\tilde{A}(\k)]\;{\rm and}\\
&\{\vec{k_1},\ldots,\vec{k_p}\}= \rot{\phi}
\cdot\{\vec{c_1},\ldots,\vec{c_n},\vec{c}_{l_{n+1}},\ldots, \vec{c}_{l_{p}}\}\\
&\\
{\epsilon}&\mbox{\rm if}\;\;\{\vec{c_1},\ldots,\vec{c_n}\}\subseteq
\tilde{A}(\k)\cap[\gamma_{\vec{q_1}}\rot{-\phi}\tilde{A}(\k)]\; {\rm and}\\
& \{\vec{k_1},\ldots,\vec{k_p}\}= \rot{\phi}\gamma_{\vec{q_1}}\cdot
\{\vec{c_1},\ldots,\vec{c_n},\vec{c}_{l_{n+1}},\ldots, \vec{c}_{l_{p}}\}\\
&\\
0&\mbox{\rm otherwise};
\end{array}\right.\label{eq:84pfpq1}
\ee
The calculations to verify (\ref{eq:84pff}) and (\ref{eq:84pfh}) are now
straightforward.

\item Let $\{\vec{c_1},\vec{c_2},\ldots,\vec{c_n}\}$ denote the wave vector set
associated with the monomial $m'(z,w)$ used to define $h(z,w)$. Then $m'(z,w)$
can be written as
\be
m'(z,w)=a(\vec{c_1})\,a(\vec{c_2})\ldots a(\vec{c_{n}})a(\vec{c}_{l_{n+1}})
\ldots a(\vec{c}_{l_{p}}).\label{eq:84mprimeb}
\ee
$P_h$ is defined as
\be
P_h(\vec{k},\vec{k_1},\ldots, \vec{k_p})=\left(\prod_{i=1}^{n}
\prod_{j=1}^{n}\,\delta_{\vec{k}\cdot\vec{k_i},\vec{p_1}\cdot\vec{c_i}}
\;\delta_{\vec{k_i}\cdot\vec{k_j},\vec{c_i}\cdot\vec{c_j}}\right)\,
\left(\prod_{j=n+1}^{p}\, \delta_{\vec{k_j}\cdot\vec{c}_{l_j},\k^2}\right).
\label{eq:84phdef1}
\ee
The verification of (\ref{eq:84phh}) and (\ref{eq:84phf}) proceeds as for
$P_f$.
\end{enumerate}
{\bf $\Box$} \end{quote}

\begin{lemma}  Let $\overline{z}_1 m(z,w)$ and $\overline{w}_1 m'(z,w)$ be
$\tor$-invariant monomials of degree p such that $f(z,w)=m(z,w)$ and
$h(z,w)=m'(z,w)+m'(\g 2\cdot(z,w))$ meet the conditions of Proposition {\rm
III.11}. There exists a single function $P$ satisfying \mbox{\rm
(\ref{eq:irrcond1})-(\ref{eq:irrcond2})} such that
\beq
f(z,w)&=&\sum_{\vec{k_1}\in \tilde{A}(\k)}\sum_{\vec{k_2}\in \tilde{A}(\k)}
\ldots\sum_{\vec{k_p}\in\tilde{A}(\k)} a(\vec{k_1})\,a(\vec{k_2})\,\ldots
a(\vec{k_p})\,P(\vec{q_1},\vec{k_1},\ldots, \vec{k_p})\label{eq:84pf}\\
&&\nonumber\\
h(z,w)&=&\sum_{\vec{k_1}\in\tilde{A}(\k)}\sum_{\vec{k_2}\in \tilde{A}(\k)}
\ldots\sum_{\vec{k_p}\in \tilde{A}(\k)} a(\vec{k_1})\,a(\vec{k_2})\,\ldots
a(\vec{k_p})\,P(\vec{p_1},\vec{k_1},\ldots, \vec{k_p})\label{eq:84ph}
\eeq
if and only if
\be
f(z_1,0,0,z_4,w_1,w_2)=
h(0,\overline{w}_2,w_1,0,z_1,\overline{z}_4).\label{eq:84rotconda}
\ee
\end{lemma}
\noindent {\em {\bf Proof}.}\begin{quote} The proof follows that for Lemma
II.2.
The necessity of (\ref{eq:84rotconda}) is established by Proposition {\rm
III.14} and the proof of necessity in Theorem {\rm III.1}. Assume that
(\ref{eq:84rotconda}) holds and let $P_f(\vec{k},\vec{k_1},\ldots, \vec{k_p})$
and $P_h(\vec{k},\vec{k_1},\ldots, \vec{k_p})$ be the functions of Lemma {\rm
III.7} for $f$ and $h$ respectively.
\begin{enumerate}
\item If $f(z_1,0,0,z_4,w_1,w_2)=0$ then define
\be
P(\vec{k},\vec{k_1},\ldots, \vec{k_p})\equiv P_f(\vec{k},\vec{k_1},\ldots,
\vec{k_p}) + P_h(\vec{k},\vec{k_1},\ldots, \vec{k_p}).
\ee
The equations (\ref{eq:84pf}) and (\ref{eq:84ph}) follow from the previous
Lemma.

\item If $f(z_1,0,0,z_4,w_1,w_2)\neq0$ or then define
\be
P(\vec{k},\vec{k_1},\ldots, \vec{k_p})\equiv P_f(\vec{k},\vec{k_1},\ldots,
\vec{k_p}).
\ee
Again (\ref{eq:84pf}) and (\ref{eq:84ph}) can be verified using  the previous
Lemma.{\bf $\Box$}
\end{enumerate}
\end{quote}

For this mode interaction, these results imply the analogue of Lemmas III.3 and
III.6.
\begin{lemma}
For a homogeneous vector field $\Ft$, equivariant with respect to the
representation of $\semi$ in \mbox{\rm (\ref{eq:84repa})-(\ref{eq:84repc})},
there exists a $\euc$-symmetric vector field $\strobe$ on $E^c(\rtwo)$ such
that
\be
\Ft=\strobe|_{E^c(\ext)}\label{eq:84Ftrestrict}
\ee
if and only if for all $\rot\theta\in\euc$
\be
\rot {\theta}\cdot\Ft(\Phi)=\Ft(\rot {\theta}\cdot\Phi)\label{eq:84iff}
\ee
for all $\Phi\in E^c(\ext)\,\cap\,[\rot{-\theta}\,E^c(\ext)]$.
\end{lemma}

\section{\hspace{0.125in}Binary Mode interactions with $\crit{\rtwo}$
reducible}

For these mode interactions we still assume the decomposition
\be
\crit{\ext}=\irreds{1}\oplus\irreds{2},\label{eq:modeintred}
\ee
but the irreducible form $E^c(\rtwo)=\eig\k\rtwo$ is
replaced by
\be
\crit{\rtwo}=\eig{\k_1}{\rtwo}\oplus\eig{\k_2}{\rtwo}\label{eq:e2modeint}
\ee
where $\irreds{i}\subset\eig{\k_i}{\rtwo}$.
No relationship is assumed between $\k_1^2=l_1^2+n_1^2$ and
$\k_2^2=l_2^2+n_2^2$; they may be equal or unequal. In (\ref{eq:modeintred})
any pair of irreducible representations is allowed; in particular [4,4] mode
interactions are now possible.

We  denote the wave vector sets $\tilde{A}(\k_1)=
\{\pm\vec{q_1},\ldots,\pm\vec{q}_{j_1}\}$ and
$\tilde{A}(\k_2)=\{\pm\vec{p_1},\ldots,\pm\vec{p}_{j_2}\}$  as before,
and write the $\semi$ representations in the usual way
\beq
\irreds {1}&=&\{z_1\gef{\vec{q_1}}(\vec{r}) +\cdots+
z_{j_1}\gef{\vec{q_{j_1}}}(\vec{r}) + cc)|\;(z_1,\ldots,z_{j_1})\in {\bf
C}^{j_1}\},\label{eq:irrede2a}\\
\irreds {2}&=&\{(w_1\gef{\vec{p_1}}(\vec{r}) +\cdots+
w_{j_2}\gef{\vec{p_{j_2}}}(\vec{r})+ cc)|\;(w_1,\ldots,w_{j_2})\in {\bf
C}^{j_2}\}.\label{eq:irrede2b}
\eeq
In these expressions $j_i=(\mbox{\rm dim}\,\irreds {i})/2$.

As in the $\euc$-irreducible examples, a vector field $\Ft$, equivariant with
respect to the representation (\ref{eq:modeintred}), is uniquely specified by
giving two functions
$\Ft=[f,h]$
where $f(z,w)$ is the $z_1$ component of $\Ft$ and determines the other
$\irreds {1}$ components and $h(z,w)$ is the $w_1$ component of $\Ft$ and
determines the other $\irreds {2}$ components. The specific properties required
of $f$ and $h$ depend on the representations considered.

There are several ways that this type of mode interaction arises in
applications. Since all the irreducible representations of $\semi$ are
absolutely irreducible, a theorem of Golubitsky and Stewart implies that the
critical eigenvalues can be imaginary only if the representations in
(\ref{eq:modeintred}) are equivalent.\cite{gss} Thus Hopf bifurcation with
$\semi$ symmetry leads to mode interactions of this type. The Hopf bifurcation
for the $(l_1,n_1)=(l,0)$ and  $(l_2,n_2)=(l,0)$ mode interaction has been
analyzed by Swift, Pismen, and Silber and Knobloch.\cite{swift}-\cite{sk} One
subtlety in the case of imaginary eigenvalues is that the eigenvectors are
complex linear combinations of the vectors $\gef{\vec{k}}(\vec{r})$ in
(\ref{eq:irrede2a})-(\ref{eq:irrede2b}); see Silber and Knobloch for an
example.\cite{sk} For real eigenvalues, the mode interaction with equivalent
representations occurs at the Takens-Bogdanov bifurcation when the Hopf
frequency vanishes.  If $\irreds{1}{}$  and $\irreds{2}{}$ are inequivalent,
then the critical eigenvalues must be real. This situation arises naturally in
codimension-two mode interactions. For example these bifurcations may be seen
in the parameter space of the Faraday experiment with square
geometry.\cite{cgl,sim}

In light of the reducible structure of $\crit{\rtwo}$ in (\ref{eq:e2modeint}),
there are no hidden rotations acting on $\crit{\ext}$ to connect the two
$\semi$ representations. Consequently for all mode interactions of this type,
the $\euc$ symmetry of $\strobe$ imposes no requirements on the normal form
$\Ft$ beyond the constraints of $\semi$-equivariance. Recall our notation
(\ref{eq:strobe})-(\ref{eq:akprime}) for the $\euc$-reducible case; $\strobe$
is given by
$\strobe=\strobec 1+\strobec 2$
where
$\strobec i:\crit{\rtwo}\rightarrow\eig{\k_i}{\rtwo}$
for $i=1,2$. We assume $\strobe$ is homogeneous of degree p. For a vector in
$\crit{\rtwo}$,
\be
\Phi(\vec{r})=\sum_{i=1}^2\sum_{\vec{k}\in A(\k_i)}
a_i(\vec{k})\,\gef{\vec{k}}(\vec{r}),\label{eq:redgenef}
\ee
the components $\strobe_i$ must have the form
\be
\strobe_i (\Phi)=\sum_{\vec{k}\in
A(\k_i)}\,a_i'(\vec{k})\,\gef{\vec{k}}(\vec{r})\label{eq:pnonlinred}
\ee
where
\be
a_i'(\vec{k})=\sum_{l_1\leq\ldots\leq l_p=1,2}
\left[\sum_{\vec{k_1}\in A(\k_{l_1})}\cdots\sum_{\vec{k_p}\in A(\k_{l_p})}
a_{l_1}(\vec{k_1})\ldots a_{l_p}(\vec{k_p})\,P_i(\vec{k},\vec{k_1},\ldots,
\vec{k_p})\right]. \label{eq:akprimered}
\ee
Note that for $P_i(\vec{k},\vec{k_1},\ldots, \vec{k_p})$ the argument $\vec{k}$
only takes values in $A(\k_i)$.

The restriction of $\strobe$ to $\crit{\ext}$
\be
\Ft=\strobe|_{E^c(\ext)}
\ee
is accomplished by specifying $a_1(\vec{k})$ and $a_2(\vec{k})$ appropriately
in (\ref{eq:redgenef}):
\beq
a_1(\vec{k})&=&z_1\;\delta_{\vec{k},\vec{q_1}} +
\overline{z}_1\;\delta_{\vec{k},\vec{-q_1}} +\cdots+
z_{j_1}\;\delta_{\vec{k},\vec{q_{j_1}}} +
\overline{z}_{j_1}\;\delta_{\vec{k},\vec{-q_{j_1}}}\\
a_2(\vec{k})&=&w_1\;\delta_{\vec{k},\vec{p_1}} +
\overline{w}_1\;\delta_{\vec{k},\vec{-p_1}} +\cdots+
w_{j_2}\;\delta_{\vec{k},\vec{p_{j_2}}} +
\overline{w}_{j_2}\;\delta_{\vec{k},\vec{-p_{j_2}}}.
\eeq
The $\semi$-equivariant vector field $\Ft=[f,h]$ on $E^c(\ext)$ defined in this
way is determined by
\beq
f(z,w)&=&\sum_{l_1\leq\ldots\leq l_p=1,2}
\left[\sum_{\vec{k_1}\in \tilde{A}(\k_{l_1})}\ldots\sum_{\vec{k_p}\in
\tilde{A}(\k_{l_p})}  a_{l_1}(\vec{k_1})\dots
a_{l_p}(\vec{k_p})\,P_1(\vec{q}_1,\vec{k_1},\ldots,
\vec{k_p})\right]\label{eq:e2redf}\\
&&\nonumber\\
h(z,w)&=&\sum_{l_1\leq\ldots\leq l_p=1,2}
\left[\sum_{\vec{k_1}\in \tilde{A}(\k_{l_1})}\ldots\sum_{\vec{k_p}\in
\tilde{A}(\k_{l_p})}  a_{l_1}(\vec{k_1})\dots
a_{l_p}(\vec{k_p})\,P_2(\vec{p}_1,\vec{k_1},\ldots,
\vec{k_p})\right].\label{eq:e2redh}
\eeq
The crucial point is that now, because $\euc$ is reducible, $f$ and $h$ are
determined from different functions $P_1$ and $P_2$, respectively.

\begin{theorem}  Assume $\crit{\ext}$ carries a reducible representation of
$\semi$ as in \mbox{\rm (\ref{eq:modeintred})}, and  $\crit{\rtwo}$ carries a
reducible representation of $\euc$ as in \mbox{\rm (\ref{eq:e2modeint})}. Let
$\Ft:{\crit{\ext}}\rightarrow{\crit{\ext}}$
denote a homogeneous $\semi$-equivariant vector field $\Ft=[f,h]$, then there
is a $\euc$-equivariant vector field
$\strobe:{\crit{\rtwo}}\rightarrow{\crit{\rtwo}}$
such that
\be
\Ft = \strobe |_{{\crit{\ext}}}.\label{eq:eucred}
\ee \end{theorem}
\noindent {\em {\bf Proof}.}
\begin{quote}The construction of functions $P_f(\vec{k},\vec{k_1},\ldots,
\vec{k_p})$ and $P_h(\vec{k},\vec{k_1},\ldots, \vec{k_p})$ satisfying
(\ref{eq:cond1}) - (\ref{eq:cond2})
and
\beq
\lefteqn{f(z,w)=}\nonumber\\
&&\sum_{l_1\leq\ldots\leq l_p=1,2}
\left[\sum_{\vec{k_1}\in \tilde{A}(\k_{l_1})}\ldots\sum_{\vec{k_p}\in
\tilde{A}(\k_{l_p})}  a_{l_1}(\vec{k_1})\dots
a_{l_p}(\vec{k_p})\,P_f(\vec{q}_1,\vec{k_1},\ldots, \vec{k_p})\right]\\
&&\nonumber\\
\lefteqn{h(z,w)=}\nonumber\\
&&\sum_{l_1\leq\ldots\leq l_p=1,2}
\left[\sum_{\vec{k_1}\in \tilde{A}(\k_{l_1})}\ldots\sum_{\vec{k_p}\in
\tilde{A}(\k_{l_p})}  a_{l_1}(\vec{k_1})\dots
a_{l_p}(\vec{k_p})\,P_h(\vec{p}_1,\vec{k_1},\ldots, \vec{k_p})\right]
\eeq
follows the same procedure as in the mode interactions with irreducible
$\crit{\ext}$. See Lemmas III.1, III.4, and III.7. Set
$P_1(\vec{k},\vec{k_1},\ldots, \vec{k_p})=P_f(\vec{k},\vec{k_1},\ldots,
\vec{k_p})$ and $P_2(\vec{k},\vec{k_1},\ldots,
\vec{k_p})=P_h(\vec{k},\vec{k_1},\ldots, \vec{k_p})$ and define
$\strobe=\strobec 1+\strobec 2$
where each $\strobec i$ is defined as in (\ref{eq:pnonlinred}) -
(\ref{eq:akprimered}). This gives an $\euc$-equivariant vector field which
restricts to $\Ft$ on $\crit{\ext}$.
{\bf $\Box$}\end{quote}

\section{\hspace{0.125in}Conclusions}
The analysis of binary mode interactions shows that the Euclidean symmetry of
the dynamical equations will constrain the $\semi$-symmetric normal form only
if the representation of $\euc$ is irreducible. This is our main qualitative
conclusion and it implies that the effects of hidden rotation symmetry on such
binary mode interactions are only important when the neutral eigenvalue is
real.
A practical result of this study are the explicit formulas indicating how the
standard $\semi$-symmetric normal form must be modified to reflect the hidden
rotations; these results are stated as constraints on the functions $[f,h]$ in
(\ref{eq:rotconda}), (\ref{eq:88rotconda}) - (\ref{eq:88rotcondb}), and
(\ref{eq:84rotconda}) for the $[4,8]$, $[8,8]$, and $[8,4]$ mode interactions
respectively. The approach used here for the binary mode interactions may be
used to construct normal forms for more complex interactions involving
additional representations of $\semi$ as in (\ref{eq:modeint}),
\be
\crit{\ext}=\irred{l_1}{n_1}\oplus\irred{l_2}{n_2}\cdots
\oplus\irred{l_j}{n_j}.\label{eq:modeintcon}
\ee
In this case $\Ft$ would have $j$ component functions and the hidden rotations
connecting the representations would induce relations between the functions.
For mode interactions (\ref{eq:modeintcon}) with reducible representations of
$\euc$,
\be
\crit{\rtwo}=\eig{\k_1}{\rtwo}\oplus\eig{\k_2}{\rtwo}\cdots
\oplus\eig{\k_\nu}{\rtwo},\label{eq:eucirrsum2con}
\ee
the analysis of Section IV generalizes immediately when $\nu=j$. Then there are
no hidden rotations connecting the representations of $\semi$ and no relations
between the $\nu$ component functions of $\Ft$. In the case $\nu<j$ then two or
more of the $\semi$ representations in (\ref{eq:modeintcon}) are contained in a
single $\euc$ representation of (\ref{eq:eucirrsum2con}). These representations
will be connected by hidden rotations and there will be constraints on the
corresponding component functions of $\Ft$.

An application of these results to construct normal forms for the bifurcation
of surface waves in the Faraday experiment will be published elsewhere. In this
experiment parametric forcing of a fluid layer excites waves through
period-doubling bifurcations.  Under suitable conditions the moving fluid
approximately obeys a Neumann boundary condition at the sidewalls and in square
containers this boundary condition allows the  corresponding fluid model to be
extended to a larger square with periodic boundary conditions and $\semi$
symmetry. Thus the hidden rotations are two steps removed from the actual
equations describing the fluid behavior, nevertheless their effects on the
normal form for the experiment may be quite significant.

\section{\hspace{0.125in}Appendix}
\subsection{Proof of Theorem II.1}

Let
\be
L\gef{\vec{k}}(\vec{r})=\sum_{\vec{k'}\in A(\k)}
c_{\vec{k}}(\vec{k'})\,\gef{\vec{k'}}(\vec{r})\label{eq:linop}
\ee
describe the action of $L$ on $\gef{\vec{k}}(\vec{r})$. When applied to
$\gef{\vec{k}}(\vec{r})$ the assumption that $L$ commutes with an arbitrary
translation ${\rm \cal T}_{\vec{p}}$ requires
$e^{-i\vec{k}\cdot\vec{p}}\, c_{\vec{k}}(\vec{k'})=
e^{-i\vec{k'}\cdot\vec{p}}\,c_{\vec{k}}(\vec{k'})$
for arbitrary $\vec{p}$. Thus $c_{\vec{k}}(\vec{k'})$ must vanish unless
$\vec{k}=\vec{k'}$:
$c_{\vec{k}}(\vec{k'})\equiv\delta_{\vec{k},\vec{k'}}\,C(\vec{k}).$
When $L$ is applied to a general element of $\eig{\k}{\rtwo}$, this
simplification implies
\beq
L\,\Phi(\vec{r})&=&\sum_{\vec{k'}\in A(\k)}
a(\vec{k'})\,L\,\gef{\vec{k'}}(\vec{r})\nonumber\\
&=&\sum_{\vec{k'}\in A(\k)} a(\vec{k'})\,C(\vec{k'})\,\gef{\vec{k'}}
\eeq
where $\Phi(\vec{r})$ is given by (\ref{eq:genef}). Acting alone, the remaining
generators $\rot\phi$ and $\g 2$ generate the subgroup $\otwo$, and the
assumption that $\gamma\,L\,\Phi(\vec{r})=L\,\gamma\,\Phi(\vec{r})$ for
$\gamma\in\otwo$ implies $C(\vec{k'})$ must be an $\otwo$ invariant function
$C(\gamma\cdot\vec{k'})=C(\vec{k'}).$
The $\otwo$ invariance of $C$ means that is only depends on the magnitude of
$\vec{k'}$ and hence is a $\k$-dependent constant
$C(\vec{k'})=\sigma(\k).$
Thus our linear operator is simply $L=\sigma(\k)\,I$.

\subsection{Proof of Theorem II.2}

The Euclidean group $\euc=\otwo\dot{+}T(2)$ is the semi-direct product of
$\otwo$ and the group of translations $T(2)$, and the conditions
(\ref{eq:cond1}) - (\ref{eq:cond2}) correspond to these two components.
\begin{enumerate}
\item The effect of an arbitrary translation ${\rm \cal T}_{\vec{x}}$ on
\be
\Phi=\sum_{i=1}^\nu\sum_{\vec{k}\in A(\k_i)}
a_i(\vec{k})\,\gef{\vec{k}}(\vec{r})\label{eq:rgenefapp}
\ee
is to replace $a_i(\vec{k})$ by $a_i(\vec{k})\,e^{-i\vec{k}\cdot\vec{x}}$. Thus
${\rm \cal T}_{\vec{x}}\cdot \strobe (\Phi)=\strobe ({\rm \cal
T}_{\vec{x}}\cdot\Phi)$ requires
\be
[e^{-i\vec{x}\cdot(\vec{k_1}+\vec{k_2}+\cdots+\vec{k_p}-\vec{k})}-1]\,P_i(\vec{k},\vec{k_1},\vec{k_2},\ldots,\vec{k_p})= 0
\ee
for $i=1,\ldots,\nu$ and arbitrary $\vec{x}$; hence if
$\vec{k}\neq\vec{k_1}+\vec{k_2}+\ldots+\vec{k_p}$, then
\be
P_i(\vec{k},\vec{k_1},\vec{k_2},\ldots,\vec{k_p})=0.
\ee

\item For $\gamma\in\otwo$, we evaluate $\gamma\cdot \strobe $ using the
invariance of the inner product
$\vec{k}\cdot\vec{r}=(\gamma\vec{k})\cdot(\gamma\vec{r})$
\beq
\gamma\cdot \strobe_i (\Phi)&=&\sum_{\vec{k}\in
A(\k_i)}\,\gef{\vec{k}}(\gamma^{-1}\vec{r})\;a_i'(\vec{k}')\nonumber\\
&=&\sum_{\vec{k}\in A(\k_i)}\,\gef{\vec{k}}(\vec{r})
\;a_i'(\gamma^{-1}\cdot\vec{k}'),
\eeq
where
\be
a_i'(\gamma^{-1}\cdot\vec{k})=\sum_{l_1\leq\ldots\leq l_p}
\left[\sum_{\vec{k'_1}\in A(\k_{l_1})}\sum_{\vec{k'_2}\in A(\k_{l_2})}
\ldots\sum_{\vec{k'_p}\in A(\k_{l_p})}
a_{l_1}(\vec{k'_1})\,a_{l_2}(\vec{k'_2})\,\ldots a_{l_p}(\vec{k'_p})\,
P_i(\gamma^{-1}\cdot\vec{k},\vec{k'_1},\ldots, \vec{k'_p})\right].
\label{eq:akprimea}
\ee
Now from
\beq
(\gamma\cdot\Phi)(\rperp)&=&\sum_{i=1}^\nu\; \sum_{\vec{k}\in A(\k_i)}
a_i(\vec{k})\,\gef{\vec{k}}(\gamma^{-1}\vec{r})\nonumber\\
&=&\sum_{i=1}^\nu\; \sum_{\vec{k}\in A(\k_i)}
a_i(\gamma^{-1}\cdot\vec{k})\,\gef{\vec{k}}(\vec{r}),
\eeq
we obtain
\be
\strobe_i (\gamma\cdot\Phi)=\sum_{\vec{k}\in A(\k_i)} a_i'(\vec{k})
\,\gef{\vec{k}}(\vec{r})
\ee
where
\beq
a_i'(\vec{k})&=&\sum_{l_1\leq\ldots\leq l_p}
\left[\sum_{\vec{k'_1}\in A(\k_{l_1})}\sum_{\vec{k'_2}\in A(\k_{l_2})}
\ldots\sum_{\vec{k'_p}\in A(\k_{l_p})} a_{l_1}(\gamma^{-1}\cdot\vec{k'_1})\,
\ldots a_{l_p}(\gamma^{-1}\cdot\vec{k'_p})\,
P_i(\vec{k},\vec{k'_1},\ldots,\vec{k'_p})\right]\nonumber\\
&&\nonumber\\
&=&\sum_{l_1\leq\ldots\leq l_p} \left[\sum_{\vec{k'_1}\in
A(\k_{l_1})}\ldots\sum_{\vec{k'_p}\in A(\k_{l_p})} a_{l_1}(\vec{k'_1})\ldots
a_{l_p}(\vec{k'_p})\, P_i(\vec{k},\gamma\cdot\vec{k'_1},\ldots,
\gamma\cdot\vec{k'_p})\right].\nonumber
\eeq
Thus $\gamma\cdot \strobe (\Phi)=\strobe (\gamma\cdot\Phi)$ requires that each
$P_i$ is an $\otwo$-invariant function:
\be
P_i(\gamma\cdot\vec{k},\gamma\cdot\vec{k_1},\gamma\cdot\vec{k_2},
\ldots,\gamma\cdot\vec{k_p})= P_i(\vec{k},\vec{k_1},\vec{k_2},\ldots,\vec{k_p})
\ee
for $i=1,\ldots,\nu$.

\item Note that for the reflection $\vec{r}\rightarrow-\vec{r}$, this gives
\be
P_i(-\vec{k},-\vec{k_1},-\vec{k_2}, \ldots,-\vec{k_p})=
P_i(\vec{k},\vec{k_1},\vec{k_2},\ldots,\vec{k_p});\label{eq:reflct}
\ee
if $\strobe$ is real-valued then $\strobe ^\ast=\strobe $ requires
\be
P_i(-\vec{k},-\vec{k_1},-\vec{k_2}, \ldots,-\vec{k_p})^\ast=
P_i(\vec{k},\vec{k_1},\vec{k_2},\ldots,\vec{k_p})
\ee
and the reflection symmetry (\ref{eq:reflct}) then implies $P_i$ is
real-valued.
\end{enumerate}

\subsection{Proof of Proposition III.1}

Let
\be
\Ft(z,w)=\left(
\begin{array}{c}
f_1(z,w)\\
\\
f_2(z,w)\\
\\
h_1(z,w)\\
\\
h_2(z,w)\\
\\
h_3(z,w)\\
\\
h_4(z,w)
\end{array}
\right)\label{eq:genmap}
\ee
denote an arbitrary vector field $\Ft:E^c(\ext)\rightarrow E^c(\ext)$. This
vector field will be $\semi$-equivariant if and only if it commutes with $\g
1$, $\g 2$ and $\tran ab$ for arbitrary $(a,b)$. The first condition $\g
1\cdot\Ft(z,w)=\Ft(\g 1\cdot(z,w))$ requires
\beq
\overline{f_1(z,w)}&=&f_1(\g 1\cdot(z,w))\label{eq:g11}\\
f_2(z,w)&=&f_2(\g 1\cdot(z,w))\label{eq:g12}\\
\overline{h_2(z,w)}&=&h_1(\g 1\cdot(z,w))\\
\overline{h_4(z,w)}&=&h_3(\g 1\cdot(z,w))
\eeq
so we can write $\Ft$ as
\be
\Ft(z,w)=\left(
\begin{array}{c}
f_1(z,w)\\
\\
f_2(z,w)\\
\\
h_1(z,w)\\
\\
\overline{h_1(\g 1\cdot(z,w))}\\
\\
h_3(z,w)\\
\\
\overline{h_3(\g 1\cdot(z,w))}
\end{array}
\right)\label{eq:g1map}
\ee
where $f_1$ and $f_2$ satisfy (\ref{eq:g11})-(\ref{eq:g12}). The second
condition $\g 2\cdot\Ft(z,w)=\Ft(\g 2\cdot(z,w))$ applied to (\ref{eq:g1map})
requires
\beq
f_2(z,w)&=&f_1(\g 2\cdot(z,w))\label{eq:g21}\\
h_3(z,w)&=&h_1(\g 2\cdot(z,w))\label{eq:g22}\\
h_3(\g 1\cdot(z,w))&=&\overline{h_1(\g 1\g 2\cdot(z,w))}.\label{eq:g23}
\eeq
By eliminating $f_2$ and $h_3$ using (\ref{eq:g21})-(\ref{eq:g22}), the
remaining conditions (\ref{eq:g11})-(\ref{eq:g12}) and (\ref{eq:g23}) become
\beq
\overline{f_1(z,w)}&=&f_1(\g 1\cdot(z,w))\label{eq:g11b}\\
f_1(\g 2\cdot(z,w))&=&f_1(\g 2\g 1\cdot(z,w))\label{eq:g12b}\\
h_1(\g 2\g 1\cdot(z,w))&=&\overline{h_1(\g 1\g 2\cdot(z,w))},\label{eq:g23b}
\eeq
and (\ref{eq:g1map}) becomes
\be
\Ft(z,w)=\left(
\begin{array}{c}
f_1(z,w)\\
\\
f_1(\g 2\cdot(z,w))\\
\\
h_1(z,w)\\
\\
\overline{h_1(\g 1\cdot(z,w))}\\
\\
h_1(\g 2\cdot(z,w))\\
\\
\overline{h_1(\g 2\g 1\cdot(z,w))}
\end{array}
\right).\label{eq:g1g2map}
\ee
Since $\g 3=\g 2\g 1\g 2$ the second condition (\ref{eq:g12b}) on $f_1$ is
equivalent to
\be
f_1(z,w)=f_1(\g 3\cdot(z,w));\label{eq:g12c}
\ee
thus $f_1$ is $\g 3$-invariant. Since $\g 1\g
3\cdot(z,w)=(\overline{z},\overline{w})$, using (\ref{eq:g12c}) the remaining
condition (\ref{eq:g11b}) on $f_1$ becomes
$\overline{f_1(z,w)}=f_1(\overline{z},\overline{w}).$
Similarly the remaining condition (\ref{eq:g23b}) on $h_1$ can be rewritten as
\be
\overline{h_1(z,w)}=h_1(\g 2\g 1\g 2\g 1\cdot(z,w))
=h_1(\overline{z},\overline{w}).\label{eq:g23c}
\ee
This property of $h_1$ allows the vector field (\ref{eq:g1g2map}) to be
re-expressed as
\be
\Ft(z,w)=\left(
\begin{array}{c}
f_1(z,w)\\
\\
f_1(\g 2\cdot(z,w))\\
\\
h_1(z,w)\\
\\
h_1(\g 3\cdot(z,w))\\
\\
h_1(\g 2\cdot(z,w))\\
\\
h_1(\g 2\g 1\cdot(z,w))
\end{array}
\right)\label{eq:g1g2mapb}
\ee
where the identity $\g 3\g 1\g 2\g 1=\g 2\g 1$ has been applied. This vector
field satisfies the requirements of $\{\g 1,\g 2\}$-equivariance and, after
dropping subscripts on $f$ and $h$, agrees with the form shown in
(\ref{eq:48map}). Finally the translation symmetry $\tran ab\cdot\Ft (z,w)=\Ft(
\tran ab\cdot(z,w))$ requires
\beq
e^{-il_1a}f_1(z,w)&=&f_1( \tran ab\cdot(z,w))\\
e^{-i(l_2a+n_2b)}h_1(z,w)&=&h_1( \tran ab\cdot(z,w)).
\eeq
This is equivalent to assuming that $\overline{z}_1 f_1(z,w)$ and
$\overline{w}_1 h_1(z,w)$ are $\tor$-invariant functions. This completes the
characterization of $\Ft$.

\subsection{Proof of Proposition III.2}

Any monomial
\be
M({\bf z})=z_1^{\alpha_1}{\overline{z}_1}^{\alpha'_1}
z_2^{\alpha_2}{\overline{z}_2}^{\alpha'_2}
w_1^{\beta_1}{\overline{w}_1}^{\beta'_1}
w_2^{\beta_2}{\overline{w}_2}^{\beta'_2}
w_3^{\beta_3}{\overline{w}_3}^{\beta'_3} w_4^{\beta_4}
{\overline{w}_4}^{\beta'_4},\label{eq:first}
\ee
can be written in the form (\ref{eq:t2mon}); this reduction only requires
extracting all factors of $|z_1|^2$, $|z_2|^2$, $|w_1|^2$, $|w_2|^2$,
$|w_3|^2$, and $|w_4|^2$. The result will be $\tor$-invariant however only if
the remaining factor $\Omega_1^{\mu_1}\Omega_2^{\mu_2}
\omega_1^{\nu_1}\omega_2^{\nu_2}\omega_3^{\nu_3}\omega_4^{\nu_4}$ is
$\tor$-invariant. From the representation (\ref{eq:48repc}), this requires the
conditions (\ref{eq:th1})-(\ref{eq:th2}).

\subsection{Proof of Proposition III.5}

\begin{enumerate}
\item For $(\vec{k},\vec{k'})=(\vec{p_1},\vec{p_1})$, the vectors
$\{\vec{k_1},\vec{k_2},\ldots,\vec{k_n}\}$ must satisfy
\beq
\vec{p_1}\cdot\vec{k_i}&=&\vec{p_1}\cdot\vec{c_i}\;\;\hspace{0.25in} \mbox{\rm
for }i=1,\ldots n\label{eq:pinner1}\\
\vec{k_i}\cdot\vec{k_j}&=&\vec{c_i}\cdot\vec{c_j}\;\;\hspace{0.25in} \mbox{\rm
for }i,j=1,\ldots n\label{eq:pinner2}
\eeq
and obviously the set (\ref{eq:pset1}) is a solution.
The $n$ conditions (\ref{eq:pinner1}) require that either $\vec{k_i}=\vec{c_i}$
or $\vec{k_i}=\gamma_{\vec{p_1}}\cdot\vec{c_i}$ where the reflection
$\gamma_{\vec{p_1}}$ is defined in (\ref{eq:ciref}). If
$\gamma_{\vec{p_1}}\cdot\vec{c_i}\not\in\tilde{A}(\k)$ for all $i$, then
(\ref{eq:pset1}) is the unique solution and there is nothing to prove. Assume
this is not the case. Note first that
$\gamma_{\vec{p_1}}\cdot\vec{c_i}\in\tilde{A}(\k)$ requires
$\vec{c_i}\in\tilde{A}(\k)\cap[\gamma_{\vec{p_1}}\tilde{A}(\k)]$ so we have
$\gamma_{\vec{p_1}}\cdot\vec{c_i}\in\tilde{A}(\k)$ for all $i$ only if
(\ref{eq:insect3}) holds. In this event the entire set of vectors
\be
\{\vec{k_1},\vec{k_2},\ldots,\vec{k_n}\}=
\gamma_{\vec{p_1}}\cdot\{\vec{c_1},\vec{c_2},\ldots,\vec{c_n}\}
\label{eq:badset2}
\ee
is contained in $\tilde{A}(\k)$ and provides a second solution. A third
solution distinct from (\ref{eq:pset1}) and (\ref{eq:pset2}) is not possible.
For such a solution we need to find two vectors $\vec{c_l}$ and $\vec{c_{l'}}$
in $\{\vec{c_1},\vec{c_2},\ldots,\vec{c_n}\}$ such that
$\vec{c_l}\neq\gamma_{\vec{p_1}}\cdot\vec{c_l}$,
$\vec{c_{l'}}\neq\gamma_{\vec{p_1}}\cdot\vec{c_{l'}}$, and
$\gamma_{\vec{p_1}}\cdot\vec{c_l}\in\tilde{A}(\k)$. From (\ref{eq:insect1})
these assumptions require $\vec{c_l}=\vec{p_4}$ or $\vec{c_l}=-\vec{p_4}$. Now
we try to modify the set (\ref{eq:pset1}) by substituting
$\vec{k_l}=\gamma_{\vec{p_1}}\cdot\vec{c_l}$ for $\vec{k_l}=\vec{c_l}$ but
leave $\vec{k_{l'}}=\vec{c_{l'}}$ unchanged. Condition (\ref{eq:pinner2}) then
implies
\be
(\gamma_{\vec{p_1}}\cdot\vec{k_{l'}})\cdot\vec{c_l}=
\vec{c_{l'}}\cdot\vec{c_{l}}.\label{eq:pinner2ll}
\ee
There are two choices in (\ref{eq:pinner2ll}):
$\gamma_{\vec{p_1}}\cdot\vec{k_{l'}}=\vec{c_{l'}}$, or
$\gamma_{\vec{p_1}}\cdot\vec{k_{l'}}=\gamma_{\vec{c_l}}\cdot\vec{c_{l'}}$.
The first choice means  $\vec{k_{l'}}=\gamma_{\vec{p_1}}\cdot\vec{c_{l'}}$
which contradicts our assumption that
$\vec{k_{l'}}=\vec{c_{l'}}\neq\gamma_{\vec{p_1}}\cdot\vec{c_{l'}}$. The second
choice means
$\vec{k_{l'}}=\gamma_{\vec{p_1}}\gamma_{\vec{c_l}}\cdot\vec{c_{l'}}$. This is
consistent with $\vec{k_{l'}}=\vec{c_{l'}}$ only if
\be
\vec{c_{l'}}=\gamma_{\vec{p_1}}\gamma_{\vec{c_l}}\cdot\vec{c_{l'}};
\label{eq:cl'}
\ee
however $\gamma_{\vec{c_l}}=\gamma_{\vec{p_4}}$ and
$\gamma_{\vec{p_1}}\gamma_{\vec{p_4}}\cdot\vec{k}=-\vec{k}$ so (\ref{eq:cl'})
cannot hold. Thus (\ref{eq:pinner2ll}) leads to a contradiction and the
attempted substitution fails; there are no additional solutions. This proves
the first assertion.

\item For the second assertion, recall that $\gamma_{\vec{q_1}}=\g 3$ so that
both (\ref{eq:set1}) and (\ref{eq:set2}) are subsets of $\tilde{A}(\k)$ and
clearly satisfy the required conditions
\beq
\vec{q_1}\cdot\vec{k_i}&=&\vec{q_1}\cdot\vec{c_i}\;\;\hspace{0.25in} \mbox{\rm
for }i=1,\ldots n\label{eq:inner1}\\
\vec{k_i}\cdot\vec{k_j}&=&\vec{c_i}\cdot\vec{c_j}\;\;\hspace{0.25in} \mbox{\rm
for }i,j=1,\ldots n.\label{eq:inner2}
\eeq
The first $n$ conditions (\ref{eq:inner1}) require that either
$\vec{k_i}=\vec{c_i}$ or $\vec{k_i}=\g 3\cdot\vec{c_i}$. If $\vec{c_i}=\g
3\cdot\vec{c_i}$ for all $i$ then there is nothing to prove. Suppose
$\vec{c_i}\neq\g 3\cdot\vec{c_i}$ for at least two values $i=l,l'$ which
requires that $\vec{c_l}\neq\pm\vec{q_1}$ and $\vec{c_{l'}}\neq\pm\vec{q_1}$.
We attempt to find a third solution by replacing $\vec{k_l}=\vec{c_l}$ in
(\ref{eq:set1}) by $\vec{k_l}=\g 3\cdot\vec{c_l}$ but leaving
$\vec{k_{l'}}=\vec{c_{l'}}$. Clearly (\ref{eq:inner1}) is still satisfied, but
for $(i,j)=(l,l')$ (\ref{eq:inner2}) now implies
\be
(\g 3\cdot\vec{k_{l'}})\cdot\vec{c_l}=\vec{c_{l'}}\cdot\vec{c_l}.
\label{eq:innerl}
\ee
This requires $\vec{k_{l'}}=\g 3\cdot\vec{c_{l'}}$ or $\vec{k_{l'}}=\g
3\gamma_{\vec{c_l}}\cdot\vec{c_{l'}}$ where the reflection $\gamma_{\vec{c_l}}$
is defined in (\ref{eq:ciref}); both possibilities lead to contradictions. The
first choice contradicts $\vec{k_{l'}}=\vec{c_{l'}}$ and the second choice is
only consistent if
\be
\vec{c_{l'}}=\g 3\gamma_{\vec{c_l}}\cdot\vec{c_{l'}}.\label{eq:l'}
\ee
Note that if $\vec{c_l}=\vec{q_2}$ or $\vec{c_l}=-\vec{q_2}$, then
$\gamma_{\vec{c_l}}=\gamma_{\vec{q_2}}=\gamma_1$ in (\ref{eq:l'}); since $\g
3\g 1\cdot\vec{k}=-\vec{k}$ this is a contradiction. Thus we must restrict
$\vec{c_l}$ to the values
\be
\vec{c_l}\in\{\pm\vec{p_1},\pm\vec{p_2},\pm\vec{p_3},\pm\vec{p_4}\}.
\label{eq:cl}
\ee
In addition (\ref{eq:l'}) implies
$\vec{c_{l'}}\in\tilde{A}(\k)\cap[\gamma_{\vec{c_l}}\tilde{A}(\k)].$
In light of (\ref{eq:cl}) and (\ref{eq:insect1})-(\ref{eq:insect2}), this
requires
\be
\vec{c_{l'}}\neq \pm\vec{q_1},\pm\vec{q_2},\label{eq:cl'2}
\ee
and implies that $\vec{c_{l'}}$ is either equal to $\pm\vec{c_{l}}$ or
$\vec{c_{l'}}$ is perpendicular to $\vec{c_{l}}$. The first possibility implies
$\vec{c_{l'}}=\g 3\cdot\vec{c_{l'}}$ from (\ref{eq:l'}) which is a
contradiction. The second possibility yields
$\gamma_{\vec{c_l}}\cdot\vec{c_{l'}}=-\vec{c_{l'}}$ so that (\ref{eq:l'})
implies $\vec{c_{l'}}=-\g 3\cdot\vec{c_{l'}}$; this in turn requires
$\vec{c_{l'}}$ to be equal to $\vec{q_2}$ or $-\vec{q_2}$ which contradicts
(\ref{eq:cl'2}). Thus condition (\ref{eq:innerl}) cannot be satisfied and the
sets (\ref{eq:set1}) and (\ref{eq:set2}) are the only solutions. This proves
the second assertion.
\item The third assertion can be reduced to the first case. Since
$\vec{q_1}=\rot{-\phi}\cdot\vec{p_1}$, the conditions
(\ref{eq:cinner1})-(\ref{eq:cinner2}) for
$(\vec{k},\vec{k'})=(\vec{q_1},\vec{p_1})$ can be re-written as
\beq
\vec{p_1}\cdot(\rot{\phi}\cdot\vec{k_i})&=&\vec{p_1}\cdot\vec{c_i}\;\;\hspace{0.25in} \mbox{\rm for }i=1,\ldots n\label{eq:qpinner1}\\
(\rot{\phi}\cdot\vec{k_i})\cdot(\rot{\phi}\cdot\vec{k_j})&=&\vec{c_i}\cdot\vec{c_j}\;\;\hspace{0.25in} \mbox{\rm for }i,j=1,\ldots n.\label{eq:qpinner2}
\eeq
Thus by our first assertion we have only two possibilities for solutions:
$\rot{\phi}\cdot\{\vec{k_1},\vec{k_2},\ldots,\vec{k_n}\}=
\{\vec{c_1},\vec{c_2},\ldots,\vec{c_n}\}$
and $\rot{\phi}\cdot\{\vec{k_1},\vec{k_2},\ldots,\vec{k_n}\}=
\gamma_{\vec{p_1}} \cdot\{\vec{c_1},\vec{c_2},\ldots,\vec{c_n}\}.$
Since $\rot{-\phi}\gamma_{\vec{p_1}}=\g 3\rot{-\phi}$, these possibilities can
be re-expressed as
\be
\{\vec{k_1},\vec{k_2},\ldots,\vec{k_n}\}=
\rot{-\phi}\cdot\{\vec{c_1},\vec{c_2},\ldots,\vec{c_n}\}\label{eq:qpset1''}
\ee
and
\be
\{\vec{k_1},\vec{k_2},\ldots,\vec{k_n}\}= \g 3\rot{-\phi}
\cdot\{\vec{c_1},\vec{c_2},\ldots,\vec{c_n}\}.\label{eq:qpset2''}
\ee
The final requirement that the vectors
$\{\vec{k_1},\vec{k_2},\ldots,\vec{k_n}\}$ belong to $\tilde{A}(\k)$ is only
met if condition (\ref{eq:qpcond}) holds.

\item The fourth assertion can be reduced to the second case. Since
$\vec{p_1}=\rot{\phi}\cdot\vec{q_1}$, the conditions
(\ref{eq:cinner1})-(\ref{eq:cinner2}) for
$(\vec{k},\vec{k'})=(\vec{p_1},\vec{q_1})$ can be re-written as
\beq
\vec{q_1}\cdot(\rot{-\phi}\cdot\vec{k_i})&=&\vec{q_1}\cdot\vec{c_i}\;\;\hspace{0.25in} \mbox{\rm for }i=1,\ldots n\label{eq:pqinner1}\\
(\rot{-\phi}\cdot\vec{k_i})\cdot\vec(\rot{-\phi}\cdot\vec{k_j})&=&\vec{c_i}\cdot\vec{c_j}\;\;\hspace{0.25in} \mbox{\rm for }i,j=1,\ldots n.\label{eq:pqinner2}
\eeq
Thus by our second assertion we have only two possibilities for solutions:
\be
\rot{-\phi}\cdot\{\vec{k_1},\vec{k_2},\ldots,\vec{k_n}\}=
\{\vec{c_1},\vec{c_2},\ldots,\vec{c_n}\}\label{eq:pqset1'}
\ee
and
\be
\rot{-\phi}\cdot\{\vec{k_1},\vec{k_2},\ldots,\vec{k_n}\}= \gamma_{\vec{q_1}}
\cdot\{\vec{c_1},\vec{c_2},\ldots,\vec{c_n}\}.\label{eq:pqset2'}
\ee
Since $\gamma_{\vec{q_1}}=\g 3$ and $\rot{\phi}\g 3=\g 3\rot{-\phi}$, these
possibilities can be re-expressed as
\be
\{\vec{k_1},\vec{k_2},\ldots,\vec{k_n}\}=
\rot{\phi}\cdot\{\vec{c_1},\vec{c_2},\ldots,\vec{c_n}\}\label{eq:pqset1''}
\ee
and
\be
\{\vec{k_1},\vec{k_2},\ldots,\vec{k_n}\}= \g 3\rot{-\phi}
\cdot\{\vec{c_1},\vec{c_2},\ldots,\vec{c_n}\}.\label{eq:pqset2''}
\ee
The additional requirement that the vectors
$\{\vec{k_1},\vec{k_2},\ldots,\vec{k_n}\}$ in (\ref{eq:pqset1''}) belong to
$\tilde{A}(\k)$ is only met if
$\{\vec{c_1},\ldots,\vec{c_n}\}\subseteq
\tilde{A}(\k)\cap[\rot{-\phi}\tilde{A}(\k)];$
this requirement is only satisfied in (\ref{eq:pqset2''}) if
$\{\vec{c_1},\ldots,\vec{c_n}\}\subseteq
\tilde{A}(\k)\cap[\rot{\phi}\tilde{A}(\k)].$

\end{enumerate}

\section{\hspace{0.125in}Acknowledgements}

I have enjoyed helpful conversations with I. Melbourne. This work  was
supported by the National Science Foundation under Grant DMS 9201028.
\newpage
\addtocontents{toc}
\newpage

\end{document}